\documentclass{article}
\usepackage[utf8]{inputenc}
\usepackage{authblk}
\usepackage[margin=1in]{geometry}
\usepackage{natbib}
\setcitestyle{numbers,square}

\usepackage{xcolor}

\usepackage{mathtools}
\usepackage{hyperref}
    
\newcommand\extrafootertext[1]{%
    \bgroup
    \renewcommand\thefootnote{\fnsymbol{footnote}}%
    \renewcommand\thempfootnote{\fnsymbol{mpfootnote}}%
    \footnotetext[0]{#1}%
    \egroup
}
    
\usepackage{cleveref}
\usepackage{graphicx}
\usepackage{caption}
\usepackage{subcaption}
\usepackage{enumerate}

\newcommand{\Binom}[2]{B\left(#1, #2\right)}
\newcommand{\Poisson}[1]{\text{Pois}\left(#1\right)}
\newcommand{\Normal}[2]{\mathcal{N}\left(#1, #2\right)}
\newcommand{\Gam}[2]{\mathcal{G}\left(#1, #2\right)}
\newcommand{\Beta}[2]{\text{Beta}\left(#1, #2\right)}

\newcommand{\pPhiDiv}{$\Phi\text{-Div}_P$}
\newcommand{\pPhiDivStar}{$\Phi\text{-Div}_P^*$}
\newcommand{\pMSE}{$\text{MSE}_P$}
\newcommand{\pAMSE}{$\text{AMSE}_P$}
\newcommand{\PGone}{$\textbf{PG}_1$}

\title{Measurement Integrity in Peer Prediction:\texorpdfstring{\\}{} A Peer Assessment Case Study}
\author{Noah Burrell}
\author{Grant Schoenebeck}
\affil{University of Michigan, Ann Arbor, USA.}
\date{September 2022}

\begin{document}

\maketitle
\thispagestyle{empty}
\setcounter{page}{0}

\begin{abstract}
We propose \textit{measurement integrity}, a property related to \textit{ex post} reward fairness, as a novel desideratum for peer prediction mechanisms in many natural applications.
Like \textit{robustness against strategic reporting}, the property that has been the primary focus of the peer prediction literature, measurement integrity is an important consideration for understanding the practical performance of peer prediction mechanisms.

We perform computational experiments, both with an agent-based model and with real data, to empirically evaluate peer prediction mechanisms according to both of these important properties. Our evaluations simulate the application of peer prediction mechanisms to peer assessment---a setting in which \textit{ex post} fairness concerns are particularly salient. We find that peer prediction mechanisms, as proposed in the literature, largely fail to demonstrate significant measurement integrity in our experiments. We also find that theoretical properties concerning robustness against strategic reporting are somewhat noisy predictors of empirical performance. Further, there is an apparent trade-off between our two dimensions of analysis. The best-performing mechanisms in terms of measurement integrity are highly susceptible to strategic reporting. Ultimately, however, we show that supplementing mechanisms with realistic parametric statistical models can, in some cases, improve performance along both dimensions of our analysis and result in mechanisms that strike the best balance between them.
\end{abstract} 

\extrafootertext{This work is supported by the National Science Foundation under award \#2007256. \vspace{1ex}}

\extrafootertext{The code for our experiments is hosted in the following GitHub repository:\\ 
\hspace*{2em} \url{https://github.com/burrelln/Measurement-Integrity-and-Peer-Assessment}}

\newpage

\section{Introduction}
\label{section:intro}
Peer prediction \cite{Miller2005}, or information elicitation without verification, is a paradigm for designing mechanisms that elicit reports from a population of agents about questions or tasks in settings where ground truth (and therefore the possibility of spot-checking) need not exist. One important dimension of evaluation for a peer prediction mechanism is the degree to which it rewards agents for their reports in a way that incentivizes truthfulness. This dimension, which we refer to as robustness against strategic reporting, has been the overwhelming focus of the theoretical peer prediction literature. We broaden this focus by introducing a new dimension of comparison, \textit{measurement integrity}.

In the peer prediction paradigm, we assume that agents receive a signal (perhaps at some cost) about each task, drawn from some joint prior distribution. In the current literature, mechanisms are typically characterized by two properties that attest to their robustness against strategic reporting:
\begin{enumerate}
    \item An equilibrium concept related to truthfulness that the mechanism induces under certain assumptions.
    
    \item The assumptions, which typically constrain the form of the joint prior distribution of signals for every agent, that are sufficient to ensure inducement of the equilibrium concept. 
\end{enumerate}

\Cref{appendix:concepts-and-assumptions} details these two properties for a representative selection of fundamental mechanisms from the peer prediction literature. However, these properties alone are insufficient for evaluating peer prediction mechanisms' suitability for a given application. Firstly, this characterization omits other important considerations. In many applications, for example, it is just as, if not more, important for rewards to be fair as to be incentive-compatible. Secondly, even for a particular setting where incentive-compatibility is a primary desiderata for peer prediction mechanisms, this characterization fails to determine the best mechanism to use. 

It is possible for a mechanism to induce a stronger equilibrium concept than another mechanism, but only under a stronger assumption. It is also possible for a mechanism to ``approximately'' induce a stronger equilibrium concept under a given assumption. In both cases, there is no clear answer to the question of which mechanism is more robust against strategic reporting. Considering secondary desiderata also discussed in the theoretical peer prediction literature does not help. Such properties, for example that mechanisms require little or no prior knowledge of the distribution of signals or that mechanisms only require simple reports from the agents, often fail to meaningfully differentiate the state-of-the-art mechanisms.

\subsection{Measurement Integrity}
\label{subsection:measurement-integrity}
In this work, we introduce and study a new property for peer prediction mechanisms in order to better understand how they operate in practice: \textit{measurement integrity}. Measurement integrity is the property that the rewards assigned by a mechanism can be interpreted as a measurement\footnote{See Stevens \cite{Stevens1946} for a discussion of the meaning of ``measurement'' that we invoke in the phrase ``measurement integrity.''} of the quality of the reports submitted by each agent in a given population. 

Fundamentally, measurement integrity concerns the nature of the mathematical relationship between the quality of an agent's reports and the payment they are awarded under a mechanism. That is, what information about report quality is encoded in a mechanism's payments? If agents whose report quality is better than that of the median agent consistently receive payments that are in the same direction relative to that of the median agent, then we can say the mechanism's payments measure report quality according to a \textit{nominal} scale---they classify the agents relative to the median. If the payments consistently rank agents according to the quality of their reports, then we can say that they measure report quality according to an \textit{ordinal} scale. If the magnitude of the payments contains some additional information about the (relative) magnitudes of the report quality, then we may even be able to say that a mechanism's payments measure report quality according to an \textit{interval} or \textit{ratio} scale. 

In general, the key principle is that a mechanism with high measurement integrity assigns rewards that reliably place the agents on some scale that quantifies the (relative) quality of their responses. However, the type of scale that is desirable (or achievable) and the appropriate definition for the quality of an agent's reports depends on the specific application of peer prediction. As a result, the general definition of measurement integrity as a property that we provide here is necessarily abstract. This is similar to the abstract property of \textit{robustness against strategic reporting}, which has been defined in many different ways, including the various equilibrium concepts discussed above. In this work, we consider \textit{peer assessment} (\Cref{section:peer-assessment}) as a case study in which to explore concrete, empirical notions of measurement integrity (\Cref{section:quantifying-mi}) and robustness against strategic reporting (\Cref{section:robustness}) and their quantification with respect to various peer prediction mechanisms.

Although measurement integrity is a somewhat abstract property, it also has practical implications, which make it a primary desideratum for mechanisms in many purported natural applications of peer prediction. It has fundamental relevance whenever fairness---in particular, \textit{ex post}\footnote{Specifically, \textit{ex post} here refers to the randomness of agents' strategies given their private signals, their uncertainty about other agents' signals, and the randomness of the mechanism given the agents' reports. In \Cref{appendix:alternative-quality}, we further consider \textit{ex post} also relative to agents' effort which partially, but not fully, determines the quality of their signal.} fairness---is a concern. In peer assessment, for example, receiving an $A$ with 80\% probability and a $F$ with 20\% probability is not the same as a $B$.  Students should receive the grade they earn and a mechanism with reliable measurement integrity is well-positioned to facilitate this. In general, when agents reflect on their experience with a mechanism and assess the fairness of that interaction, they are far more likely to ask an \textit{ex post} question like ``Was my reward fair compensation for my effort?'' than an \textit{ex ante} one like ``Was my \textit{expected} reward fair compensation for my effort?'' Measurement integrity speaks directly to questions of the former kind. As a result, taking measurement integrity seriously is an important step in transforming peer prediction mechanisms from intellectual curiosities into practical tools for eliciting information in the real world. 

\subsection{Computational Experiments: Pros \& Cons}
\label{subsection:benefits-and-limitations}
In this work, we consider empirical notions of measurement integrity and of robustness against strategic reporting by conducting computational experiments, first with an agent-based model (ABM) of our peer assessment setting and then with real peer grading data. 

Computational experiments offer important advantages over theoretical work in analyzing measurement integrity. Computational experiments are more naturally outcome-oriented. Simulated outcomes can be generated cheaply, frequently, and reliably under a wide range of parameter specifications. In contrast, making general theoretical statements quantifying \textit{ex post} payments from a mechanism in this setting is cumbersome and difficult. While theorems have the advantage of potentially applying to a larger range of settings, such theorems are likely to either not give tight bounds, be hard to interpret, or both.  In contrast, computational experiments readily provide interpretable results, albeit on a chosen set of inputs.

On the other hand, computational experiments have their own limitations. When using an ABM, for example, the need to specify a complete model of peer assessment---instead of making a few comparatively mild assumptions about a generic underlying model, as in theoretical work---potentially limits the generalizability of our results beyond our specific model of peer assessment. When using real data, experiments can be noisy and difficult to interpret. However, our combination of these approaches helps to mitigate these concerns. Each kind of experiment complements the other and we find that the key results from our experiments with ABM are corroborated by the results from analogous experiments with real data.

\subsection{Our Results}
\label{subsection:results}
\begin{figure*}
    \centering
    \begin{subfigure}[t]{.495\textwidth}
        \includegraphics[width=\linewidth]{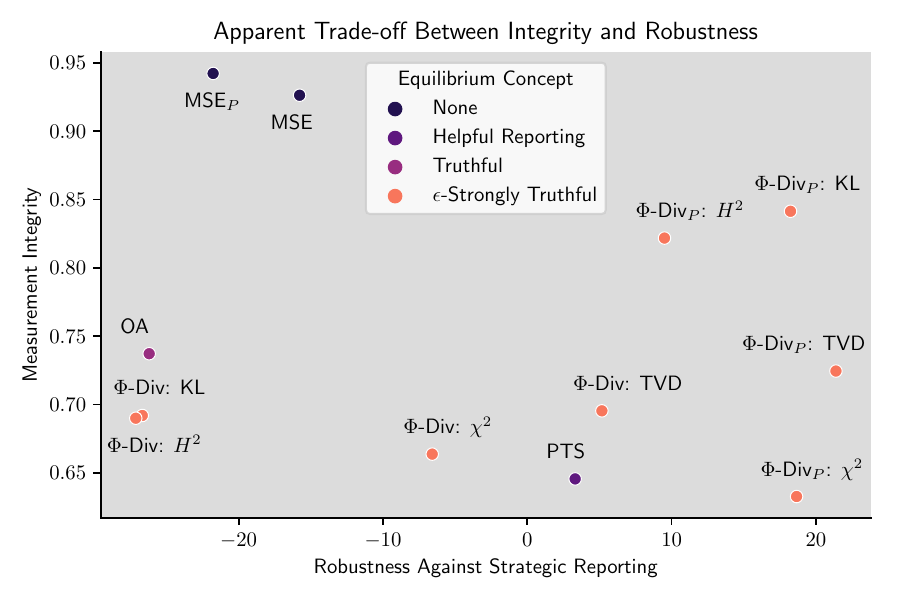} 
        \caption{\textit{Experiments with ABM.}}
        \label{subfig:2D-tradeoff-ABM}
    \end{subfigure} \hfill%
    \begin{subfigure}[t]{.495\textwidth}
        \includegraphics[width=\linewidth]{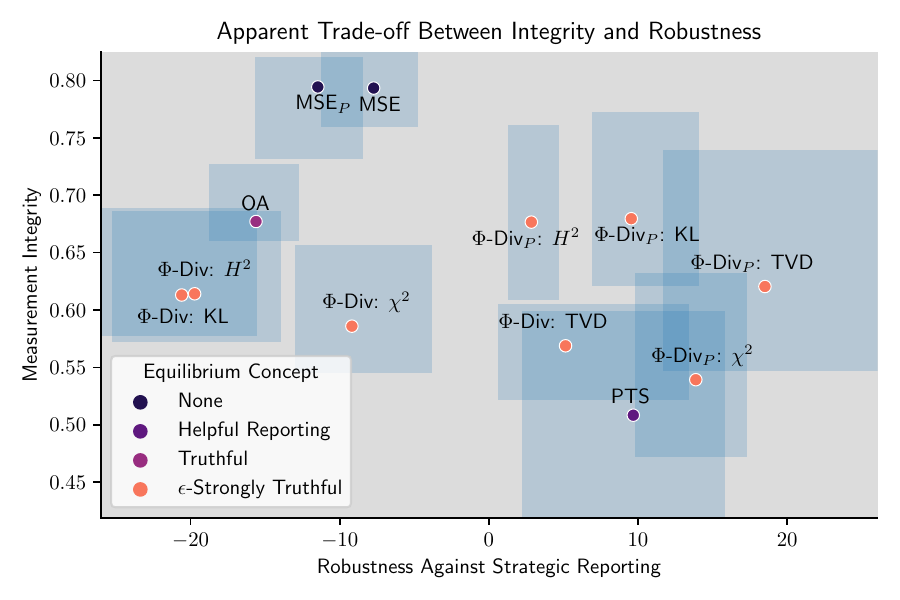} 
        \caption{\textit{Experiments with Real Data.}}
        \label{subfig:2D-tradeoff-real}
    \end{subfigure}
    
    \caption[Direct, two-dimensional comparisons of peer prediction mechanisms based on aggregating their performance within our experiments.]{Direct, two-dimensional comparisons of peer prediction mechanisms based on aggregating their performance in our experiments. \textit{Measurement integrity} is computed as the average area under the curve (AUC, binary) recorded over all numbers of assignments in our experiments for quantifying measurement integrity (\Cref{section:quantifying-mi}). Empirical \textit{robustness against strategic reporting} is computed as the negative of the average mean gain recorded over all considered numbers of strategic graders, for all considered strategies in our experiments for quantifying robustness against strategic reporting (\Cref{section:robustness}). In the experiments with real data, the final point values are computed by taking the average results over all four semesters in the dataset (\Cref{section:data}). Uncertainty is estimated using the maximum and minimum values of the relevant quantities across the individual semesters.
    
    \hspace{1 em}Mechanisms are colored according to their theoretical robustness against strategic reporting, as described by the relevant equilibrium concept (\Cref{subappendix:equilibrium-concepts}). Comparing the color scale to the $x$-axis, it is clear that theoretical robustness against strategic reporting is a somewhat noisy predictor of empirical robustness in our experiments.}
    \label{fig:2D-tradeoff}
\end{figure*}

Our experimental results, summarized in \Cref{fig:2D-tradeoff}\footnote{\Cref{fig:2D-tradeoff} summarizes our experimental results by aggregating them across experiments, but is generally representative of the non-aggregated results, which are presented in \Cref{section:quantifying-mi,section:robustness}.}, robustly differentiate existing peer prediction mechanisms according to their empirical performance. Moreover, they indicate an apparent trade-off inherent in seeking to simultaneously optimize measurement integrity and empirical or theoretical robustness against strategic reporting.

They also reveal the following lessons:

\begin{itemize}
    \item Generic peer prediction mechanisms from the literature largely fail to demonstrate significant measurement integrity compared with simple baselines.
    
    \item Evaluating empirical notions of measurement integrity and robustness against strategic reporting helps to facilitate a more fine-grained comparison between mechanisms than solely considering theoretical notions of robustness against strategic reporting, which are often not directly comparable. For example, we find that implementation choices for some mechanisms affect their empirical performance in our experiments. In contrast, theoretical properties are typically defined to be agnostic about implementation choices.
    
    \item Certain peer prediction mechanisms can be augmented with parametric statistical models to improve their empirical performance with respect to measurement integrity and robustness against strategic reporting. As a consequence, parametric mechanisms should receive more attention in the peer prediction literature.
\end{itemize}

\section{Peer Assessment}
\label{section:peer-assessment}
Peer assessment, where students provide feedback on the work of their peers, is associated with myriad pedagogical benefits. Peer grading---a specific form of peer assessment in which the grades awarded to students are computed directly from their peers' feedback---has the added benefit of easing the workload of the course staff. This consideration is important in MOOCs, where it can be intractable for the course staff alone to provide timely feedback.

In peer assessment, it is typically necessary to specify a mechanism by which students will receive credit for the task of assessing or grading their peers. In distributed, largely anonymous settings like MOOCs, hedonic rewards for expending effort to provide good feedback to peers are likely to be less motivating. Thus, assigning credit in the form of a grade that reflects the quality of the peer assessment is an important tool for encouraging quality feedback. We call this task---assessing the quality of peer assessments or ``grading the graders''---\textit{meta-grading}. 

In peer grading, the primary challenge is to correctly aggregate peer reports into grades that reflect the quality of a submission for an assignment. In meta-grading, incentive concerns are more salient. A student's grade for an assignment depends on other students' feedback (and the quality of their own submission). A student's meta-grade depends directly on their own feedback for other students. As a result, mechanisms for which the incentives are poorly designed will discourage rather than encourage high-quality feedback. For that reason, the peer prediction paradigm is a natural fit for the meta-grading task. 
Meta-grading also presents some unique challenges for the peer prediction paradigm that make it a particularly interesting case study:

\vspace{1 ex}
\noindent
\textbf{Heterogeneous Quality.} 
Agents may have different skills, exert different effort levels, and may not be fully calibrated. Mechanisms may vary in their ability to handle such differences.

\vspace{1 ex}
\noindent
\textbf{Scarcity of Data.}
Which mechanism performs best may be highly dependent on how much data is available. In the peer assessment setting, the amount of data is severely constrained. There are limits on the number of assignments students can be expected to complete in a course (typically on the order of 10) and the number of submissions students can be expected to grade for each assignment (typically between 3 and 6). Also, each student should receive the same amount of feedback on an assignment as their peers.

\vspace{1 ex}
\noindent
\textbf{Fairness Constraints.}
Grades, by design, are intended to reflect the quality of students' work, so the meta-grades assigned by any peer assessment mechanism should be expected to reflect the quality of each student's performance in their assigned peer assessment tasks. 
Further, grades are intended to be fair \textit{ex post}, not just in expectation. Together, these constraints indicate that peer assessment mechanisms, in order to be suitable for assigning meta-grades, need to demonstrate significant \textit{measurement integrity}. 

In this work, we focus on measurement itself and leave the final mapping to grades as a free parameter that is best chosen in a specific peer assessment context. If a mechanism has high measurement integrity, then the rewards assigned by the mechanism can be mapped to fair grades, so long as the mapping is monotone.

\subsection{Related Work}
\label{subsection:related-work}

\subsubsection{Peer Prediction and Information Elicitation}
\label{subsubsection:prediction-and-elicitation}
Our goal in this work is to better understand the practical performance of state-of-the-art peer prediction mechanisms by evaluating them according to \textit{ex post} empirical quantities. However, in the information elicitation literature, peer prediction mechanisms have been proposed in a variety settings. The particularities of different settings necessitate different strategies for designing effective mechanisms. As a result, mechanisms for different settings will plausibly exhibit different empirical behaviors. Thus, in this work, we focus on one particular setting---the elicitation of categorical signals without verification on multiple tasks---which was first explored (independently) by Dasgupta and Ghosh \cite{Dasgupta2013} and Witkowski and Parkes \cite{Witkowski2013}. For simplicity, we generally use the more generic term ``peer prediction'' as a shorthand for this specific setting and note when this is not the case. We leave the extension of our core ideas and methodology to other settings for future work. For an introduction to the other settings for peer prediction and information elicitation, we refer the interested reader to Faltings et al. \cite{Faltings2017book}.

The existing peer prediction literature, both in our setting and more broadly, is focused primarily on theoretical properties related to robustness against strategic behavior. However, there has been some work that explores incentive-compatibility from other perspectives. Gao et al. \cite{Gao2014} take an experimental approach and find evidence that agents are willing and able to exploit peer prediction mechanisms by coordinating on uninformative reports instead of truthful reports. Shnayder et al. \cite{Shnayder2016measuring} use replicator dynamics to quantify desirable incentive properties of peer prediction mechanisms. Replicator dynamics---a simulation-based method that incorporates a notion of learning in an agent ``continuum'''---are quite different from our more-realistic, discrete ABM approach. We leave incorporating agent learning into our framework as an extension for future work. We conjecture that mechanisms that strike a good balance in the trade-off between measurement integrity and robustness against strategic reporting are most likely to efficiently promote truthful reporting as an effective strategy for learning agents to adopt.

Some attention has been paid to other properties of peer prediction mechanisms. Kong's ``Information Evaluation'' is the notion most closely related to measurement integrity \cite{Kong2020}. Like measurement integrity, information evaluation relates to a mechanism's tendency to assign higher rewards to agents with higher-quality reports, but unlike measurement integrity, it is an \textit{ex ante} property.

Properties beyond robustness against strategic reporting have also recently garnered attention in other domains of information elicitation. In many cases, these domains are distinct from the broader peer prediction paradigm because they assume that it is possible to access to ground truth information. In the setting of general crowdsourcing with limited access to ground truth, Goel and Faltings \cite{Goel2019} advance a novel notion of fairness---that the expected payment of each agent be directly proportional to the accuracy of their reports and independent from the strategy and accuracy of the other agents. The philosophical point embedded in this definition---that fair mechanisms must reward agents independently from the reports of other agents---would imply that any peer prediction mechanism is necessarily unfair. We do not accept this premise. In this work we demonstrate that, in practice and under certain circumstances, some peer prediction mechanisms have the ability to reliably reward agents fairly \textit{ex post}, even though they rely on the reports of other agents. In the forecasting setting, Hartline et al. \cite{Hartline2020} and Neyman et al. \cite{Neyman2021} both consider optimizing for properties related to incentivizing effort when selecting a \textit{proper scoring rule} with which to score forecasts.

Another important work related to proper scoring rules is Liu et al. \cite{Liu2020}, which considers the elicitation of forecasts without access to ground truth information. They propose a family of mechanisms, surrogate scoring rules (SSRs), which extend useful properties of proper scoring rules, including robustness against strategic behavior and ``quantifying the value of information'' (an \textit{ex ante} property similar to Kong's ``Information Evaluation'', above) to the setting without verification. In addition to theoretical exploration of this \textit{ex ante} property, they conduct experiments that quantify the extent to which the scores assigned by various mechanisms---including SSRs and certain peer prediction mechanisms from our setting (adapted to elicit forecasts instead of categorical reports)---correlate empirically (i.e. \textit{ex post}) with various metrics of forecast quality. These experiments are interesting precursors of our experiments, however there are crucial differences. First, Liu et al. use their empirical results as a way to attest to the salience of the \textit{ex ante} theoretical property of ``quantification of the value of information.'' They do not emphasize or explain the importance of \textit{ex post} properties as important on their own. As a result, they further do not propose a methodology for choosing an appropriate evaluation metric of forecast quality in accordance with particular preferences of the mechanism designer. In this sense, our discussion of measurement integrity generalizes the particular correlations they consider and provides a framework for thinking about them in a principled way. Second, their experiments do not particularly differentiate the (adapted) peer prediction mechanisms they consider, nor do they suggest strategies by which to improve their performance. Our analysis does both. Lastly, they do not consider the interaction between the properties of robustness against strategic behavior and quantifying the value of information in practice. We find that, empirically, there is an apparent trade-off between the analogous properties in our setting.  

\subsubsection{Peer Prediction and Peer Assessment}
\label{subsubsection:assessment-prediction}
There has also been work that explores the specific application of peer prediction to peer assessment. Shnayder and Parkes \cite{Shnayder2016practical} empirically analyze peer prediction mechanisms using real MOOC data. Importantly, they show that some of the underlying assumptions made in the literature---the self-dominating and self-predicting assumptions in \Cref{subappendix:sufficient-assumptions}---are likely to be violated in the peer assessment setting, especially when the report space is large. They also discuss the importance of considering the variance of rewards in settings where fairness concerns are salient. However, they do not consider the concept of measurement integrity, which is original to this work, or empirical robustness against strategic reporting. They also consider a different set of mechanisms. Some of the best-performing mechanisms in our experiments were not proposed until after the publication of their work

Radanovic et al.  \cite{Radanovic2016} also apply peer prediction to peer assessment (and other applications). In particular, they propose a particular mechanism, the Peer Truth Serum for Crowdsourcing (PTSC) mechanism, and conduct an experiment where that mechanism and some baselines are used to reward peer graders with extra credit when grading a set of quizzes in a course on artificial intelligence at EPFL. They find that students who are rewarded using the PTSC mechanism grade more accurately than those rewarded using the baselines. However, they do not consider any notion of measurement integrity (i.e., the relationship between the rewards and some measure of grading accuracy). They also do not consider whether the improvement in grading accuracy was the result of decreased strategic behavior or increased effort in grading. We discuss the importance of this distinction in \Cref{appendix:alternative-quality}. Lastly, note that the PTSC mechanism is essentially a prototype of the PTS mechanism we consider; the latter is more suited to our setting. 

More broadly, peer assessment is often touted as a natural application for the peer prediction paradigm. However, the typical approach in the literature has been to design mechanisms that are as generic as possible. The resulting mechanisms can thus be ill-suited to the challenges of a specific application, like those we identify for peer assessment above. Indeed, existing mechanisms often rely on collecting lots of data and compensating agents with rewards that exhibit high variance. Both of these characteristics foreshadow the tendency of out-of-the-box peer prediction mechanisms to largely demonstrate low measurement integrity in our experiments. 

On the other hand, certain works have been more skeptical of the application of peer grading to peer assessment. Gao et al. \cite{Gao2016} and Zarkoob et al. \cite{Zarkoob2019} explore how limited spot-checking (in the form of ``ground-truth'' grading by teaching assistants on certain assignments) can (theoretically) incentivize truthful reporting in grading in a simple model. Surprisingly, Gao et al. \cite{Gao2016} find that, compared with spot checking alone, supplementing spot-checking with peer prediction increases the number of spot-checks required to obtain the desired incentive properties. However, the goal of minimizing the number of spot checks necessary to achieve theoretical incentive guarantees in a specific model is somewhat orthogonal to our goals of identifying mechanisms on the frontier of the trade-off between measurement integrity and robustness against strategic behavior (in more realistic settings). As a result, these negative results concerning the use of peer prediction do not preclude the ideas that we present here. In more practical explorations of the utility of spot-checking, Wright et al. \cite{Wright2015} and Zarkoob et al. \cite{Zarkoob2021} propose and refine, respectively, a peer grading system that is centered around the deployment of teaching assistants to improve the quality of feedback. We leave the application of our ideas to mechanisms that incorporate spot checking to future work.

Lastly, Gao et al. \cite{Gao2016} introduce the idea that coordination on partially-informative, ``cheap'' signals that take less effort to observe than a fully-informative signal can also present a challenge to the truthful equilibria of peer prediction mechanisms. This issue is particularly salient in peer assessment, where students might learn to coordinate on superficial qualities of a submission (e.g. frequency of typos or grammatical errors, quality of only the introduction, etc.) that may be misaligned with its holistic quality. We leave the questions of incorporating the existence of cheap signals open for future extensions of our model and framework. Kong and Schoenebeck \cite{Kong2018} make progress toward addressing this concern.

\subsubsection{Modeling Peer Assessment}
\label{subsubsection:modeling-peer-assessment}
The construction of our agent-based model is most influenced by the analysis of MOOC data on the platform Coursera conducted by Piech et al. \cite{Piech2013}. Piech et al. propose a sequence of increasingly complex parametric statistical models of peer assessment. They show that estimating the parameters of each of their models is useful for estimating the true grades of student submissions that have been evaluated by peers. Grade estimates computed using their models are found to outperform grade estimates computed by the algorithm used by Coursera at the time. 
The inclusion of grader biases in each of their models is found to be the most significant single factor underlying this result.

The model we adopt subsequently is not one proposed by Piech et al., but it is structurally similar to their model \PGone, which strikes a good balance between simplicity and performance in their analysis. 

The decision to propose a new model, instead of adopting one of theirs, stems from a few important points:
\begin{enumerate}
    \item Their models are continuous. In practice, though, essentially all assignment scores, rubrics, etc. are discrete. Thus, the ``true grade'' of a submission, and each grader's signal, in a peer assessment model should be discrete.
    
    \item Nearly all peer prediction mechanisms require a discrete report space.
    
    \item It allows us to use model \PGone\,in the implementation of our parametric peer prediction mechanisms (\Cref{subsection:parametric}) without giving the mechanisms unrealistically accurate information about the underlying process by which true grades and reports are generated.
    
\end{enumerate} 

\section{Peer Assessment Agent-Based Model}
\label{section:model}
Our ABM simulates a class of students enrolled in a semester-long course for which there is at least one graded assignment. For each assignment, each student turns in one submission and is randomly assigned submissions from four other students to grade.

\vspace{1 ex}
\noindent
\textbf{Submissions.} 
For each assignment $j$, each student $i$ turns in a submission $s_{i, j}$. Each submission $s_{i, j}$ has a true integer score $g_{i, j}^* \in [0, 10]$, drawn (independently at random) from the binomial distribution $\Binom{10}{\frac{7}{10}}$.
\vspace{1 ex}

The process of grading is modeled by an agent receiving a \textit{signal} about the true score of a submission. The signal is a function of some number of draws from a \textit{latent distribution} that depends on the true score of the submission being graded and the \textit{bias} of the agent. 

\vspace{1 ex}
\noindent
\textbf{Bias.}
In practice, agents may have some bias in grading assignments. That is, the latent distribution from which draws are used to construct their signal for each assignment could have a mean that is slightly higher or lower than that assignment's ground truth score.
An agent $k$'s bias $b_k$ is sampled uniformly at random from the normal distribution $\Normal{0}{1}$. Their signal for submission $s_{i, j}$ is a function of draws from the latent distribution $\Binom{10}{\frac{g_{i, j}^* + b_k^{}}{10}}$. If $g_{i, j}^* + b_k^{}$ is less than 0 or greater than 10, then the value is truncated to be 0 or 10, respectively.
\vspace{1 ex}

The number of draws from an agent's latent distribution that are used to create their signal---which determines the variance of the distribution of their signals---is a function of their \textit{effort}; greater effort corresponds to lower variance.

When the signal is created using a single draw, defining the signal is trivial---the signal is equal to the outcome. When the signal is created using more than one draw from the latent distribution, the signal is defined as the simple average of the outcomes of the draws rounded to the nearest integer. This convention ensures that the space of signals is equal to the space of reports, so the notion of a ``truthful report'' is straightforward and well-defined. 
Our model of effort is as follows:

\vspace{1 ex}
\noindent
\textbf{Continuous Effort.}
Effort is parameterized by a continuous value $\lambda \in (0, 2]$ drawn uniformly at random. The number of draws from the latent distribution used to create an agent's signal is equal to $1 + X$, where $X \sim \Poisson{\lambda}$ is drawn according to the Poisson distribution. 
\vspace{1 ex}

Lastly, we need to introduce an appropriate notion of the quality of agent's reports in order to reason about measurement integrity:

\vspace{1 ex}
\noindent
\textbf{Report Quality.}
Here, we use a simple, intuitive notion of report quality---the squared distance between the report value and the true grade of the corresponding submission. In \Cref{appendix:alternative-quality}, we consider alternative conceptions of report quality and discuss the relative merits of the different approaches in detail.

\section{Peer Assessment Data}
\label{section:data}
Our real peer grading data set---which was collected by others for other projects \cite{Yuan2020} and graciously shared with us for this work---contains grading information from an undergraduate-level course on the design and analysis of algorithms taught at Northwestern University in both the Spring and Fall semesters of 2017 and 2019. For each student enrolled in the course, the data set contains information about the submissions they turned in during the course of the semester and the grades that they provided for submissions from other students. For each submission, the data set identifies the assignment that the submission corresponds to, specifies the grade that was ultimately awarded for that submission---which we treat as its true grade---and some number of peer grades. These awarded grades are a mix of grades assigned by instructors and grades assigned by the (non-parametric) \texttt{vancouver} algorithm \cite{deAlfaro2014}, which takes into account instructor grades, peer grades, and the accuracy of peer graders. Because this combination of methods was deemed sufficient for fairly assigning grades to real students in the courses from which the data were collected, we are comfortable treating the assigned grades as unbiased estimates of the ground truth. For each peer grade that a student provided, the data set includes an identifier for the corresponding assignment, a numerical score, and written comments. 

Many of the peer prediction mechanisms that we consider impose restrictions on the form of the data set. Certain mechanisms require that students grade at least two submissions for each assignment in which they will be evaluated by the mechanism. Other mechanisms require that at least two students grade each submission. Accordingly, we (iteratively) remove peer grades from students for assignments for which they graded fewer than 2 submissions and submissions with fewer than 2 graders until the modified data set meets these specifications. After this pre-processing, we simplify the grading and report space. The numerical scores---true grades and peer grades---from 2017 are out of 100 and from 2019 are out of 30. We coarsen these raw grades into the integer range $[0, 10]$. This simplifies the implementation of the peer prediction mechanisms and our experiments by keeping the space of possible reports the same for our ABM and all 4 semesters in the data set and helps to make the empirical distribution of reports less sparse in the space of all possible reports. It also lends our analysis some robustness to the method of assigning true grades, since small changes to the value of a true grade will tend not to change the value to which it is mapped during the pre-processing procedure.

Lastly, our parametric mechanisms (\Cref{subsection:parametric}) and certain reporting strategies (\Cref{subsection:strategies}) employ information about a (continuous) prior distribution for the true grades. Thus, we use maximum likelihood estimation to fit normal distributions to the empirical distribution of true grades for each semester. 

The relevant information about each semester after all of the pre-processing is given in \Cref{tab:data}.

\begin{table}
    \renewcommand{\arraystretch}{1.2}
        \centering
        \begin{tabular}{|c|c|c|c|c|c|c|c|}
            \hline
            & Students & Assignments & Submissions & Peer Grades & Retained & $\mu$ & $\sigma^2$ \\
            \hline
            Spring 2017 & 94 & 16 & 758 & 4080 & 99.9\% & 8.71 & 3.8025 \\
            \hline
            Fall 2017 & 86 & 16 & 577 & 3102 & 99.9\% & 7.57 & 4.9729 \\
            \hline
            Spring 2019 & 49 & 13 & 313 & 1586 & 91.9\% & 7.68 & 3.6864 \\
            \hline
            Fall 2019 & 65 & 14 & 389 & 2089 & 98.1\% & 8.25 & 2.8561 \\
            \hline
        \end{tabular}
        \caption{Summary of the pre-processed peer grading data, including the total number of students, assignments, submissions, and peer grades for each semester. The percentage of grades in the raw data that are retained after pre-processing is also shown. For example, in Spring 2019, the 1586 grades left after pre-processing represent 91.9\% of the total grades in the raw data. Lastly, the parameters $\mu$ and $\sigma^2$ are the mean and variance, respectively, of the normal prior distribution fit to the empirical distribution of true grades. These values are used in the implementations of parametric peer prediction mechanisms.}
        \label{tab:data}
\end{table}

\section{Peer Prediction Mechanisms}
\label{section:mechanisms}
In this work, we consider a representative selection of fundamental mechanisms from the peer prediction literature. In what follows, we describe the intuition behind the mechanisms that we evaluate using our agent-based modeling framework. We also discuss the challenges that the particularities of the peer assessment setting and our peer assessment model pose to the implementation of the mechanisms. For a more specific discussion of the actual implementation of the various mechanisms and how we overcome these challenges, see \Cref{appendix:implementation}. 

\subsection{Baseline}
\label{subsection:baseline}
In theoretical work, simple baselines would typically be excluded, due to the existence of trivial, non-truthful reporting equilibria. Despite this concern, however, such simple mechanisms are used in practice. Bachelet et al. \cite{Bachelet2015}, for example, recommend using the following mechanism to assign grades (when the reports have been appropriately pre-processed):

\vspace{1 ex}
\noindent
\textbf{Mean Squared Error (MSE) Mechanism.}
On each submission that they grade, agents are paid according to the mean squared error of their reports from the \textit{consensus grade} of each submission. The consensus grade of a submission---a basic estimate of its unobservable true grade---is defined to be the simple average of the reports of all 4 agents that graded it. To maintain the convention that the higher rewards correspond to higher quality agents, the payments are equal to the negative of the mean squared error.

\subsection{Non-Parametric Mechanisms}
\label{subsection:non-parametric}
Our first category of peer prediction mechanisms reflects the options that a novice mechanism designer would find in an initial search for peer prediction mechanisms to deploy in some application. In keeping with this, we implement these mechanisms as faithfully as possible to the descriptions given in the works in which they were proposed. We make changes only when necessary to ensure basic functionality within the setting of our model.

Note that for all mechanisms that involve pairing an agent with another agent in order to compute their scores on a grading task (i.e. generating a report for one submission), we take the expectation over all of the possible pairings to reduce the variance of the scores. 

\vspace{1 ex}
\noindent
\textbf{Output Agreement (OA) Mechanism.}
The simplest type of peer prediction mechanism, common in the literature \cite{Faltings2017}, is an \textit{output agreement} mechanism, which serves as another simple baseline with which to compare state-of-the-art peer prediction mechanisms. To compute payments for a task in the OA mechanism, agents are paired and their reports are compared. Agents are paid 1 if their reports match and 0 otherwise.  

\vspace{1 ex}
\noindent
\textbf{Peer Truth Serum (PTS) Mechanism.}
Developed by Faltings et al. \cite{Faltings2017}, the PTS mechanism pays agents if their report for a task agrees with the report of a randomly selected peer on the same task. The magnitude of the payment is proportional to the inverse of the frequency of their report according to a distribution $R$ over the report space.
    
\vspace{1 ex}
\noindent
\textbf{$\Phi$-Divergence Pairing ($\Phi$-Div) Mechanism.}
This mechanism was proposed by Schoenebeck and Yu \cite{Schoenebeck2021} and is based on the application of an information-theoretic framework for designing peer prediction mechanisms described by Kong and Schoenebeck \cite{Kong2019}. Like the OA mechanism, this mechanism pairs agents with peers. The pairs are rewarded for submitting correlated reports on a \textit{bonus task} and penalized for submitting correlated reports on a pair (one for each agent) of \textit{penalty tasks} that are distinct from each other and from the bonus tasks. 

The magnitudes of the respective reward and penalty depend on a convex function $\Phi$ chosen by the mechanism designer and on $\text{JP}(x, y)$, the joint-to-marginal-product ratio of random variables $X$ and $Y$ drawn, respectively, from each agent's distribution of reports:
\[
    \text{JP}(x, y) = \frac{P_{X, Y}(x, y)}{P_X(x) P_Y(y)},
\]
where $P_{X, Y}(x, y)$ is the probability of observing reports $x$ and $y$ as answers to the same question under the joint distribution of reports $X$ and $Y$, and $P_X(x) P_Y(y)$ is that probability according to the product of the marginal report distributions. Their ratio can be understood as measuring how much more likely a pair of reports $x$ and $y$ are to occur on the same question versus different questions. Note that each quantity is a function of the random variables $X$ and $Y$, which depend both on the agents' strategies and the joint prior. In general, \text{JP} is unknown to the mechanism and will need to be estimated.   

For a given pair of agents, a bonus task $b$, and a pair of \textit{penalty tasks}, $p \neq q$, the payment is :
\[
    \partial \Phi(\text{JP}(x_b, y_b)) - \Phi^*(\partial \Phi(\text{JP}(x_{p}, y_{q}))),
\]
where $\partial \Phi$ is the \textit{subgradient} of $\Phi$, $\Phi^*$ denotes the \textit{convex conjugate} of $\Phi$, and $x_i$ and $y_j$ denote the first agent's report on task $i$ and the second agent's report on task $j$, respectively. 
The intuition is that the first term rewards an agent based on the likelihood of their report for the bonus question $b$ given the other agents' reports on $b$.  The second term penalizes agents for reporting generically likely answers by considering the likelihood of an agent's report for the penalty question $p$ given the other agents' reports on the distinct penalty question $q$. See Schoenebeck and Yu \cite{Schoenebeck2021} for a complete discussion of each component of this mechanism, including definitions for all of the relevant terms above.

Ideally, we would want to estimate the joint-to-marginal-product ratio of the reports for each pair of agents, but given the limited availability of data in this setting, the best we can do is treat the agents anonymously and compute one estimate, $\hat{\text{JP}}$, that applies to the entire agent population. See \Cref{subsubappedix:implementation-nonparametric} for the details of this estimation procedure.
    
In our experiments, we consider 4 common $\Phi$-divergences, which are described in \Cref{tab:phi-divs}. One important note with respect to the choice of $\Phi$-divergence is that when the total variation distance (TVD) is chosen, the $\Phi$-Div mechanism emulates the Correlated Agreement (CA) mechanism from Shnayder et al. \cite{Shnayder2016informed}.

\vspace{1 ex}
\noindent
One significant omission from our selection of state-of-the-art mechanisms is Kong's Determinant-based Mutual Information (DMI) mechanism \cite{Kong2020}, which has impressive theoretical properties. In our computational experiments, the mechanism requires significant modifications to the setting or else it will simply assign every agent a reward of 0. With the necessary modifications, the mechanism does not perform particularly well with respect to measurement integrity and the results for robustness against strategic reporting are not a fair comparison to the other mechanisms. As a result, we omit DMI from consideration in the body of this work. However, we discuss our implementation and the necessary modifications in \Cref{appendix:implementation} and include DMI, when possible, in the additional experimental results presented in \Cref{appendix:alternative-quality,appendix:additional-results}.

\begin{table}
    \renewcommand{\arraystretch}{1.2}
        \centering
        \begin{tabular}{|c|c|c|}
            \hline
            $\Phi$(x) & $\Phi$-divergence & Notation\\
            \hline
            $\frac{1}{2} \lvert x - 1 \rvert$ & Total Variation Distance & TVD \\[1 pt]
            \hline
            $x \log x$ & Kullback-Leibler divergence & KL \\[1 pt]
            \hline
            $x^2 - 1$ & $\chi^2$-divergence & $\chi^2$ \\[1 pt]
            \hline
            $(1 - \sqrt{x})^2$ & Squared Hellinger distance & $H^2$ \\[1 pt]
            \hline
        \end{tabular}
        \caption{Common choices for $\Phi$ and their associated $\Phi$-divergences.}
        \label{tab:phi-divs}
\end{table}
 
\subsection{Parametric Mechanisms}
\label{subsection:parametric}
Anticipating the challenges that generic mechanisms might encounter when deployed in a specific setting, we also explore how certain peer prediction mechanisms can be supplemented with domain-specific, parametric statistical models. To implement these, we adopt the perspective of a real-life mechanism designer. In the real world the ``true'' distributions and parameter values that ``govern'' the behavior of students participating in peer assessment are inaccessible. Instead, a mechanism designer can examine the peer assessment literature to find a model of peer assessment inspired by and validated on real data for which the hyperparameters of the model can be tuned to fit their particular application. Here, model \PGone\,from Piech et al. \cite{Piech2013} meets both criteria. It constitutes a reasonable continuous approximation to our primarily discrete underlying model (in which ``reliability'' serves as a proxy effort). 
The model, with hyperparameters that are appropriate for our setting, is described below:
\begin{align*}
    \text{True Score}&: \quad g_{i, j}^* \sim \Normal{7}{2.1} \text{for each submission $s_{i,j}$},\\
    \text{Reliability}&: \quad \tau_{i} \sim \Gam{10/1.05}{10} \text{for each agent $i$}\protect\footnotemark,\\
    \text{Bias}&: \quad b_{i} \sim \Normal{0}{1} \text{for each agent $i$}\\
    \text{Signal}&: \quad z_{i, j}^k \sim \Normal{g_{i, j}^* + b_k}{\tau_k^{-1}} \text{for a grader $k$, who is grading submission $s_{i, j}$},.
\end{align*}

\footnotetext{$\mathcal{G}$ denotes a Gamma distribution. The hyperparameters $\alpha_0 = 10/1.05$ and $\beta_0 = 10$ for $\mathcal{G}$ were chosen by inspection, subject to having the correct expected value for a continuous effort agent.}

To reiterate, the simulated data in our experiments is always generated according to the model described previously in \Cref{section:model}. But instead of estimating parameters of that underlying model, we estimate the parameters of model \PGone. We then use those estimates in deploying the parametric peer prediction mechanisms described below. This simulates the situation faced by a mechanism designer in a real deployment. They would be unable to know the ``true'' underlying model, but would be able to tune a reasonable statistical model for their application using past data.

Using model \PGone\,is also useful because existing work from the peer assessment literature shows how to estimate its parameters. Chakraborty et al. \cite{Chakraborty2020} propose a method for estimating the parameters of model \PGone\,(and computing meta-grades) using limited access to ground truth. In the absence of ground truth, their estimation method (though not their meta-grading method) can be adapted to estimate the parameters of the model using an expectation-maximization-style algorithm with Bayesian priors for the bias (when applicable) and reliability of each agent. The details of our estimation procedure are available in \Cref{subsubappendix:implementation-parametric}.
    
\vspace{1 ex}
\noindent
\textbf{Parametric MSE (\pMSE) Mechanism.}
Under this mechanism, each agent is awarded according to the mean squared error of their reports (corrected for estimated biases when appropriate) from the estimated true scores. As with the baseline, the payments are equal to the negative of the mean squared error.
    
\vspace{1 ex}
\noindent
\textbf{Parametric $\Phi$-Divergence Pairing (\pPhiDiv) Mechanism.}
Instead of using empirical estimates of the joint-to-marginal-product ratio of reports, we can pre-compute the joint-to-marginal-product ratio $\text{JP}(x, y)$ analytically under model \PGone\,and score the tasks after estimating the parameters of model \PGone\,using the estimation procedure described above. See \Cref{appendix:optimal-scoring-rules} for the calculation. This allows us to individualize the joint-to-marginal-product ratio for each pair of agents, which as we noted earlier is desirable but intractable for the non-parametric version of this mechanism, given the scarcity of data.
    
For each task, agents are paired and scored according to the same procedure used for the non-parametric $\Phi$-Div mechanism, but using the closed-form expression we derived for $\text{JP}(x, y)$ instead of an empirical estimate.
    
\section{Quantifying Measurement Integrity}
\label{section:quantifying-mi}
In our experiments, we seek to empirically evaluate the above mechanisms according to their measurement integrity and robustness against strategic reporting. Evaluating mechanisms for both properties simultaneously would make it difficult to isolate which features were beneficial for which property. As a result, we first quantify measurement integrity in isolation, assuming that agents report their signals honestly.

\subsection{Computational Experiments with ABM}
\label{subsection:quantifying-mi-abm}
One key advantage of the agent-based modeling approach is access to latent quantities, e.g., a submission's true score, that are generally not observable, without noise, in the real world.  Another is the ability to readily repeat experiments over a range of parameter specifications. We leverage these advantages in order to analyze the relationship between agents' payments assigned under the various mechanisms and the squared error of their reports to the ground truth scores, considering various intuitive notions of measurement integrity.

\subsubsection{Methods}
\label{subsubsection:quantifying-mi-abm-methods}
For each mechanism, we perform the same procedure of simulating ``semesters,'' which consist of 500 students submitting and grading some number of simulated assignments. The number of assignments is varied from $i = 1, 2, \ldots, 15$. For every value of $i$, we simulate 50 semesters, which each proceed as follows:
\begin{enumerate}
    \item For each assignment:
    \begin{enumerate}[i.]
        \item All students turn in a submission whose true grade is drawn from the true grade distribution.
        
        \item A random 4-regular graph of agents is constructed.
        
        \item Students grade the submissions of their neighbors in the graph according to our peer assessment model. The squared errors of their grades to the true grades are recorded.
        
        \item The grades are reported to the mechanism, which assigns a reward to each student for their performance in peer assessment for that assignment. 
    \end{enumerate}
    
    \item Students' total rewards for the semester---the sums of their rewards for each of the $i$ individual assignments---and their cumulative squared error to the true scores are used to calculate the value of the relevant evaluation metrics.
\end{enumerate}

\subsubsection{Evaluation}
\label{subsubsection:quantifying-mi-abm-evaluation}
Abstractly, the appropriate metric for evaluating measurement integrity for a peer prediction application, can depend on a mechanism designer's tolerance for making different kinds of errors in that application. In peer grading, for example, a practitioner may consider it worse to fail a borderline student that should have passed than to pass a borderline student that should have failed. An appropriate evaluation metric, then, may take these preferences into account. The concept of measurement integrity, and the setup of our computational experiments, is flexible enough to allow for such preferences to be incorporated. 

To illustrate this principle, in our experiments, we consider a few relatively general evaluation metrics that faithfully correspond to (increasingly strict) notions of measurement described by Stevens \cite{Stevens1946}:

\vspace{1 ex}
\noindent
\textbf{Coarse Ordinal Measurement}. Nominal measurement is akin to standard classification, but in our setting, it is more instructive to consider classification with ordered classes, which, when the number of classes is small, is a coarse kind of ordinal measurement. We consider two such relevant classification tasks. First, we consider binary classification of agents as being above or below the median in terms of squared error to ground truth scores. The area-under-the-curve (AUC) of the receiver operating characteristic (ROC) curve is well-suited to evaluating performance at this task. AUC summarizes how useful the rewards assigned by the mechanism are for being translated into a classification via the selection of a threshold such that all students with rewards above the threshold are classified as above-median and all students with rewards below the threshold are classified as below-median. 
    
To look for higher-resolution information contained in the rewards, while still remaining at the level of classification, we also consider the task of placing agents into quintiles in terms of squared error to ground truth scores. This resembles the assignment of traditional letter grades---A, B, C, D, or F.\footnote{Viewed through this lens, this quinary classification is also a good candidate to be evaluated with a custom loss function that encodes the preferences of the mechanism designer. However, we use the quinary AUC metric, which is more generally applicable and faithfully corresponds to the notion of nominal measurement.} AUC does not have a generalization beyond binary classification that preserves its attractive properties as an evaluation metric in our setting. However, we can use the average \textit{pairwise} AUC of the rewards in classifying agents from two quintiles as belonging to one or the other, over all possible pairs of quintiles. We refer to this quantity as quinary AUC. Intuitively, it gives a holistic appraisal of how well the payments separate agents into five distinct categories according to their squared error from the ground truth.
    
\vspace{1 ex}
\noindent
\textbf{Fine Ordinal Measurement}. The most fine-grained ordinal measurement is ranking. For rankings, the Kendall rank correlation coefficient ($\tau_B$) is an effective evaluation metric. The value of $\tau_B$ is related to the number of pairs that appear in the same order (concordant pairs) and in the opposite order (discordant pairs) in the two rankings being compared. In the case that neither ranking has ties, $\tau_B$ is equal to the proportion of pairs that are concordant minus the proportion of pairs that are discordant.
    
\vspace{1 ex}
\noindent
\textbf{Beyond Ordinal Measurement}. It is conceivable that the rewards from some mechanisms might contain even more information about agents' squared error from the ground truth than just the ordinal notion that higher payments correspond to lower squared error. The \textit{magnitude} of the difference in payments between agents, for example, may contain information about the magnitude of their difference in squared error. At the very least, this should be the case for the baseline MSE and \pMSE\,mechanisms, if they are computing good estimates of the unobserved (by the mechanisms) ground truth scores. To explore this possibility, we use the Pearson correlation coefficient ($\rho$), which measures the strength of the \textit{linear} relationship between two variables, as an evaluation metric.\footnote{Inspecting scatterplots of squared error vs. payments for each mechanism does not provide any compelling evidence of a clear \textit{non-linear} relationship between the quantities of interest.}   

\vspace{1 ex}
Both correlation coefficients ($\tau_B$ and $\rho$) vary between -1 and 1, with more positive values indicating a stronger positive relationship between the quantities of interest, 0 indicating no relationship, and more negative values indicating a stronger negative relationship between the quantities of interest.

\subsubsection{Results}
\label{subsubsection:quantifying-mi-abm-results}
\begin{figure}
\begin{subfigure}{\textwidth}
  \centering
  \includegraphics[width=0.4\linewidth]{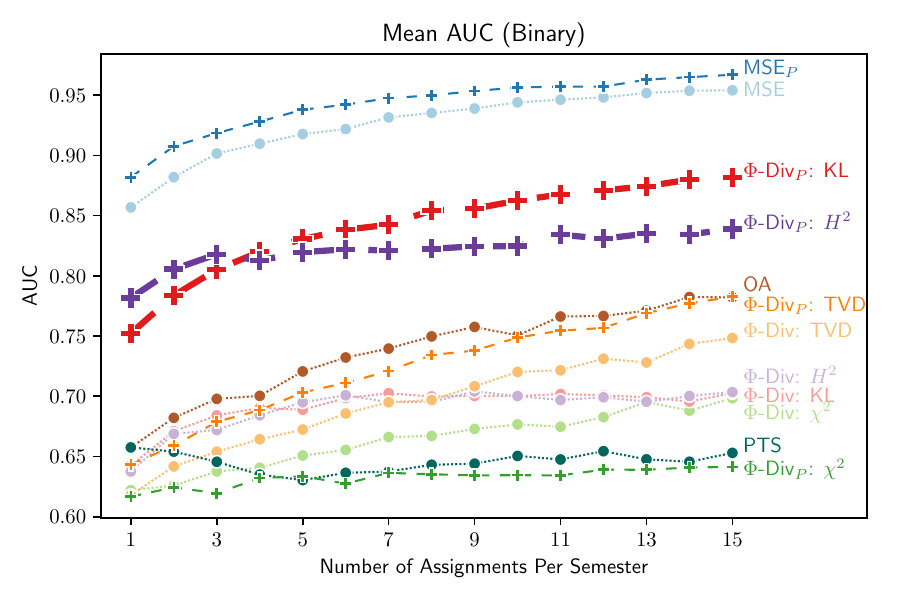}
  \includegraphics[width=0.4\linewidth]{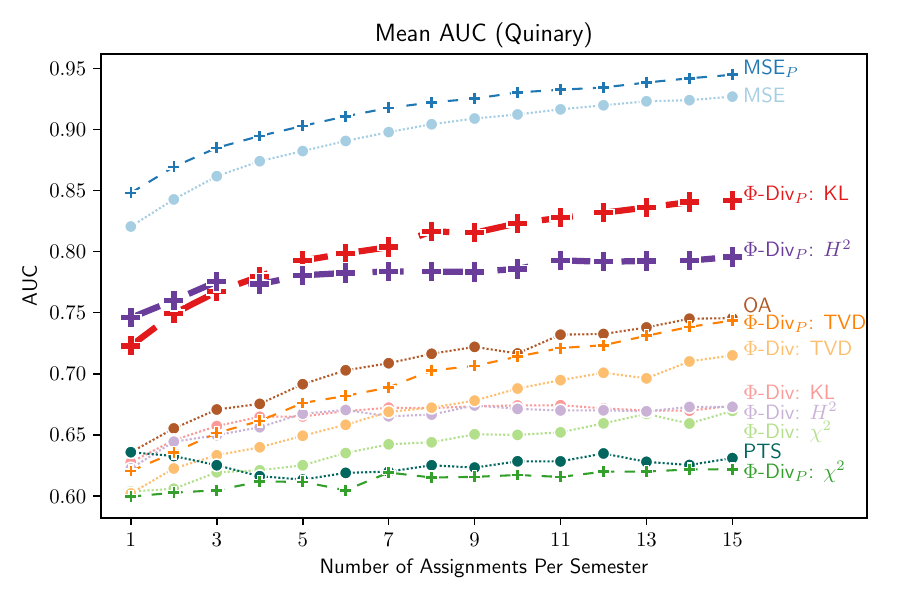}
  \hfill
  \includegraphics[width=0.4\linewidth]{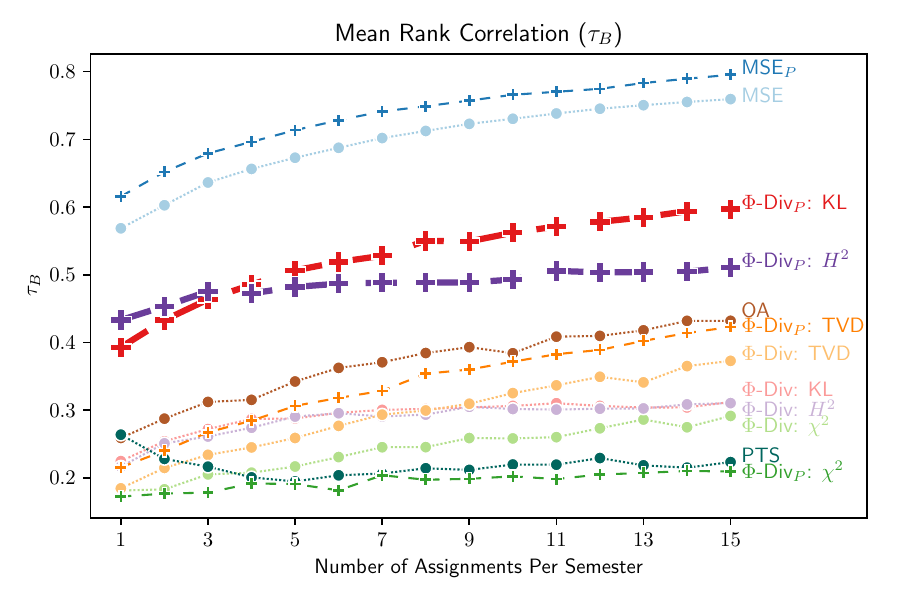}
  \includegraphics[width=0.4\linewidth]{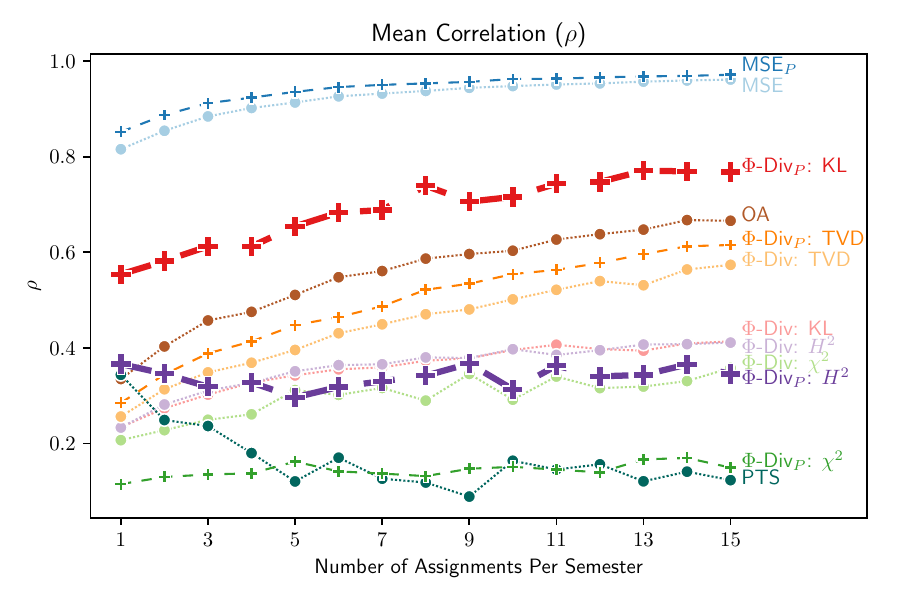}
\end{subfigure}
\caption{\textit{Quantifying Measurement Integrity with ABM.} Average values of evaluation metrics for measurement integrity corresponding to different intuitive notions of measurement as the number of assignments in a simulated semester grows. The average for each number of assignments is taken over 50 simulated semesters.}
\label{fig:quantifying-mi-abm}
\end{figure}

The values of the various evaluation metrics, averaged over the 50 semesters simulated for each value of the number of assignments in a semester, are plotted in \Cref{fig:quantifying-mi-abm}.

The first result that stands out is that it is feasible to achieve high levels of measurement integrity, even according to strict notions of measurement. As the number of assignments in a semester increases, thereby increasing the amount of information available to the mechanisms, the baseline MSE and \pMSE\,mechanisms score highly according to each of the evaluation metrics, including near-perfect Pearson correlation. This indicates that, unsurprisingly,  it is possible to estimate true scores that are not observed by the mechanism highly reliably in our model when agents report truthfully. However, despite this possibility, peer prediction mechanisms generally do not appear to take advantage of this. As a result, they largely perform relatively poorly according to each evaluation metric compared to simple baseline mechanisms. The exceptions are two parametric mechanisms, \pPhiDiv: KL and \pPhiDiv: $H^2$, which mostly outperform the OA baseline.

Interestingly, the pattern established in the plots of the first three evaluation metrics, which are qualitatively nearly identical, is disrupted by the plot of the Pearson correlation ($\rho$). In particular, the \pPhiDiv: $H^2$ mechanism performs much less well according to Pearson correlation than the other evaluation metrics, indicating that the payments it assigns contain useful \textit{ordinal} information about the squared error of agents' reports, but not very useful linear information. Importantly, this shows that level of evaluation for measurement integrity that is relevant to a particular application---the notion of measurement that is desirable---can matter significantly with respect to how a mechanism performs. In this case, inspection reveals that the difference in performance is due to a tendency of the \pPhiDiv: $H^2$ mechanism to occasionally assign very negative outlier payments. These outliers interfere with the linear relationship between the payments and agents' squared error, but not the ordinal relationship. The payments from the PTS mechanism are also notably less useful with respect to linear than ordinal information. However, the ordinal information conveyed via the PTS mechanism was already poor relative to the other mechanisms.

\subsection{Computational Experiments with Real Data}
\label{subsection:quantifying-mi-real}

\subsubsection{Methods}
\label{subsubsection:quantifying-mi-real-methods}
We replicate the experiments with simulated data described in \Cref{subsubsection:quantifying-mi-abm-methods}, substituting our four semesters worth of real data for the 50 semesters worth of data that we simulated via our peer grading ABM. For each semester, as in our experiments with our ABM, we assign rewards according to each mechanism for each assignment, one at a time. However, due to limitations in the data and the necessary pre-processing (\Cref{section:data}), not every student is associated with peer grades for submissions on every assignment. In fact, different students are associated with different numbers of peer grades, which complicates our analysis. 

We address this complication in two steps. First, we divide each student's squared error of reports from true scores and their payments by the number of peer grades with which they are associated. Thus, we consider the average squared error and average payment of each student when computing the evaluation metrics. Second, to control for the fact that the average squared error and payment for students associated with few peer grades may be much noisier than those of students associated with many peer grades, we focus on students associated with many peer grades when computing the evaluation metrics. To accomplish this, for each semester, we split the assignments into four blocks of roughly equal size and consider only students who are associated with at least one peer grade in each assignment block when computing the evaluation metrics.\footnote{In general, this rule need not exclude students associated with few peer grades. In practice, however, it strikes a good balance between being as inclusive as possible while excluding students associated with very few grades. In particular, it seems to do better than using, for example, a threshold based on a percentage of the maximum number of peer grades for the given semester.} Under this rule, the number of students considered when computing the evaluation metrics is 84, 49, 42, and 54 for the Spring 2017, Fall 2017, Spring 2019, and Fall 2019 semesters, respectively. Overall, the evaluation metrics are computed 4 times for each semester---once after each assignment block has been processed. Note that information accumulates as each block is processed---the computation of the evaluation metrics uses all the information obtained in the current block of assignments and its predecessors. Thus, a nice consequence of this procedure is that new information for every student is incorporated every time the evaluation metrics are recomputed.

To reduce the variance of the results for non-deterministic mechanisms, we perform 50 iterations of the procedure described above and record the average of the evaluation metrics over the iterations. 

\subsubsection{Results}
\begin{figure}
\begin{subfigure}{\textwidth}
  \centering
  \includegraphics[width=0.4\linewidth]{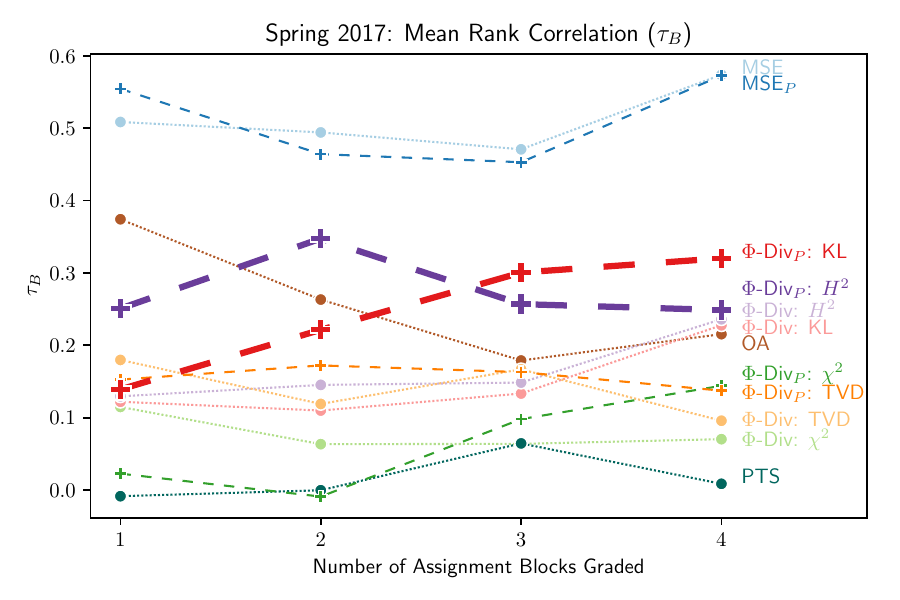}
  \includegraphics[width=0.4\linewidth]{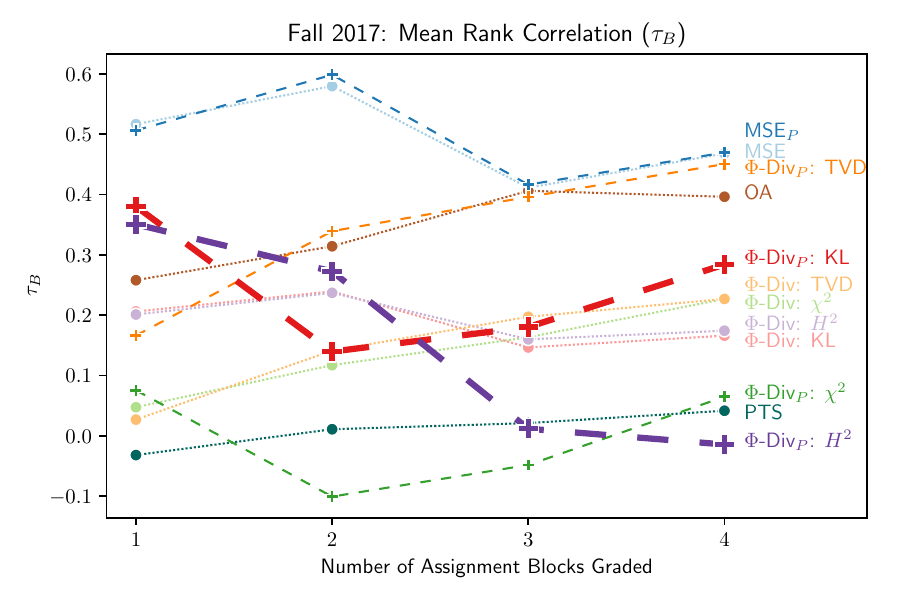}
  \hfill
  \includegraphics[width=0.4\linewidth]{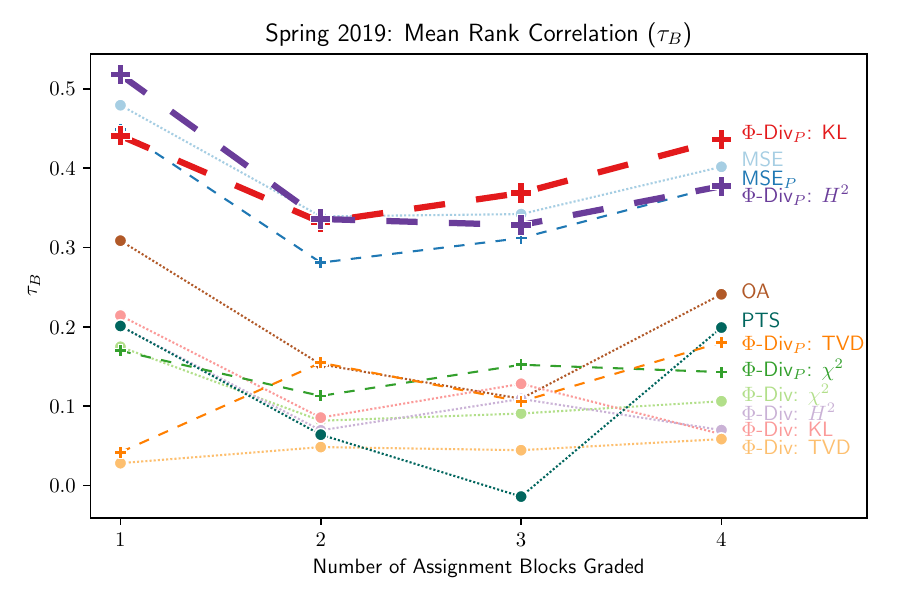}
  \includegraphics[width=0.4\linewidth]{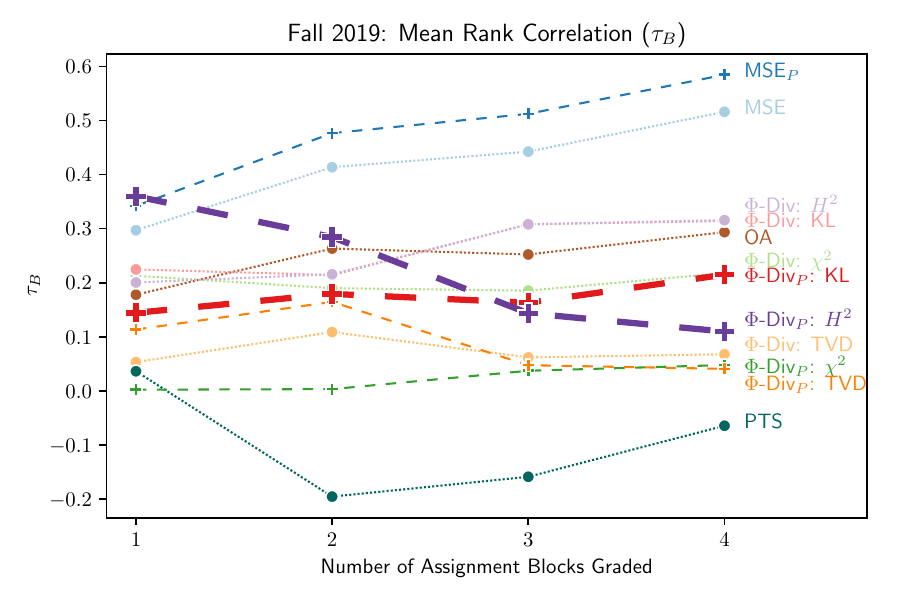}
\end{subfigure}
\caption{\textit{Quantifying Measurement Integrity with Real Data.} For each semester, the average Kendall rank correlation coefficient ($\tau_B$) between students' payments and the average squared error of their reports from the true scores over 50 iterations of each mechanism. In each iteration, the metric $\tau_B$ is calculated within each quartile---according to the number of peer grades with which they are associated---of students.}
\label{fig:quantifying-mi-real}
\end{figure}

We focus on the results--shown in \Cref{fig:quantifying-mi-real}---for which the evaluation metric is the Kendall rank correlation coefficient ($\tau_B$), since, as in our simulated experiments, those results are qualitatively similar to those for which the evaluation metric is (binary or quinary) AUC and since $\tau_B$ connects with our experiments for quantifying robustness against strategic reporting, which involve rankings. The results for the other evaluation metrics are given in \Cref{subsubappendix:mi-results-real}.

Unsurprisingly, the results for the real data are much noisier than those for the simulated data. However, there are some patterns that emerge, which corroborate observations from our experiments with ABM. In particular, there is a general consensus about the best-performing mechanisms. The MSE and \pMSE\,baselines are among the best-performing mechanisms at nearly every point. The best-performing peer prediction mechanisms from the simulated experiments, \pPhiDiv: KL and \pPhiDiv: $H^2$, are also often among the best mechanisms, albeit less consistently. Indeed, if we average the value of $\tau_B$ for each mechanism across all quartiles and all semesters, we find that, the 5 best-performing mechanisms on average are \pMSE, MSE, \pPhiDiv: KL, OA, and \pPhiDiv: $H^2$, in that order. These are exactly the same 5 mechanisms that results from a similar analysis of the simulated data, albeit in a slightly different order.\footnote{In the experiments with real data, the differences between the values for \pPhiDiv: KL, OA, and \pPhiDiv: $H^2$ are not different enough to make reliable statements about their relative ordering.} Notably, these two parametric mechanisms perform especially well relative to the other mechanisms in the first block of assignments, when the least amount of information is available to a mechanism. In many courses, grades are assigned after each assignment and using only information from that particular assignment. In those cases, the amount of information available to a grading mechanism would be most similar to the amount of information available to the mechanisms in the first block of assignments in our experiments. This result mirrors the results from the experiments with ABM, in which the relative advantage of these two mechanisms over certain other peer prediction mechanisms (e.g. OA) tends to decrease as the number of assignments per semester increases (\Cref{fig:quantifying-mi-abm}).
 
\label{subsubsection:quantifying-mi-real-results}

\section{Quantifying Robustness Against Strategic Reporting}
\label{section:robustness}
We now turn to the second dimension of our analysis---robustness against strategic reporting---which has traditionally been the focus of the peer prediction literature. The key question we seek to answer is: to what extent can an individual agent improve their outcome, according to the rewards assigned by a given mechanism, by strategically manipulating their reports? We explore this question in the context of our ABM below and in the context of the real data in \Cref{subsubappendix:strategic-results-real}.

In the setting of peer assessment, we expect that it is unlikely for students to expend the effort required to compute optimal deviations or play particularly complex strategies. Further, in educational settings, there are other ways to motivate honest, effortful grading, e.g. providing instruction and practice in grading accurately, that can complement the incentive properties of a peer assessment mechanism. Thus, in this line of inquiry, we seek to depart from the typical theoretical approach of considering all possible strategies. Instead, we focus on the relative performance of a few intuitive, easy-to-compute strategies.  

\subsection{Strategies}
\label{subsection:strategies}
In addition to truthful reporting, we consider the following types of strategies:

\vspace{1 ex}

First, there are \textit{uninformed strategies}---strategies that do not depend on an agent's signal---which we consider primarily as robustness checks:

\vspace{1 ex}
\noindent
\textbf{Report All 10s.}
Agents following this strategy constantly report 10, the highest possible score.

\vspace{1 ex}
\noindent
\textbf{Revert to the Prior.}
Agents following this strategy constantly report $\mu$, the expectation of the prior distribution of true scores (rounded to the nearest integer, $\lfloor \mu \rceil$, when applicable).
\vspace{1 ex}

The more interesting strategies are \textit{informed strategies}, which define a procedure by which agents manipulate their signals to generate their reports. We consider simple strategies for which there is some intuition as to why they might be present in or perform well in a peer assessment application:

\vspace{1 ex}
\noindent
\textbf{Hedge.}
A more realistic strategy for incorporating prior beliefs than fully reverting to the prior is to hedge reports back toward the prior mean, $\mu$. Piech et al. find some evidence for this tendency in their MOOC peer grading data \cite{Piech2013}. Agents following this strategy apply Bayesian reasoning by adopting a prior $\Beta{\mu}{10-\mu}$ on $p$, the value of the ground truth score divided by $10$. After receiving their signal, they update their prior and report 10 times the mean of their posterior distribution for $p$, which is given by $\frac{\mu + \text{signal}}{2}$, rounded to the nearest integer.

\vspace{1 ex}
\noindent
\textbf{Fix Bias.}
Real students may have some indication of the direction of their bias---whether they tend to assign grades that are too high or too low---and attempt to correct for that bias. To model this, agents following this strategy are given the sign of their bias. At the beginning of a semester, they each draw a constant ``bias correction'' term $\beta$ from the half-normal distribution that models the magnitude of biases drawn according to the bias distribution $\Normal{0}{1}$. For each submission that they grade, they report their signal plus or minus $\beta$---depending on the sign of their bias---rounded to the nearest integer in the report space.

\vspace{1 ex}
\noindent
\textbf{Add Noise.}
On the other hand, students who do not have some indication of the direction of their bias, or who think the direction of their bias varies from submission to submission, might still try to perform some correction. Similarly, students might try to guess (without any outside information) whether their signal is above or below the average or their peers and try to adjust their report accordingly. The result of either of these actions would look a lot like adding noise to their signal to generate their report. To model this, agents following this strategy draw a value $\nu \sim \Normal{0}{1}$ for each submission that they grade and report the sum of their signal and $\nu$, rounded to the nearest integer in the report space.

\vspace{1 ex}
\noindent
\textbf{Merge Signals.}
There is some evidence that when the report space is sufficiently large, students tend to under-utilize certain report values (particularly values that are low, but non-zero) \cite{Shnayder2016practical}. To model this, agents following this strategy map the signal space to a lower-dimensional report space and report the outcome of applying that map to their signal. The map from signals to reports is detailed in \Cref{tab:merge-map}.

\begin{table}
\renewcommand{\arraystretch}{1.2}
    \centering
    \begin{tabular}{|c|c|c|c|c|c|}
         \hline
         \textbf{Signal} & $0$ & $1,2,3$ & $4,5,6$ & $7,8,9$ & $10$\\
         \hline
         \textbf{Report} & $0$ & $3$ & $6$ & $\lfloor \mu \rceil$ & $10$\\
         \hline
    \end{tabular}
    \caption{Summary of the mapping of signals to reports for agents following the \textit{Merge Signals} strategy. \\ 
    $\lfloor \mu \rceil$ is the mean of the prior distribution, $\mu$, rounded to the nearest integer.}
    \label{tab:merge-map}
\end{table}

\subsection{Computational Experiments with ABM}
\label{subsection:strategic-abm}
In these experiments, we continue to focus on the best-performing mechanisms from the measurement integrity experiments, including the baselines, with the goal of identifying mechanisms that perform well according to both dimensions of our analysis. However, the degree to which the remaining mechanisms create incentives for deviating from truthful reporting is still of interest, so we record the results of our first experiment in this section with those mechanisms in \Cref{subsubappendix:strategic-results-abm}. We also corroborate the results of our experiments with ABM in experiments with the real data, for which results are shown in \Cref{subsubappendix:strategic-results-real}.

\subsubsection{Methods}
\label{subsubsection:strategic-abm-methods}
We explore how the incentives for deviating from truthful reporting change as the number of agents adopting some non-truthful strategy grows. We perform the following: 
\begin{enumerate}
        \item The number of strategic agents is varied from 10 to 90, in steps of size 10.
    At each step, we perform 100 iterations:
    \begin{enumerate}[i.]
        \item A population of 100 agents is initialized and a semester's worth of submissions and reports for grading them are generated, as in \Cref{section:quantifying-mi}.
        
        \item Rewards are assigned twice according to the given mechanism\footnote{Note that for parametric mechanisms, which compute a bias estimate for each agent, we subtract off the estimated biases to ``correct'' agents' reports in these experiments.} with a fixed random seed.
        In the first assignment, the number of truthful and strategic agents is as given by the current step.
        In the second assignment, one agent that reported truthfully in the first assignment is randomly selected. That agent modifies their reports according to the prescribed strategy. Due to the fixed random seed, every other factor is consistent with the first reward assignment.
        
        \item The gain in rank achieved by that single agent, i.e difference in the ranks according to the two reward assignments computed by the mechanism, is recorded.\footnote{For example, if the agent was given the 5th highest payment when they reported truthfully and the 10th highest payment when they reported strategically, a gain of -5, the difference in the ranks, would be recorded.}
    
    \end{enumerate}
        
    \item The mean gain in rank over the 100 iterations is computed for each step.
    
\end{enumerate}
A student's rank is calculated by counting the number of payments in the population of students that is greater than or equal to their own payment.

Note, in these experiments, we consider strategy profiles that are unlikely to arise organically from agents learning via repeated interactions with a mechanism and thus unlikely to be observed in real-world or laboratory data. Some of these, e.g. where nearly every student uses the \textit{Report All 10s} strategy, strain the incentive properties of many mechanisms, even those that perform well against many of the other strategy profiles that we consider. This more comprehensive exploration of the space of strategy profiles that is possible in our experiments---though not as exhaustive as theoretical results, which often consider the entire space---is a useful advantage of applying ABMs. 

\subsubsection{Results}
\label{subsubsection:strategic-abm-results}
\begin{figure}
\begin{subfigure}{\textwidth}
  \centering
  \includegraphics[width=0.32\linewidth]{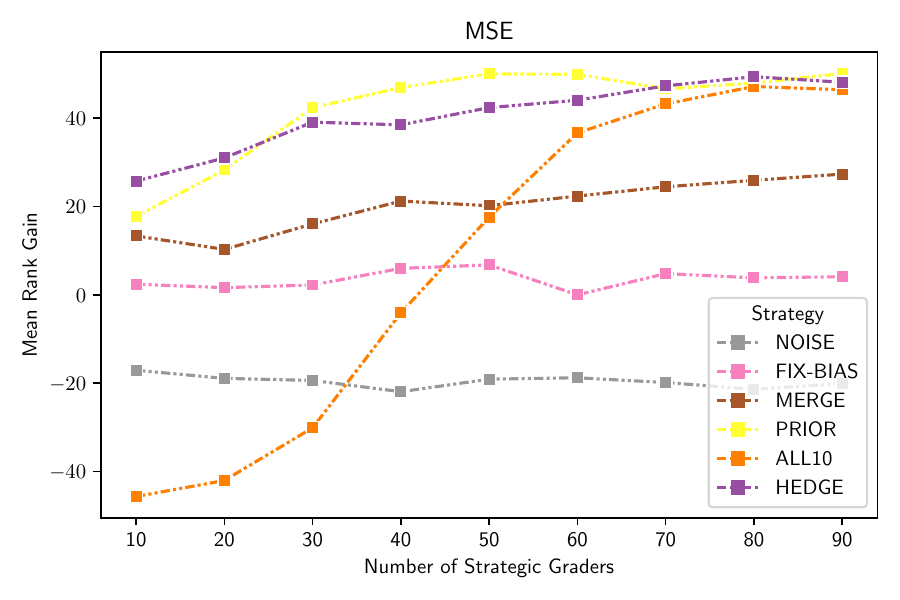}
  \hfill
  \includegraphics[width=0.32\linewidth]{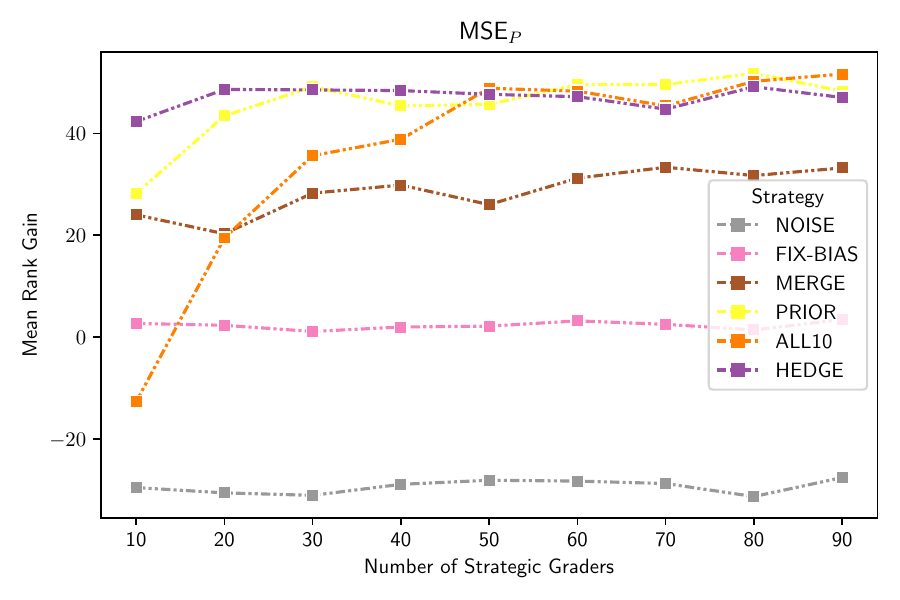}
  \hfill
  \includegraphics[width=0.32\linewidth]{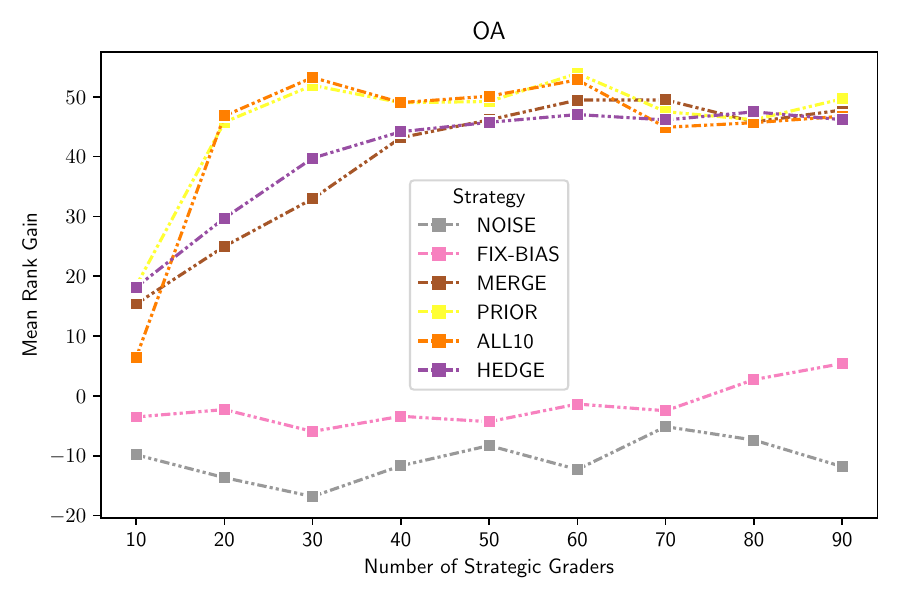}
  \hfill
  \includegraphics[width=0.32\linewidth]{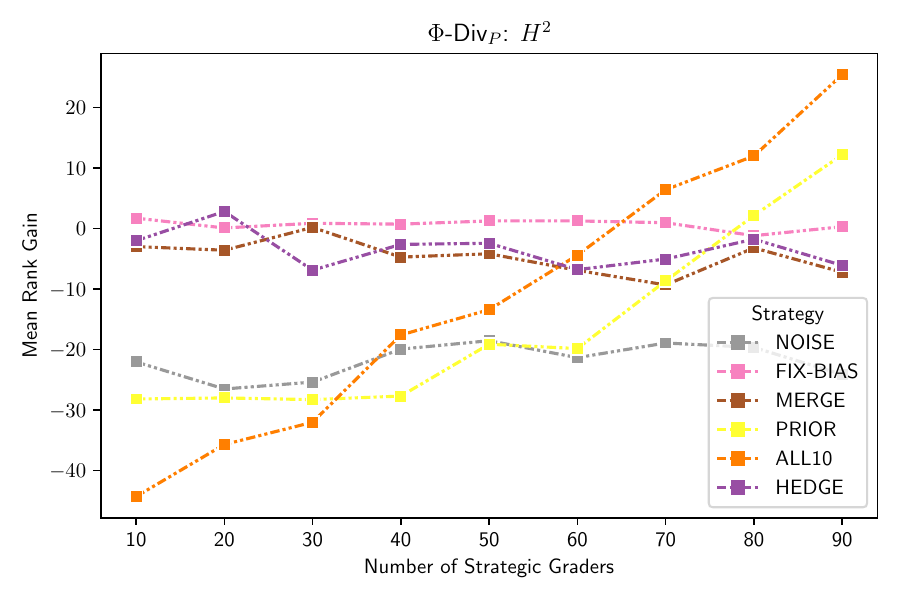}
  \includegraphics[width=0.32\linewidth]{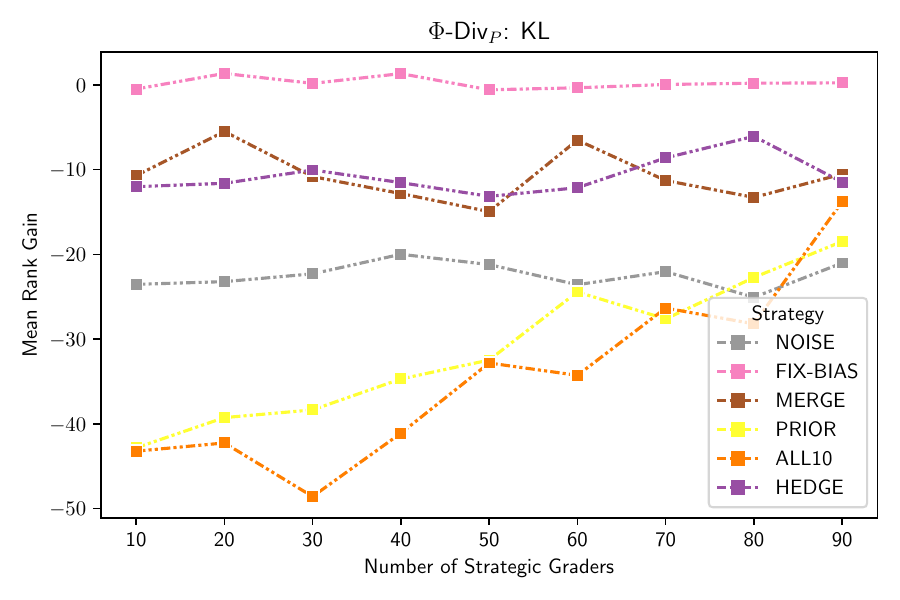}
  \label{subfig:deviation-mean-rank-gain}
\end{subfigure}
\caption{\textit{Quantifying Robustness with ABM.} For each mechanism and each strategy, the mean gain in rank achieved, \textit{ceteris paribus}, by a single agent changing their reports from truthful to strategic. The mean is taken over the outcomes of 100 simulated semesters as the number of other strategic agents varies in steps of size 10.
}
\label{fig:deviation-gain}
\end{figure}

\noindent
As shown in \Cref{fig:deviation-gain}, we find that the individual incentives for deviating from truthful reporting are strong under the baseline mechanisms, whereas \pPhiDiv: $H^2$ and \pPhiDiv: $KL$ are much more robust against strategic reporting. \pPhiDiv: $KL$, in particular, is robust against all the strategies and strategy profiles that we consider, in the sense that the mean rank gain achieved by deviating to strategic reporting is essentially never positive.

Predictably, not all strategies are equally effective against the less robust mechanisms. There are some noteworthy strategies that stand out: \textit{Hedge}, for example, exhibits both high average rank gain (that does not depend too much on the number of other strategic agents) and relatively low variance (see \Cref{appendix-fig:deviation-rank-variance} in \Cref{appendix:additional-results}) under the baseline mechanisms, which makes it a very attractive strategy for students deviating from truthful reporting. \textit{Report All 10s} is also a very potent strategy, under all the mechanisms shown except for \pPhiDiv: $KL$ , when the number of strategic agents is high. On the other hand, \textit{Add Noise} is universally ineffective and \textit{Fix Bias} is approximately neutral in all cases.

\subsection{Computational Experiments with Real Data} 
\label{subsection:strategic-results-real}

\subsubsection{Methods}
\label{subsubsection:strategic-real-methods}
In our experiments with simulated data, we explored how incentives for deviating from truthful reporting changed as the number of agents adopting a particular strategy increased. In the real data, however, we do not need to impose a strategy profile on the population of agents---the data were already generated according to some actual strategy profile adopted by the real students. As a result of this key difference, we modify the experiment described in \Cref{subsubsection:strategic-abm-methods} in the following manner.

For each semester in the real data, for each strategy, and for each mechanism:
    \begin{enumerate}
        \item The students, submissions, and reports for that semester are loaded from the data.
        
        \item For each student $s$:
        
        \begin{enumerate}[i.]
        
            \item Rewards are assigned twice according to the mechanism with a fixed random seed.
            In the first assignment, assignment of payments occurs without any modifications to the data.
            In the second assignment reports from student $s$ are modified according to the prescribed strategy. Due to the fixed random seed, every other factor is consistent with the first reward assignment.
        
            \item The gain in rank achieved by student $s$, i.e difference in the ranks according to the two reward assignments computed by the mechanism, is recorded.
        \end{enumerate}
        
    \item The mean and variance of the gain in rank over all students is computed for each mechanism, for each semester.
\end{enumerate}

Since we don't have access to a student's latent bias in the real data (and the true scores are noisy as a reference point) we do not consider the \textit{Fix Bias} strategy in these experiments.

\subsubsection{Results}
\label{subsubsection:strategic-real-results}
\begin{figure}
    \begin{subfigure}{\textwidth}
      \centering
      \includegraphics[width=0.32\linewidth]{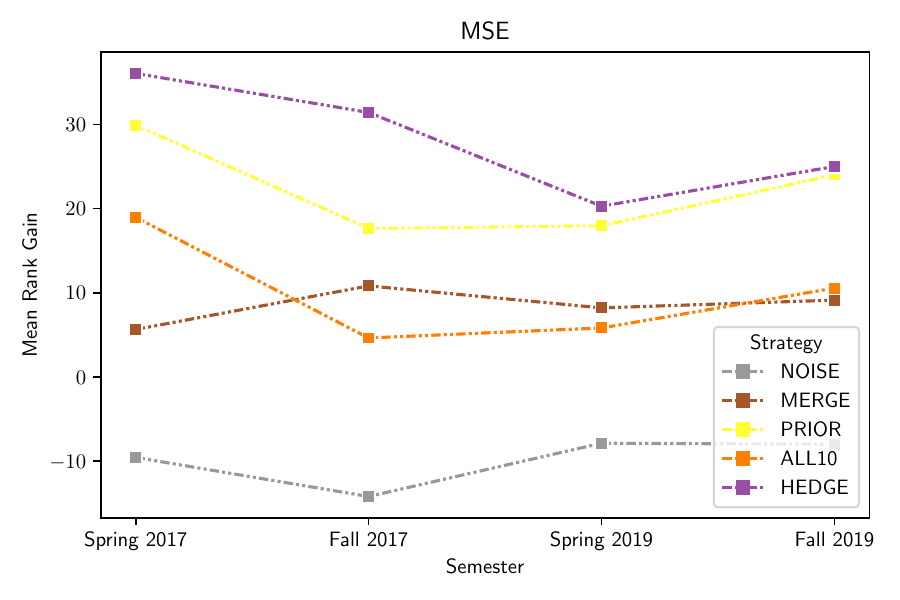}
      \hfill
      \includegraphics[width=0.32\linewidth]{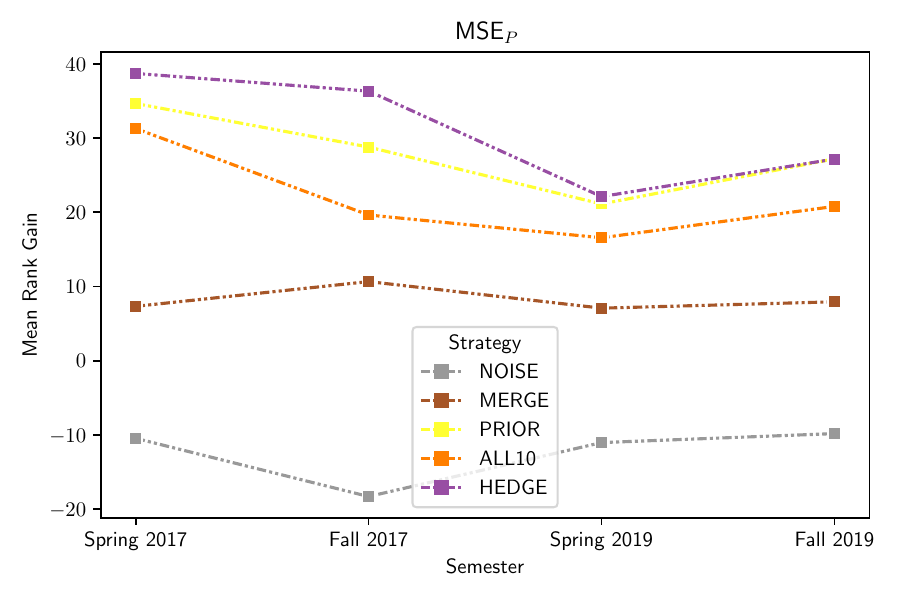}
      \hfill
      \includegraphics[width=0.32\linewidth]{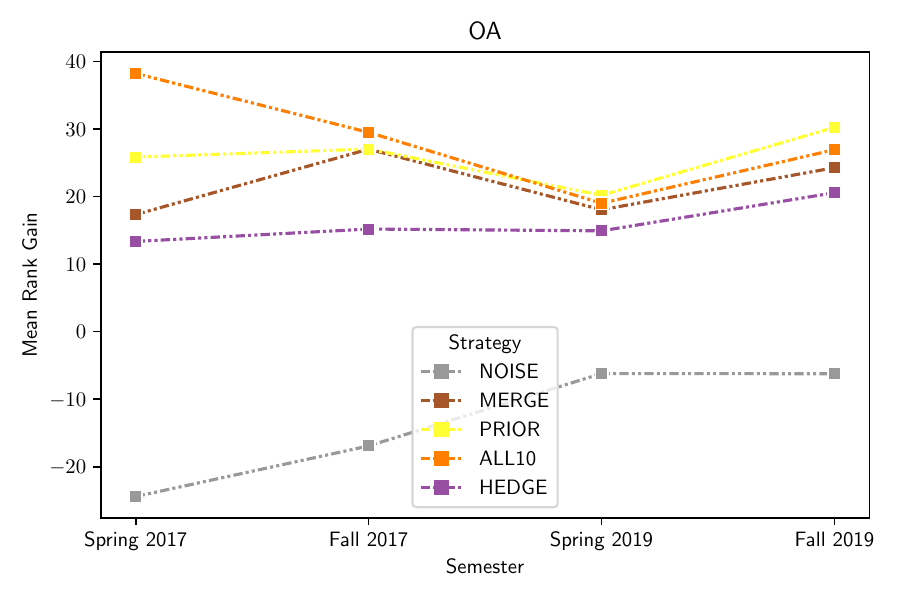}
      \hfill
      \includegraphics[width=0.32\linewidth]{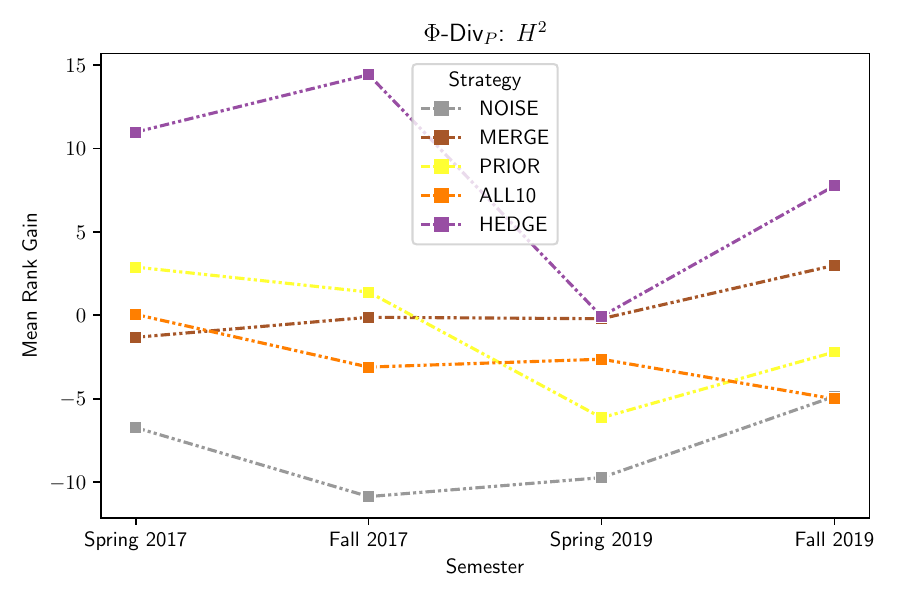}
      \includegraphics[width=0.32\linewidth]{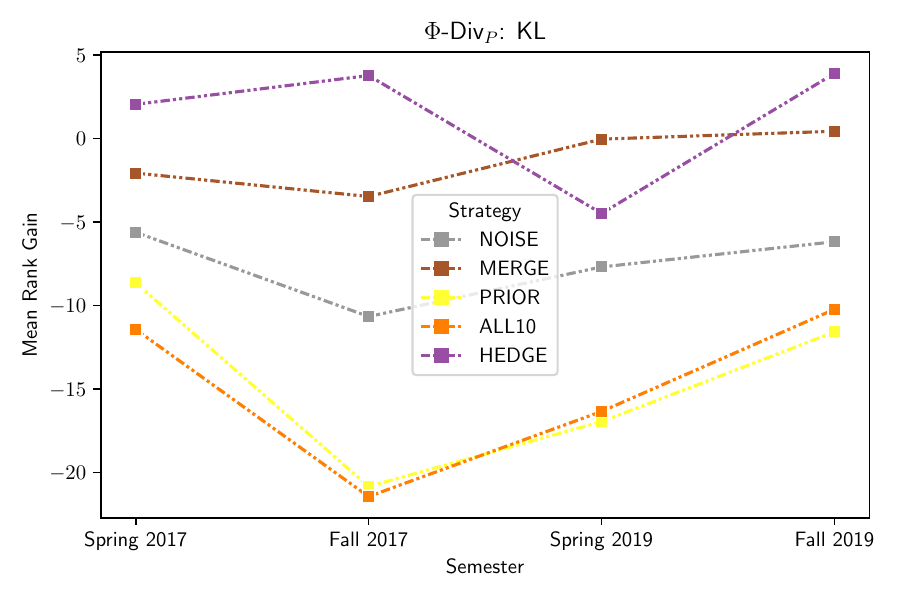} 
    \end{subfigure}
  \caption{\textit{Quantifying Robustness with Real Data.} For each mechanism, each strategy, and each semester (excluding \textit{Fix Bias}), the mean gain in rank achieved, \textit{ceteris paribus}, by a single agent changing their reports from truthful to strategic. The mean is taken over the population of students in the given semester.
}
  \label{fig:mean-rank-gain-real}
\end{figure}

The results of this experiment involve the mean gain (\Cref{fig:mean-rank-gain-real,appendix-fig:mean-rank-gain-real}) and the variance of the gain (\Cref{appendix-fig:rank-gain-variance-real}) over the population of students achieved by each student deviating (one at a time) to each strategy in each semester.

As in our experiments with measurement integrity, this experiment with the real data largely corroborates our analogous experiments with ABM. Each \pPhiDiv\,mechanism, the PTS mechanism, and to some extent the non-parametric $\Phi$-Div: TVD mechanism, are robust against strategic reporting for many or all semesters and strategies, whereas the remaining mechanisms are consistently susceptible to various kinds of strategic behavior.

Unlike in our computational experiments with ABM, though, certain seemingly unlikely strategy profiles (e.g. where nearly every student reports 10 regardless of their signal), do not come into play in these experiments with the real data, since there is only one ``strategic'' agent considered at a time. All of the other students reports (though they may also have resulted from some unknown strategy) are taken as given. As a result, certain strategies like \textit{Report All 10s} are somewhat less potent than in the experiments with ABM.

\section{Discussion}
\label{section:discussion}
We introduced measurement integrity as a novel property to consider in the design and analysis of peer prediction mechanisms. Alongside the more well-studied property of robustness against strategic reporting, measurement integrity plays an important role in understanding their practical performance. We focused on quantifying these properties empirically, using computational experiments with both an ABM and with real data. As a result, we were able to meaningfully differentiate mechanisms from the peer prediction literature in a way that has not been possible with theoretical analysis alone. Ultimately, we identified an apparent trade-off between our two dimensions of analysis (\Cref{fig:2D-tradeoff}) and found that parametric peer prediction mechanisms were best able to balance the two properties that characterize those dimensions.

Our unambiguous results suggest that our methodology itself---performing computational experiments to quantify mechanisms' empirical properties---is useful for investigating desiderata, including but not limited to measurement integrity and robustness against strategic reporting. In particular, that our approach facilitates more direct comparisons between mechanisms and uncovers consequences of implementation choices that are often abstracted away in theoretical analysis may be useful to practitioners looking select a peer prediction mechanism to deploy in a particular application. 

Although our results are grounded in the contexts of our experiments, we expect many of those results to readily generalize. Our ABM is structurally similar to a model from the peer assessment literature that was validated using a large peer grading dataset from Massive Open Online Courses (MOOCs) by Piech et al. \cite{Piech2013}. As a result, it is reasonable to expect that the results of our experiments with that ABM will, at the very least, apply to that important class of peer assessment settings. The corroboration of the results of our experiments with ABM by similar experiments with real peer grading data (not from a MOOC) suggests even broader applicability. Additionally, the relative scarcity of data for any particular grading task or agent in the peer assessment setting (\Cref{section:peer-assessment}) appears to be a significant driver of many of our results. This suggests that much of the intuition that we gain from our case study in peer assessment is likely to generalize to other settings where data is similarly sparse. Future theoretical work should explore the design of mechanisms---including parametric mechanisms, which seem well-suited for this purpose---that are less reliant on an abundance of data than the current mechanisms from the peer prediction literature. 

We also expect that the trade-off between measurement integrity and robustness against strategic behavior will remain in other settings. In \Cref{appendix:additional-mechanisms}, we discuss experiments with additional mechanisms that have not yet been studied in the peer prediction literature. We find the trade-off to be persistent---none of the novel mechanisms were able to significantly extend the Pareto frontier established by the mechanisms from the current peer prediction literature. On the other hand, particular challenges of peer assessment, like the scarcity of data, are not a concern in other settings of interest. This indicates that a mechanism's performance in our experiments will not necessarily predict its performance universally. 

\section*{Acknowledgements}
The authors would like to thank Professor Jason Hartline at Northwestern University and his student Michalis Mamakos for sharing peer grading data with us. Thanks also to Fang-Yi Yu and Yichi Zhang for helpful discussions in developing the ideas explored in this work. 

\bibliographystyle{ACM-Reference-Format}
\bibliography{arXiv}

\newpage

\appendix
\section*{{\Large Supplementary Material}}
\vspace{1 ex}

\renewcommand\thefigure{\thesubsection\alph{figure}} 
\renewcommand\thesubfigure{\roman{subfigure}}
\makeatletter
\renewcommand\p@subfigure{\thefigure.}
\makeatother

\begin{table}[h]
    \centering
    \begin{tabular}{| c || c | c |}
        \hline
         Mechanism & Equilibrium Concept & Sufficient Assumption \\
         \hline
         DMI  & Dominantly Truthful & Strictly Correlated \\
         MSE & None & - \\
         \pMSE & None & - \\
         OA & Truthful & Self-Dominating \\
         PTS & Helpful Reporting & Self-Predicting \\
         $\Phi$-Div & $\epsilon$-Strongly Truthful & Stochastically Relevant \\
         \pPhiDiv & $\epsilon$-Strongly Truthful & Stochastically Relevant \\
         \hline
    \end{tabular}
    \caption{The equilibrium concepts related to truthful reporting that are induced by each mechanism and the weakest known assumption on the joint prior distribution of signals that is sufficient to guarantee that inducement.}
    \label{tab:theoretical-props}
\end{table}

\section{Equilibrium Concepts and Sufficient Assumptions}
\label{appendix:concepts-and-assumptions}

\subsection{Equilibrium Concepts}
\label{subappendix:equilibrium-concepts}

\noindent
\textbf{Helpful Reporting} \cite{Faltings2017}. An agent with prior belief $\Pr[\cdot]$ is said to follow a \textit{helpful reporting} strategy with respect to a publicly-known distribution $R$ if both:
\begin{enumerate}
    \item The agent reports truthfully if $R$ is ``close enough'' to $\Pr[\cdot]$.
    
    \item When the agent is not truthful, their report is never ``over-represented'' in $R$. That is, if
        $R[x] \geq \Pr[x]$, given that their signal is $x' \neq x$, the agent does not report $x$.
\end{enumerate}

A strategy profile $\sigma$ is an  \textit{ex post subjective equilibrium} if no agent can improve their expected payoff by deviating from $\sigma$, given that all other agents' posterior beliefs given their signals (and their observations of the publicly-known distribution of reports $R$) respect any assumption made (e.g. the self-predicting assumption) that constrains the form of the joint prior distribution of signals.

A \textit{helpful reporting equilibrium} is an \textit{ex post subjective equilibrium} in which each agent adopts a helpful reporting strategy. A \textit{helpful reporting} mechanism admits a helpful reporting equilibrium.

\vspace{1 ex}
\noindent
\textbf{Truthful} \cite{Faltings2017}. A \textit{truthful equilibrium} is an \textit{ex post subjective equilibrium} in which each agent adopts a truthful reporting strategy, i.e reports their observed signal $s$. A \textit{truthful} mechanism admits a truthful equilibrium.

Note that while this definition of \textit{truthful} corresponds to its usage above in \Cref{tab:theoretical-props}, it takes on a larger range of meanings in the peer prediction literature as a whole.
    
\vspace{1 ex}
\noindent
\textbf{Strongly Truthful} \cite{Schoenebeck2021}. A \textit{strongly truthful} mechanism admits a Bayes-Nash equilibrium in which agents report truthfully and in which the following properties hold:
\begin{enumerate}
    \item The expected payment to each agent is maximized over the set of payments to that agent in any Bayes-Nash equilibrium.
    
    \item The expected payment to each agent is \textit{strictly} higher than their payment in any Bayes-Nash equilibrium induced by a strategy profile that is not a \textit{permutation strategy profile}.
\end{enumerate}
A \textit{permutation strategy} is a strategy in which an agent fixes a permutation of the signal space and then, for each task, reports the image of their signal for that task under the permutation. A \textit{permutation strategy profile} is a strategy profile in which each agent adopts a permutation strategy. 

\vspace{1 ex}
\noindent
$\epsilon$-\textbf{Strongly Truthful} \cite{Schoenebeck2021}. An $\epsilon$-\textit{strongly truthful} mechanism is approximately strongly truthful, in the sense that there exists a strongly truthful payment scheme such that 1) the expected payment to each agent in the truthful Bayes-Nash equilibrium is at most $\epsilon$ away from this strongly truthful payment scheme; and 2) the expected payment to each agent in any strategy profile is bounded above by the corresponding payment in this strongly truthful payment scheme.     

For the $\Phi$-Div and \pPhiDiv\,mechanisms, the optimal value of $\epsilon$ for which $\epsilon$-strong truthful-ness is achieved depends on how closely the estimated joint-to-marginal-product ratio $\hat{\text{JP}}$ (see \Cref{section:mechanisms}) approximates the true joint-to-marginal-product ratio $\text{JP}$.

\vspace{1 ex}
\noindent
\textbf{Dominantly Truthful} \cite{Kong2020}. A \textit{dominantly truthful} mechanism admits a \textit{dominant strategy} equilibrium in which agents report truthfully. That is, it admits an equilibrium in which both:
\begin{enumerate}
    \item For every agent, truthful reporting maximizes their expected payment no matter what strategies other agents play (i.e. truthful reporting is a \textit{dominant strategy}.)
    
    \item For every agent, if they believe that at least one of their informative peers\footnote{Peers with whom the agent forms a pair with \textit{strictly correlated} signals.} will tell the truth, then reporting truthfully pays \textit{strictly} higher, in expectation, than any non-permutation strategy (see above).
\end{enumerate}

\subsection{Sufficient Assumptions}
\label{subappendix:sufficient-assumptions}
\textbf{Self-Dominating} \cite{Faltings2017}. The joint prior distribution of signals is \textit{self-dominating} if and only if, for each agent, their observed signal $s$ is the most-probable outcome under the posterior distribution conditioned on having observed $s$ (for each possible signal $s$):
\[
    \Pr[s|s] > \Pr[x|s], \; \forall x \neq s. 
\]
The posterior distribution mentioned in the latter expression of the definition (after the ``iff'') is a distribution for the signal of a peer whose signal is independent (conditioned on the ground truth) from the agent's own signal
 
\vspace{1 ex}
\noindent
\textbf{Self-Predicting} \cite{Faltings2017}. The joint prior distribution of signals is \textit{self-predicting} if and only if, for each agent, the relative increase in probability for their observed signal $s$  from the agent's prior distribution to their posterior distribution is greater than for any other possible outcome:
\[
    \frac{\Pr[s|s]}{\Pr[s]} > \frac{\Pr[x|s]}{\Pr[x]}, \; \forall x \neq s. 
\]
As above, the prior and posterior distributions in the latter expression of this definition (after the ``iff'') are distributions for the signal of a peer whose signal is independent (conditioned on the ground truth) from the agent's own signal.

\vspace{1 ex}
\noindent
\textbf{Strictly Correlated }\cite{Kong2020}, \cite{Schoenebeck2021}. A pair of agents $(a_1, a_2)$ have \textit{strictly correlated} signals (represented by random variables $S_1$ and $S_2$, respectively) if the determinant of the agents' joint probability distribution of signals (written as a matrix \textbf{P} with entries $\textbf{P}_{ij} = \Pr[S_1 = s_i, S_2 =s_j]$ for each pair of possible signals $(s_i, s_j)$) is non-zero. The \textit{strictly correlated} assumption for the DMI mechanism is that each agent has at least one \textit{informative peer}---a peer with whom the agent forms a pair with \textit{strictly correlated} signals.

\vspace{1 ex}
\noindent
\textbf{Stochastically Relevant} \cite{Schoenebeck2021}. The joint prior distribution of signals is \textit{stochastically relevant} if each agent's posterior distribution given their own signal $s$ is unique for each possible signal. That is,
\[
    \Pr[S | s] \neq \Pr[S | s'] \text{ for each pair } s, s' \text{ such that } s' \neq s,
\]
where $\Pr[S | \cdot]$ denotes the entire distribution over the possible signals of a peer whose signal is independent (conditioned on the ground truth) from the agent's own signal. This is in contrast to the expression $\Pr[s|\cdot]$, used above, which denotes the specific probability $\Pr[S=s|\cdot]$ for a possible signal $s$ under that distribution.

\section{Implementation Details}
\label{appendix:implementation}
\setcounter{figure}{0}

\subsection{Implementing Peer Prediction Mechanisms}
\label{subappedix:implementation-mechanisms}
Note that in mechanisms that involve pairing an agent with another agent in order to compute their scores on a grading task (i.e. generating a report for one submission), we take the expectation over all of the possible pairings to reduce the variance of the scores. 

\subsubsection{Non-Parametric Mechanisms} \hfill
\label{subsubappedix:implementation-nonparametric}

\vspace{1 ex}
\noindent
\textbf{Output Agreement (OA) Mechanism.}
The implementation is trivial.

\vspace{1 ex}
\noindent
\textbf{Peer Truth Serum (PTS) Mechanism.} 
To score a grading task for a given submission, a pair of agents that completed that task is selected and their reports are compared. If their reports are equal, then they are awarded $\frac{1}{R[\text{report}]}$, where $R[\text{report}]$ is the probability of the given report under the distribution $R$.
    
$R$ is repeatedly updated over the course of a simulated semester via a histogram $H$ of report values. After the payments for an assignment are computed by the mechanism, the report values submitted for that assignment are added to $H$. Then, $R$ is recomputed by normalizing $H$ with Laplace (add-one) smoothing.  In particular, this means that $R$ is initialized to the uniform distribution.
    
\vspace{1 ex}
\noindent
\textbf{$\Phi$-Divergence Pairing ($\Phi$-Div) Mechanism.} 
The mechanism randomly divides the tasks in half and uses each half to compute $\hat{\text{JP}}$, an estimate of $\text{JP}$, the joint-to-marginal-product ratio of the reports\footnote{The value of the joint distribution of reports evaluated at the given pair of reports divided by the value of the product of the marginal distributions of reports evaluated at the given pair of report.} for the other half by counting the frequency of pairs of report values given by pairs of agents grading the same submission (to estimate the joint distribution of reports) and counting the overall frequency of report values (to estimate the marginal distribution of reports).
    
Using these estimates, scores are computed for each grading task, as follows: For each task $b$, referred to as the \textit{bonus task}, agents are paired with another agent who completed $b$. Then, a pair of \textit{penalty tasks} $p$ and $q$, distinct from each other and from the bonus task, are randomly chosen (one for each agent). The pair of agents is awarded the quantity $\partial \Phi(\hat{\text{JP}}(x_b, y_b)) - \Phi^*(\partial \Phi(\hat{\text{JP}}(x_{p}, y_{q})))$, where $\Phi^*$ denotes the \textit{convex conjugate} of $\Phi$, $x_i$ and $y_j$ denote the first agent's reports on task $i$ and the second agent's report on task $j$, respectively, and $\hat{\text{JP}}$ is the empirical estimate of $\text{JP}$ computed using the other half of the tasks.

Recall that for mechanisms that compute payments in pairs, we assign payments according to the average payment over all possible pairings.  When working with the real peer grading data, it is occasionally the case that a pair of agents does not have a valid pair of penalty tasks to use. That is, there are some pairs of agents who graded only the same pair of submissions for a given assignment, so distinct penalty tasks cannot be chosen. In those cases, we simply skip over that pairing when computing the average payments over all possible pairings. If that is the only possible pairing (i.e. only those two agents graded that submission), then we ignore those peer grades altogether when computing payments according to this mechanism.

\vspace{1 ex}
\noindent
\textbf{Determinant-based Mutual Information (DMI) Mechanism.} 
The DMI mechanism was developed by Kong \cite{Kong2020}. It pays each agent according to a sum of unbiased estimates of the square of the Determinant-based Mutual Information between a random variable drawn from the distribution of their reports and random variables drawn from the distributions of the reports of each other agent who completed the same tasks. 

The estimate of the square of the Determinant-based Mutual Information between two random variables depends on the product of the determinants of two matrices, each encoding the frequency of pairs of reports on one part of a bifurcation of the tasks being considered for scoring. 
As a result, it is easy to see that in our ABM, where there are 11 possible report values and agents complete only 4 grading tasks, the DMI mechanism will always pay every agent 0. Further, the DMI mechanism benefits from having the sets of tasks that agents complete overlap as much as possible. This stands in contrast to the other mechanisms we consider, where the number of tasks mutually completed by a pair of agents---as long as it is at least one---is not very significant. To make the mechanism as functional as possible while still remaining faithful to the original description of the mechanism, we make the following adjustments in our implementation: 
\begin{enumerate}
    \item We project the report space down to a space of 2 possible reports, so that each report is either 0---indicating that a submission has below average quality ($< 7$)---or 1---indicating that its quality is at least average ($\geq 7$).
    
    \item We partition the agents into clusters of 4 agents so that each cluster grades the same 4 submissions (namely, the submissions from another cluster). 
\end{enumerate}
Note that these modifications have a significant impact on the effect of strategic behavior in our experiments. Since the report space is simplified to be binary, oftentimes, a strategy applied to a signal will produce a report equal to the signal (in the binary report space), so in many cases, there is no difference between strategic and truthful reporting. Also, these modifications are not possible in experiments with the real data, so the DMI mechanism is excluded completely from those experiments.
    
\subsubsection{Parametric Mechanisms}
\label{subsubappendix:implementation-parametric}
Recall model \PGone\,from Piech et al. \cite{Piech2013}, which, with the appropriate parameters, provides a reasonable continuous approximation to our mostly discrete model from \Cref{section:model} in which reliability is a proxy for effort:
\begin{alignat*}{2}
    (\text{True Score}&) \quad g_{i, j}^* && \sim \Normal{7}{2.1} \text{for each submission $s_{i,j}$},\\
    (\text{Reliability}&) \quad \tau_{i} && \sim \Gam{10/1.05}{10} \text{for each agent $i$},\\
    (\text{Bias}&) \quad b_{i} && \sim \Normal{0}{1} \text{for each biased agent $i$ (and 0 for each unbiased agent)},\\
    (\text{Observed Score}&) \quad z_{i, j}^k && \sim \Normal{g_{i, j}^* + b_k}{\tau_k^{-1}} \text{for each submission $s_{i, j}$ and assigned grader $k$},
\end{alignat*}
where $\mathcal{G}$ is a Gamma distribution. Note that the hyperparameters $\alpha_0 = 10/1.05$ and $\beta_0 = 10$ for the Gamma distribution used to model reliability were chosen by inspection subject to having the correct expected value for a continuous effort agent (and for a uniformly random agent chosen from a population of binary effort agents with an equal number of active and passive graders). The PDF of the Gamma function with those hyperparameters is plotted in \Cref{appendix-fig:gamma-pdf}.

\begin{figure}
    \centering
    \includegraphics[width=0.333\textwidth]{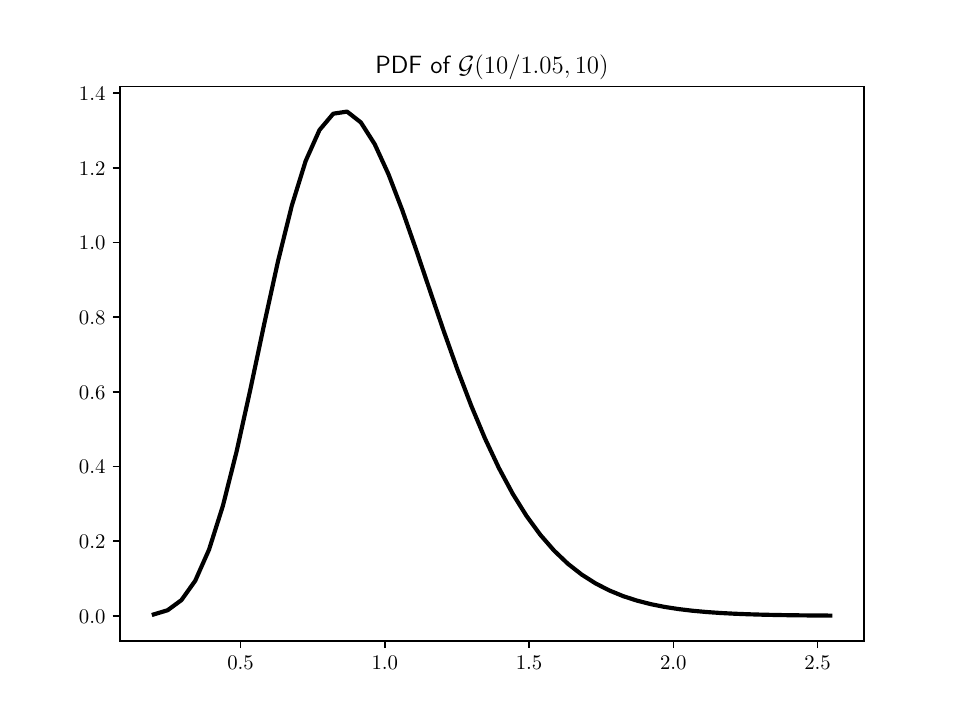}
    \caption{PDF of $\Gam{10/1.05}{10}$ on values for which its support is non-negligible.}
    \label{appendix-fig:gamma-pdf}
\end{figure}

Our procedure for estimating the parameters of model \PGone, inspired by Chakraborty et al. \cite{Chakraborty2020}, is as follows:
    \begin{itemize}
    \item[]
    \begin{itemize}
        \item[]
    \begin{itemize}
        \item[]
    \begin{itemize}
        \item[\textbf{Initialize:}] Set the bias and reliability of each agent and the score for each submission equal to their expectation (0, $1/1.05$, and 10 respectively). 
        
        \vspace{1 ex}
        
        \item[\textbf{Update:}] For each submission $s_{i, j}$, update the true score estimate $\hat{g}_{i, j}$ as in equation (1) from Chakraborty et al. \cite{Chakraborty2020}:
        \[
            \hat{g}_{i, j} = \frac{7 \cdot \sqrt{(2.1)^{-1}} + \sum_{k} \sqrt{\hat{\tau}_k}(r_{i,j}^k - \hat{b}_k)}{\sqrt{(2.1)^{-1}} + \sum_{k} \sqrt{\hat{\tau}_k}},
        \]
        where the $k$ in both sums varies over the graders of submission $s_{i, j}$, $\hat{\tau}_k$ and $\hat{b}_k$ are the estimated reliability and estimated bias, respectively, of grader $k$, and $r_{i,j}^k$ is grader $k$'s report for submission $s_{i, j}$.
        
        \vspace{1 ex}
        
        For model settings with unbiased agents, we skip the step of estimating the bias and fix all the biases as 0. Otherwise, for each agent $k$, update the bias estimate $\hat{b}_{k}$ as the mean of the posterior distribution of a Bayesian update from the conjugate prior $\Normal{0}{1}$ given the agent's reports and the estimated true scores:
        \[
            \hat{b}_{k} = \frac{\hat{\tau}_k \cdot \sum_{s_{i, j}}(r_{i,j}^k - \hat{g}_{i, j}) }{1 + n \cdot \hat{\tau}_k},
        \]
        where the sum in the numerator varies over the submissions graded by agent $k$ for the fixed assignment $j$ and $n$ is the number terms in that sum.
        
        \vspace{1 ex}
        
        For each agent $k$, update the reliability estimate $\hat{\tau}_k$ as the mean of the posterior distribution of a Bayesian update from the conjugate prior $\Gam{10/1.05}{10}$ given the agent's reports, the agent's estimated bias, and the estimated true scores:
        \[
            \hat{\tau}_k = \frac{\frac{10}{1.05} + \frac{n}{2}}{10 + \frac{1}{2} \cdot \sum_{s_{i, j}}(r_{i,j}^k - (\hat{g}_{i, j} + \hat{b}_k))^2},
        \]
        where the sum in the denominator varies over the submissions graded by agent $k$ for the fixed assignment $j$ and $n$ is the number of terms in that sum.
        
        \vspace{1 ex}
        
        \item[\textbf{Terminate:}] When the $\ell_2$ norm of the difference between the vector of estimated true scores after the previous round and the vector of estimated true scores after the current round is less than or equal to $0.0001$, terminate. 
        
        In our implementation, we also imposed a maximum of 1000 iterations of the update procedure, after which the procedure would terminate even if the $\ell_2$ norm condition were not met, but after adding the priors to the bias and reliability estimates, this extra termination condition was never applicable.
    \end{itemize}
    \end{itemize}
    \end{itemize}
    \end{itemize}

\vspace{1 ex}
\noindent
\textbf{Parametric MSE (\pMSE) Mechanism.} Given the estimation procedure outlined above and the descriptions from \Cref{subsection:parametric}, the implementation of the this mechanism is trivial.

\vspace{1 ex}
\noindent
\textbf{Parametric $\Phi$-Divergence Pairing (\pPhiDiv) Mechanism.}
Given the estimation procedure outlined above and the descriptions from \Cref{subsection:parametric} (and the derivation from \Cref{appendix:optimal-scoring-rules}), the implementation of the parametric mechanisms is straightforward. However, we make some adjustments to improve its performance. In simulations, we find that this mechanism performs best when the reliability estimates are heavily regularized and the simplest way to achieve close-to-optimal (if not optimal) performance is to set all the reliability estimates equal to $\frac{1}{0.7}$, the approximate reliability of an ``active grader'' in our setting with binary effort (see \Cref{appendix:alternative-quality}). This is akin to replacing model \PGone\,with model \PGone-\textbf{bias} \cite{Piech2013}. Consequently, only the bias estimates given by the estimation procedure are used (and only in settings with biased agents).

As with the non-parametric version of this mechanism, in the real data, it is occasionally the case that a pair of agents does not have a valid pair of penalty tasks to use. We handle those cases in the same way for the parametric and non-parametric versions of the mechanism (see \Cref{subsubappedix:implementation-nonparametric}).

\subsection{Software}
\label{subappedix:software}
Our ABM and all experiments are implemented Python 3 and use the NetworkX \cite{NetworkX}, NumPy \cite{NumPy}, SciPy \cite{SciPy}, and Scikit-learn \cite{Scikit-learn} packages. Results of the experiments are plotted using the pandas \cite{pandas1.2.1, pandas2010}, Matplotlib \cite{Matplotlib}, and seaborn \cite{seaborn0.11.1} packages.

\section{Computing the Joint-to-Marginal-Product Ratio under model \texorpdfstring{\PGone}{\textbf{PG}-1}}
\label{appendix:optimal-scoring-rules}
\setcounter{figure}{0}

Recall that in model \PGone, the true score for any submission is $g^* \sim \Normal{\mu}{\sigma^2}$, and each grader $i$ has an underlying bias score $b_i$ and reliability score $\tau_i$. Then, $i$'s observed grade is $s_i \sim \Normal{g_{i}^* + b_i}{\tau_i^{-1}}$.\footnote{Here we use $g_{i}^*$ instead of $g^*$, because the submissions graded by $i$ and $j$ are sometimes different, e.g. when considering the penalty tasks.}

Now, consider two agents $i$ and $j$ receiving signals $x = s_i$ and $y = s_j$. For the purposes of the \pPhiDiv\,mechanism, we need to compute the joint-to-marginal-product ratio of $x$ and $y$: 
\[
    \text{JP}(x, y) = \frac{P_{X, Y}(x, y)}{P_X(x) P_Y(y)}.
\]

Considering random variables $X$ and $Y$\footnote{The distributions of $X$ and $Y$ follow from writing $X$ and $Y$ as the sum of two normally-distributed random variables $G \sim \Normal{\mu}{\sigma^2}$ and $B_X \sim \Normal{b_i}{\tau_i^{-1}}$ for $X$ or $B_Y \sim \Normal{b_j}{\tau_j^{-1}}$ for $Y$.} such that
\begin{align*}
    X \sim \Normal{\mu + b_i}{\sigma^2 + \tau_i^{-1}} \text{ and } Y \sim \Normal{\mu + b_j}{\sigma^2 + \tau_j^{-1}}
\end{align*}
 and setting $\mu_i = \mu + b_i$ and $\mu_j = \mu + b_j$, we have
\begin{align*}
    P_X(x) = \frac{\exp\left(-\frac{1}{2} \left(\frac{x - \mu_i}{\sqrt{\sigma^2 + \tau_i^{-1}}}\right)^2\right)}{\sqrt{\sigma^2 + \tau_i^{-1}} \sqrt{2 \pi}} 
    \text{ and } 
    P_Y(y) = \frac{\exp\left(-\frac{1}{2} \left(\frac{y - \mu_j}{\sqrt{\sigma^2 + \tau_j^{-1}}}\right)^2\right)}{\sqrt{\sigma^2 + \tau_j^{-1}} \sqrt{2 \pi}},
\end{align*}
and so we can write the product of the marginal distributions of $X$ and $Y$ as
\begin{align*}
    P_X(x) P_Y(y) = \frac{\exp\left(-\frac{1}{2} 
    \left[x - \mu_i, y - \mu_j \right] \cdot 
    \begin{bmatrix}
    \frac{1}{\sigma^2 + \tau_i^{-1}} & 0\\
    0 & \frac{1}{\sigma^2 + \tau_j^{-1}}
    \end{bmatrix} \cdot 
    \begin{bmatrix}
    x - \mu_i\\
    y - \mu_j
    \end{bmatrix} \right)}{2\pi \sqrt{\sigma^2 + \tau_i^{-1}} \sqrt{\sigma^2 + \tau_j^{-1}}}.
\end{align*}

Then, we can compute that the joint distribution of $X$ and $Y$ is the following multivariate Gaussian:
\[
    \begin{bmatrix}
    X\\
    Y
    \end{bmatrix}
     \sim \Normal{
    \begin{bmatrix}
    \mu_i\\
    \mu_j
    \end{bmatrix}}
    {\begin{bmatrix}
    \sigma^2 + \tau_i^{-1} & \sigma^2\\
    \sigma^2 & \sigma^2 + \tau_j^{-1}
    \end{bmatrix}
    }.
\]

Let 
\[
\Sigma = 
    \begin{bmatrix}
    \sigma^2 + \tau_i^{-1} & \sigma^2\\
    \sigma^2 & \sigma^2 + \tau_j^{-1}
    \end{bmatrix},
\]
and as a result 
\begin{align*}
    & |\Sigma| = (\sigma^2 + \tau_i^{-1})(\sigma^2 + \tau_j^{-1}) - (\sigma^2)^2 = \frac{\sigma^2\tau_i + \sigma^2 \tau_j + 1}{\tau_i \tau_j},
    \quad \Sigma^{-1} = 
    \frac{1}{|\Sigma|} 
    \begin{bmatrix}
    \sigma^2 + \tau_j^{-1} & -\sigma^2\\
    - \sigma^2 & \sigma^2 + \tau_i^{-1}
    \end{bmatrix}.
\end{align*}

We can write the joint distribution of $X$ and $Y$ as
\[
    P_{X,Y}(x, y) = \frac{\exp\left(-\frac{1}{2} \left[x - \mu_i, y - \mu_j \right] \cdot \Sigma^{-1} \cdot
    \begin{bmatrix}
    x - \mu_i\\
    y - \mu_j
    \end{bmatrix}
    \right)}{\sqrt{(2\pi)^2 |\Sigma|}}.
\]

Then, we can write out the joint-to-marginal-product ratio explicitly and simplify:
\begin{align*}
    & \text{JP}(x, y) = \frac{P_{X, Y}(x, y)}{P_X(x) P_Y(y)}
    = 
    \frac{
        \frac{\exp\left(-\frac{1}{2} \left[x - \mu_i, y - \mu_j \right] \cdot \Sigma^{-1} \cdot
        \begin{bmatrix}
            x - \mu_i\\
            y - \mu_j
        \end{bmatrix}
        \right)}
        {\sqrt{(2\pi)^2 |\Sigma|}}}
    {
        \frac{\exp\left(-\frac{1}{2} \left[x - \mu_i, y - \mu_j \right] \cdot 
        \begin{bmatrix}
            \frac{1}{\sigma^2 + \tau_i^{-1}} & 0\\
            0 & \frac{1}{\sigma^2 + \tau_j^{-1}}
        \end{bmatrix} \cdot 
        \begin{bmatrix}
            x - \mu_i\\
            y - \mu_j
        \end{bmatrix} \right)}
    {2\pi \sqrt{\sigma^2 + \tau_i^{-1}} \sqrt{\sigma^2 + \tau_j^{-1}}}}\\
    & = 
    \frac{
        \exp\left(-\frac{1}{2} \left[x - \mu_i, y - \mu_j \right] \Sigma^{-1}
        \begin{bmatrix}
            x - \mu_i\\
            y - \mu_j
        \end{bmatrix}
        \right)
    }
    {2\pi \sqrt{|\Sigma|}}
    \frac{2\pi \sqrt{\sigma^2 + \tau_i^{-1}} \sqrt{\sigma^2 + \tau_j^{-1}}}
    {
        \exp\left(-\frac{1}{2} \left[x - \mu_i, y - \mu_j \right]
        \begin{bmatrix}
            \frac{1}{\sigma^2 + \tau_i^{-1}} & 0\\
            0 & \frac{1}{\sigma^2 + \tau_j^{-1}}
        \end{bmatrix}
        \begin{bmatrix}
            x - \mu_i\\
            y - \mu_j
        \end{bmatrix} \right)
    }\\
    & = \sqrt{\frac{(\sigma^2 + \tau_i^{-1})(\sigma^2 + \tau_j^{-1})}{|\Sigma|}}
    \exp\left(
        -\frac{1}{2} \left[x - \mu_i, y - \mu_j \right]
        \left( \Sigma^{-1} -
        \begin{bmatrix}
            \frac{1}{\sigma^2 + \tau_i^{-1}} & 0\\
            0 & \frac{1}{\sigma^2 + \tau_j^{-1}}
        \end{bmatrix}
        \right)
        \begin{bmatrix}
            x - \mu_i\\
            y - \mu_j
        \end{bmatrix}
    \right)\\
    & \propto 
    \exp\Bigg(
        -\frac{1}{2} \frac{\sigma^2}{|\Sigma|(\sigma^2 + \tau_i^{-1})(\sigma^2 + \tau_j^{-1})}\\
        \begin{split}
        \qquad \qquad \left[x - \mu_i, y - \mu_j \right]
        \begin{bmatrix}
            \sigma^2(\sigma^2 + \tau_j^{-1}) & -(\sigma^2 + \tau_i^{-1})(\sigma^2 + \tau_j^{-1})\\
            -(\sigma^2 + \tau_i^{-1})(\sigma^2 + \tau_j^{-1}) & \sigma^2(\sigma^2 + \tau_i^{-1})
        \end{bmatrix}
        \begin{bmatrix}
            x - \mu_i\\
            y - \mu_j
        \end{bmatrix}
        \end{split}
    \Bigg).
\end{align*}
Finally, setting
\[
    G(x, y) = \left[x - \mu_i, y - \mu_j \right]
        \begin{bmatrix}
            \sigma^2(\sigma^2 + \tau_j^{-1}) & -(\sigma^2 + \tau_i^{-1})(\sigma^2 + \tau_j^{-1})\\
            -(\sigma^2 + \tau_i^{-1})(\sigma^2 + \tau_j^{-1}) & \sigma^2(\sigma^2 + \tau_i^{-1})
        \end{bmatrix}
        \begin{bmatrix}
            x - \mu_i\\
            y - \mu_j
        \end{bmatrix},
\]
and simplifying, we can write  
\[
    \text{JP}(x, y) = \sqrt{\frac{(\sigma^2 + \tau_i^{-1})(\sigma^2 + \tau_j^{-1})}{(\sigma^2 + \tau_i^{-1})(\sigma^2 + \tau_j^{-1}) - \sigma^4}}
    \exp\left(
        -\frac{1}{2}\frac{\sigma^2 \tau_i \tau_j}{(\sigma^2\tau_i + \sigma^2\tau_j + 1)(\sigma^2 + \tau_i^{-1})(\sigma^2 + \tau_j^{-1})} G(x, y)
    \right).
\]

\vspace{1 em}
\noindent
For our experiments with ABM, this gives:
\[
    \text{JP}(x, y) = \sqrt{\frac{(2.1 + \tau_i^{-1})(2.1 + \tau_j^{-1})}{(2.1 + \tau_i^{-1})(2.1 + \tau_j^{-1}) - 4.41}}
    \exp\left(
        -\frac{1.05 \tau_i \tau_j}{(2.1\tau_i + 2.1\tau_j + 1)(2.1 + \tau_i^{-1})(2.1 + \tau_j^{-1})} G(x, y)
    \right),
\] 

\vspace{1 em}
\noindent
when the substitutions $\mu = 7$ (implicitly in the definitions of $\mu_i$ and $\mu_j$) and $\sigma^2 = 2.1$ are also made in the definition of $G(x, y)$. 
For each semester in the real data, we just substitute the corresponding estimates of $\mu$ and $\sigma^2$ given in \Cref{tab:data}.

\section{Considering Novel Mechanisms}
\label{appendix:additional-mechanisms}
\setcounter{figure}{0}
In this section, we give a brief description of some novel (to our knowledge) mechanisms that we used in an effort to push the Pareto frontier delineated by the established mechanisms from the peer prediction literature. For these experiments, we adopt the alternative notion of report quality described in \Cref{appendix:alternative-quality}.

Although these novel mechanisms do not significantly expand the Pareto frontier in our setting, we expect that many of them may be useful in other settings where data is more readily available. In particular, as with some of the established parametric mechanisms discussed in the body of the paper (\Cref{subsection:parametric}), the robustness against strategic reporting of certain mechanisms may be compromised by the fact that agent's reports, through the estimation procedure, have an effect on the ground truth estimates that are later used in scoring their reports. In settings with more data, it may be possible to get good ground truth score estimates that are independent of the reports of the particular agent being scored by the mechanism.

Lastly, note that each of our novel mechanisms is parametric. Thus, each mechanism begins by estimating the parameters of model \PGone\,according to our estimation procedure.

\vspace{1 ex}
\noindent
\textbf{Coefficient of Determination ($R^2$) Mechanism.}
This mechanism pays each agent the \textit{coefficient of determination} (denoted $R^2$) between the set of their bias-corrected reports (which constitute the ``predicted values'' in the definition of $R^2$) and the set of true score estimates on those same tasks (which constitute the ``observed data'' in the definition of $R^2$).

The intuition behind this mechanism is that $R^2$ can be interpreted as the proportion of the variance in the observed data that can be explained from the predicted values. Thus, agents who report accurately ought to do well---their reports explain a high fraction of the variance in the ground truth, because most of the variance in the reports of a reliable grader comes from variation in the ground truth scores (as opposed to coming from noise in the process of generating their signals of the ground truth scores.) Further, conditioned on their signal, any strategy that an agent applies cannot depend on the ground truth, since the signal contains all of an agent's private information about the ground truth. As a result, the coefficient of determination between an agent's signals and the ground truth scores cannot be increased by applying a strategy to the signals to generate non-truthful reports. (Note, however, that this statement is a simplification in our setting, since it ignores the role that reports play in the estimation procedure.)

\vspace{1 ex}
\noindent
\textbf{Correlation (CORR) Mechanism.}
This mechanism pays agents the (sample) Pearson correlation coefficient, $r$, between the set of their bias-corrected reports for the submissions that they graded and the set of true score estimates computed for those submissions.

The intuition here is similar to that of the previous mechanism. Accurate reports ought to be more highly correlated with the ground truth. Further, as described above, the correlation between an agent's signals and the ground truth scores cannot be increased by applying a strategy to the signals to generate non-truthful reports because---given their signal---any strategy that an agent applies cannot depend on additional private information about the ground truth score. (As above, though, this statement is a simplification in our setting, since it ignores the role that reports play in the estimation procedure.)

\vspace{1 ex}
\noindent
\textbf{Leave-One-Out (LOO) Mechanism.}
The idea of this mechanism is to capture the value of an agent's report by determining how much the quality of the true score estimate deteriorates (for each submission that they grade) when they are omitted from the population of agents. The mechanism estimates the parameters of model \PGone\,using our estimation procedure once with the entire population of $n$ agents, then $n$ more times with a population of $n-1$ agents, leaving out a different agent each time. Note that, as a result, this mechanism takes significantly longer to run than the other mechanisms. 

In implementing this mechanisms, especially in settings with more data, reliable ground truth estimates for each submission should not be difficult to compute, even when leaving one agent at a time out of the estimation procedure. In our setting, however, we found that reliable ground truth estimates would not be sufficient to make this mechanism worthwhile to run (particularly in light of the significantly increased computational resources it requires compared to the other mechanisms).

To gauge the potential of this mechanism, we gave it access to the underlying ground truth scores. Each agent was paid according to the reduction in squared error that resulted from including them in the agent population. That is, for each submission that they graded, each agent was paid the squared error of the true score estimate with them left out (with respect to the ground truth) minus the squared error of the true score estimate with them included (also with respect to the ground truth). 

Even with access to the ground truth scores, this mechanism did not demonstrate significant measurement integrity compared to the best-performing mechanisms. Moreover---because of the access to the true scores---it would not be fair to compare it to the other mechanisms (especially with respect to robustness against strategic reporting). Consequently, this mechanism is omitted from \Cref{appendix-fig:2D-extensions}.

\vspace{1 ex}
\noindent
\textbf{Maximum Correlation Coefficient (MCC) Mechanism.}
This mechanism, along with the intuition behind it, was suggested by Fang-Yi Yu.
Given a pair of random variables $(X,Y)\sim P_{X,Y}$, the \emph{maximum correlation coefficient} between $X,Y$ is 
\[
    \rho^*(X, Y) = \max_{f, g}\left\{\mathrm{E}[f(X)g(Y)]:\mathrm{E}[f(X)]=\mathrm{E}[g(Y)]=0, \mathrm{E}[f(X)^2]=\mathrm{E}[g(Y)^2]=1\right\}.
\]

For a bivariate normal distribution, it is known that $\rho^*(X, Y) = |\rho(X, Y)|$, where $\rho$ is the typical correlation coefficient \cite{Yu2008}. Since a pair of signals in model \PGone\,follows a bivariate normal distribution, we can apply that principle to a mechanism and pay each pair of agents that completed the same task according to the maximum correlation coefficient between their reports. 

According to model \PGone, for a pair of agents $i$ and $j$ who receive signals given by the random variables $X$ and $Y$ (respectively):
\begin{align}
\label{eqn:mcc}
  \rho^*(X, Y) = \left|{\frac{2.1}{\sqrt{2.1 + \tau_i^{-1}}\sqrt{2.1 + \tau_j^{-1}}}}  \right|
\end{align}

In general, the intuition for the incentive-compatibility of this mechanism follows from the revelation principle. The mechanism, in maximizing the correlation coefficient, applies the ``optimal strategy'' for the agents once it receives their reports. Therefore, agents need not perform their own strategic manipulations. 

In our setting specifically, there is an even more straightforward argument: Each agent's reports do not play a direct role in determining their payments. That is, the reports do not appear in \cref{eqn:mcc}, above. Note, however, that the reports do have an indirect effect, since they are used for computing estimates of $\tau_i$ and $\tau_j$ according to our estimation procedure. 

\vspace{1 ex}
\noindent
\textbf{Parametric $\Phi$-Divergence$^*$ Pairing (\pPhiDivStar) Mechanism.}
This mechanism, as implied by the name, is quite similar to the \pPhiDiv\,mechanism. The only difference is, in the \pPhiDivStar\,mechanism, agents are ``paired'' with the ground truth instead of with other agents. For the bonus task, the second report is an estimated true score that is computed (using the formula and parameters in the estimation procedure) with only the other three agents' reports on that task. Thus, the agent who is being scored by the mechanism is not taken into consideration when computing the estimated true score with which they are ``paired.'' For the penalty task, the second report is an estimated true score (given by the estimation procedure, i.e. computed using all 4 agents that submitted a report) for a task that was not completed by the agent who is being scored.

\vspace{1 ex}
\noindent
\textbf{Parametric Adjusted Mean Squared Error (\pAMSE) Mechanism.}
To discourage agents from hedging, this mechanism introduces a penalty for being too close to the mean of the prior distribution (i.e. 7) into the scores computed in the established \pMSE\,mechanism. For each submission $s_{i,j}$, each agent $k$ who graded that submission is assigned the following reward:
\[
    -((r_{i, j}^k-\hat{b}_k) - \hat{g}_{i,j})^2 +0.1 \cdot ((r_{i, j}^k-\hat{b}_k) - 7)^2,
\]
where $\hat{b}_k$ is the estimated bias of grader $k$, and $r_{i,j}^k$ is grader $k$'s report for submission $s_{i, j}$, and $\hat{g}_{i,j}$ is the estimated true score for submission $s_{i,j}$.

\subsection{Revisiting the Pareto Frontier}
\label{subappendix:revisiting}
Recall that in these experiments, we consider the alternative notion of report quality described in \Cref{appendix:alternative-quality}, so the Pareto frontier in \Cref{appendix-fig:2D-extensions} is slightly different than that of \Cref{fig:2D-tradeoff}.
\begin{figure}[ht]
    \centering
    \includegraphics[width=0.6\textwidth]{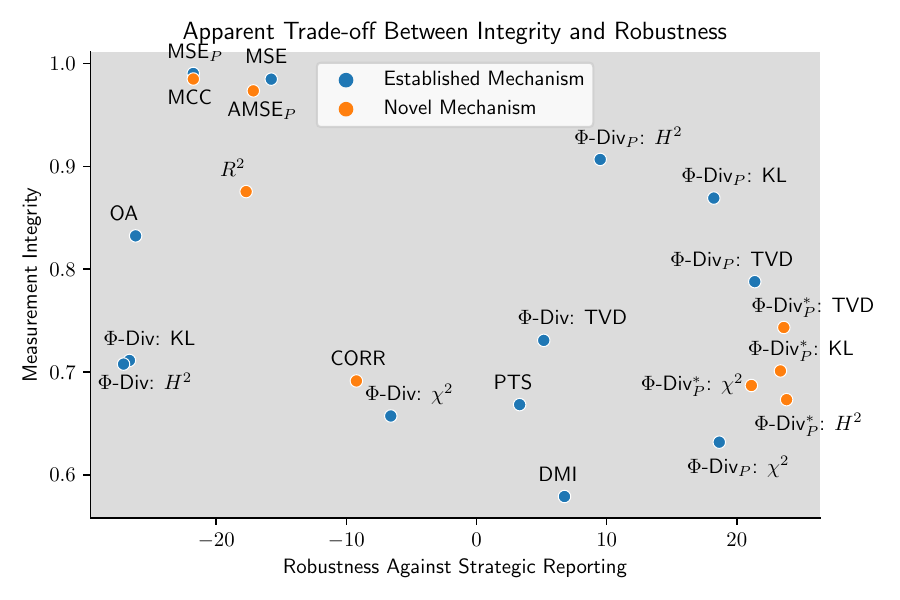}
    \caption{The novel mechanisms that we consider do not significantly expand the Pareto frontier delineated by the established mechanisms from the peer prediction literature. The most notable change is that some of the \pPhiDivStar\,mechanisms fill in the space in the lower right section of the figure, narrowly supplanting their counterpart \pPhiDiv\,mechanisms as the mechanisms that are most robust against strategic reporting in our experiments, but at the cost of lower measurement integrity.}
    \label{appendix-fig:2D-extensions}
\end{figure} 

\section{Alternative Conceptions of Report Quality}
\label{appendix:alternative-quality}
\setcounter{figure}{0}
In \Cref{section:model}, we adopt perhaps the simplest notion of report quality: the squared distance between a report value and the true grade of the corresponding submission. Thus, the quality of a report is determined with respect to the report viewed as a fixed quantity. However, this fully \textit{ex post} notion of report quality means that random chance plays a role in determining the quality of an agent's report. An agent who chooses a score uniformly at random may, by chance, submit a report that is close to or even exactly equal to a submission's true grade. In certain cases, it may be desirable to remove the role of chance in our definition, although it may result in a notion of report quality that is less intuitive. One way to accomplish this is to define report quality with respect to the report viewed as a random variable, where the randomness comes from uncertainty in the observation of a signal (and possible randomness used in the generation of a report after the observation of a signal). Then, a report can be considered high-quality if it is informative---i.e., unlikely to generate an outcome that is very far from the underlying true grade.

In our model, assuming that agents report truthfully, \textit{effort} determines the informativeness of an agent's report viewed as a random variable. As a result, another interesting notion of measurement integrity to consider is one where the quality of each agent's report is considered to be equivalent to their effort. Under this alternative regime, then, measurement integrity helps to quantify the degree to which mechanisms incentivize effort (assuming that increased effort translates into obtaining more reliable signals). Note that biased agents present a complication to this notion of quality, so in what follows we consider settings with and without agent bias. Ultimately, consistent with the peer assessment literature \cite{Piech2013}, we find that it is useful to model bias and effort separately and correct for bias whenever possible (i.e. when using parametric mechanisms, which estimate agent bias). 

We also introduce an alternative, simpler model of effort and unbiased agents (in addition to the Continuous Effort, Biased Agents model introduced in \Cref{section:model}) that are useful in building intuition about the performance of peer prediction mechanisms according to this alternative notion of measurement integrity:

\vspace{1 ex}
\noindent
\textbf{Binary Effort.}
Here, there are two types of agents: \textit{active graders} exert high effort, \textit{passive graders} exert low effort. Active graders receive signals created using three draws from their latent distribution. Passive graders receive signals created using a single draw from their latent distribution.

\vspace{1 ex}
\noindent
\textbf{Unbiased Agents.}
For an unbiased agent, the expectation of the latent distribution is equal to the ground truth score. Their signal for an assignment $s_{i, j}$ is a function of draws from the latent distribution $\Binom{10}{\frac{g_{i, j}^*}{10}}$.

\subsection{Computational Experiments with ABM}
\label{subappendix:alternative-quality-experiments}
Under this alternative notion of report quality, we conduct additional computational experiments using our peer assessment ABM. As in the body of the paper, in order to explore measurement integrity in isolation, we require agents to report their signals truthfully. This is perhaps even more important in this context. In the experiments in \Cref{section:quantifying-mi}, requiring agents to report truthfully affects the distribution of reports, but does not affect the notion of quality---whether or not a report was strategically manipulated, it is a high-quality report if it is close to the ground truth score. Here, strategic reporting confounds the relationship between effort and report quality, so it is especially important to require honest reporting in quantifying measurement integrity.

\subsubsection{Methods}
\label{subsubappendix:alternative-quality-methods}
First, we evaluate the ability of mechanisms to measure the agents according to their quality. For each mechanism under consideration, the experiment consists of a number of simulated semesters, each of which proceeds as follows: 
\begin{enumerate}
    \item A population of 100 students is initialized.
    
    \item For each of 10 assignments:
    
    \begin{enumerate}[i.]
        \item Each student turns in a submission with a true grade drawn from the true grade distribution.
        
        \item A random 4-regular graph is constructed with a vertex for each agent.\footnote{The DMI mechanism is not functional using this procedure, so for that mechanism the agents are instead randomly partitioned into disjoint 4-cliques, and each agent in a clique grades all 4 submissions from another clique.}
        
        \item Each agent grades the submissions of their neighbors in the graph according to our peer assessment model.
        
        \item The reported grades are collected by the mechanism, which assigns a reward to each student for their performance in peer assessment for that assignment.
    \end{enumerate}
    
    \item The total reward accrued by each student, which is the sum of their rewards for each of the 10 individual assignments, is used to calculate the value of the relevant evaluation metric.
    
    \item At the end of all simulated semesters, the mean, median, and variance of the evaluation metrics are calculated.
\end{enumerate}

For both the binary effort and continuous effort cases, we simulate 500 semesters without varying any underlying parameters. In the binary effort case, the performance of the mechanisms as the number of active graders varies is also of interest. For those experiments, we iterate the above procedure as we vary the number of active graders from 10 to 90 in increments of 10, simulating 100 semesters each time.

Second, we quantify the variance of the quality of the measurements (given the reports) in the continuous effort setting with biased agents. We perform the following procedure, for each of 50 iterations: 
\begin{enumerate}
    \item An agent population (100 students) and the submissions for a semester's worth of assignments (10) are fixed.
    
    \item For each of 50 iterations:
    \begin{enumerate}[i.]
        \item A semester, with the fixed agent population and fixed submission pool, is simulated---graders are assigned to submissions, reports are collected, and rewards for each assignment are computed according to the given mechanism, as described above.
        
        \item The evaluation metric is computed.
    \end{enumerate}
    
    \item The variance of the evaluation metric over the 50 simulations of that semester is computed.
\end{enumerate}

\subsubsection{Evaluation}
\label{subsubappendix:alternative-quality-evaluation}
In this setting, we consider notions of measurement in a more focused way than in the analysis we conduct in \Cref{section:quantifying-mi}. Here, the way that we model agent effort leads to an intuitive appropriate notion of measurement to adopt and a corresponding evaluation metric. In settings where effort is binary---i.e. when there are only two types of agents---the natural goal is to be able to accurately classify each agent according to their type. This establishes a clear notion of measurement. We need to quantify the ability of the mechanisms to place agents on a \textit{nominal} scale, where their assigned value is indicative only of their placement in one of the two categories. The area-under-the-curve (AUC) of the receiver operating characteristic (ROC) curve is well-suited to this task. AUC can be interpreted in several ways.  First, it is the probability that an active grader chosen uniformly at random is scored higher by a mechanism than a passive grader chosen uniformly at random.  Second, it summarizes how useful the (real-valued) rewards assigned by the mechanism are in being translated into a binary-value nominal scale (i.e. a binary classification) via the selection of a threshold such that all students with rewards above the threshold are classified as active graders and all students with rewards below the threshold are classified as passive graders. An AUC score of 1 indicates a perfect classifier, an AUC score of 0.5 is the expected value of a random classifier, and an AUC score of 0 indicates a fully incorrect classifier.

In the settings where effort is continuous, the constraints of our model similarly establish a clear notion of measurement. In particular, the absolute magnitude of an agent's effort parameter $\lambda$, as well as the differences between two agents' effort parameters, does not have a straightforward interpretation in our model. Can we expect an agent with $\lambda=1$ to be ``twice as good'' as an agent with $\lambda = 0.5$ in some clearly appreciable way? Or similarly, can we expect an agent with $\lambda=1$ to be the same amount ``better'' than an agent with $\lambda = 0.5$ as an agent with $\lambda=1.5$ would be ``better'' than an agent with $\lambda = 1.0$ in some clearly appreciable way? It seems unlikely that either question has an affirmative answer. Rather, it is most natural in our model to assign no meaning to the magnitude of the effort parameter beyond that higher values for the effort parameter should tend to correspond to more accurate reports. This interpretation implies that the appropriate notion of measurement for the continuous effort settings is an \textit{ordinal} scale, i.e. a ranking. For rankings, the Kendall rank correlation coefficient (Kendall's $\tau_B$) is an effective evaluation metric. Kendall's $\tau_B$ is related to the number of pairs that appear in the same order (concordant pairs) and in the opposite order (discordant pairs) in the two rankings being compared. In the case that there are no ties in either ranking, $\tau_B$ (or simply $\tau$ in that case) is equal to the proportion of pairs that are concordant minus the proportion of pairs that are discordant.

\subsubsection{Results}
\label{subsubappendix:alternative-quality-results}
\hfill
\vspace{1 ex}

\noindent
\textbf{Binary Effort, Unbiased Agents.}
In this simple case, for which the evaluation metric has the clearest interpretation, we can gain intuition about the relative performance of various mechanisms. The results are shown in \Cref{appendix-fig:unbias-be}.

First, we compare the choices of $\Phi$ among the various well-known candidates in the $\Phi$-divergence pairing mechanism (\Cref{subfig:unbias-be-phidiv-box,subfig:unbias-be-phidiv}) and its parametric counterpart (\Cref{subfig:unbias-be-pphidiv-box,subfig:unbias-be-pphidiv}). Among the choices for $\Phi$-divergences, TVD performs best for the non-parametric mechanism, especially as the number of active graders increases. For the parametric, the best choice of $\Phi$-divergence ($H^2$) is not the same. Additionally, the best choice for the parametric mechanism is different for this notion of measurement integrity than for the notion we consider in \Cref{section:quantifying-mi}.

Then, we compare among the non-parametric mechanisms from the peer prediction literature. One interesting note is that, in this case, we see an example of the disconnect between theoretical properties, for which the DMI mechanism is perhaps the most exemplary and the OA mechanism perhaps the least, and empirical performance, where the roles are starkly reversed (\Cref{subfig:unbias-be-nonparam-box,subfig:unbias-be-nonparam}). Lastly, looking at the best-performing mechanisms overall (\Cref{subfig:unbias-be-best-box,subfig:unbias-be-best}), we see that, as in \Cref{section:quantifying-mi}, it is largely the baseline mechanisms, not mechanisms from the peer prediction literature, that demonstrate high measurement integrity most reliably.

\begin{figure}
\centering
\begin{subfigure}{.5\textwidth}
  \centering
  \includegraphics[width=0.75\linewidth]{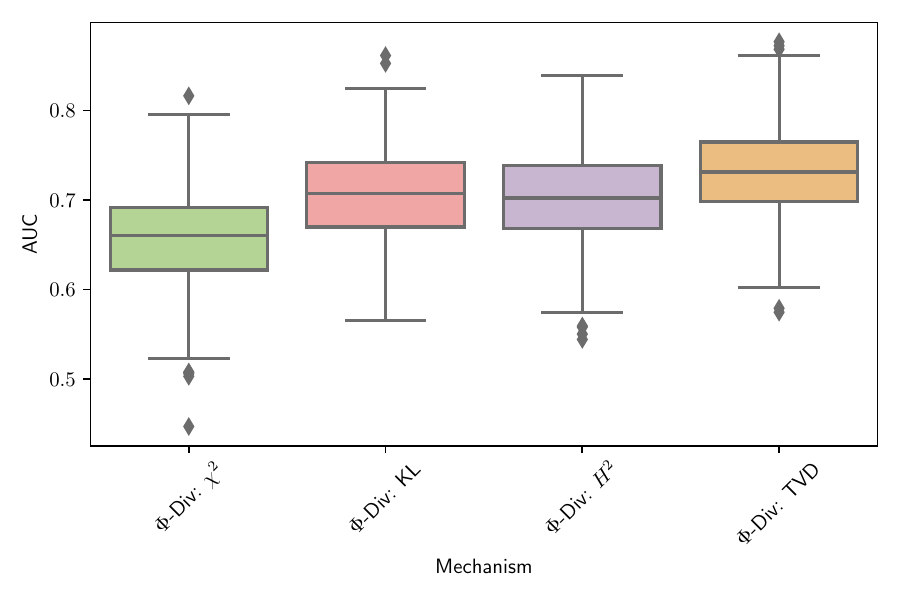}
  \captionsetup{width=0.8\textwidth}
  \caption{AUC scores for $\Phi$-Div mechanisms;\\ 500 semesters, 50 active graders.}
  \label{subfig:unbias-be-phidiv-box}
\end{subfigure}%
\begin{subfigure}{.5\textwidth}
  \centering
  \includegraphics[width=0.75\linewidth]{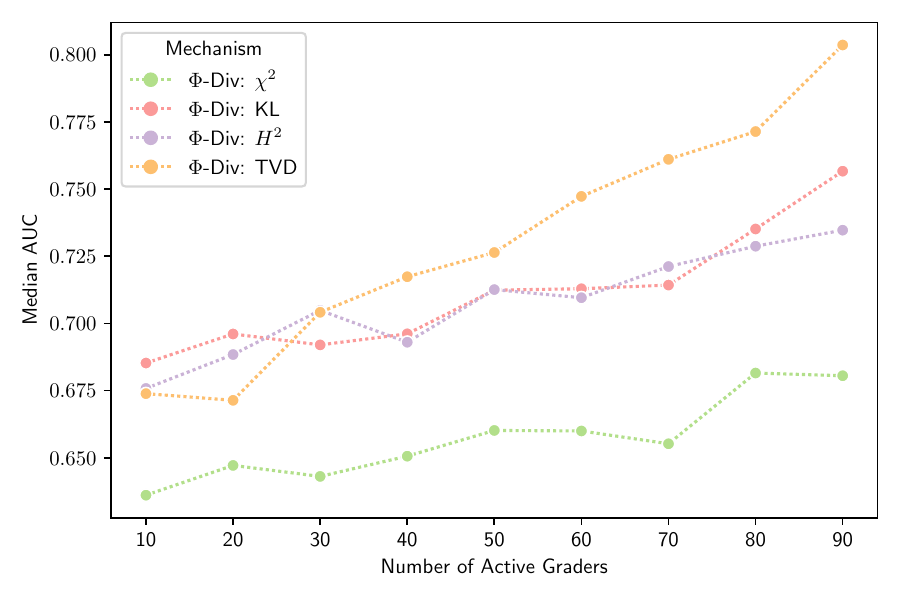}
  \captionsetup{width=0.8\textwidth}
  \caption{Median AUC scores for $\Phi$-Div mechanisms as the number of active graders varies.}
  \label{subfig:unbias-be-phidiv}
\end{subfigure}
\begin{subfigure}{.5\textwidth}
  \centering
  \includegraphics[width=0.75\linewidth]{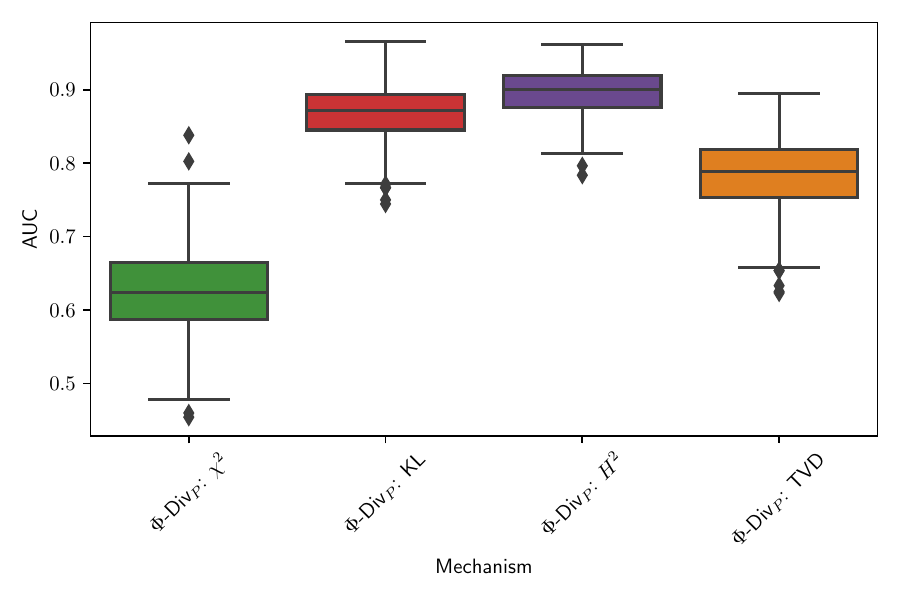}
  \captionsetup{width=0.8\textwidth}
  \caption{AUC scores for \pPhiDiv\,mechanisms;\\ 500 semesters, 50 active graders.}
  \label{subfig:unbias-be-pphidiv-box}
\end{subfigure}%
\begin{subfigure}{.5\textwidth}
  \centering
  \includegraphics[width=0.75\linewidth]{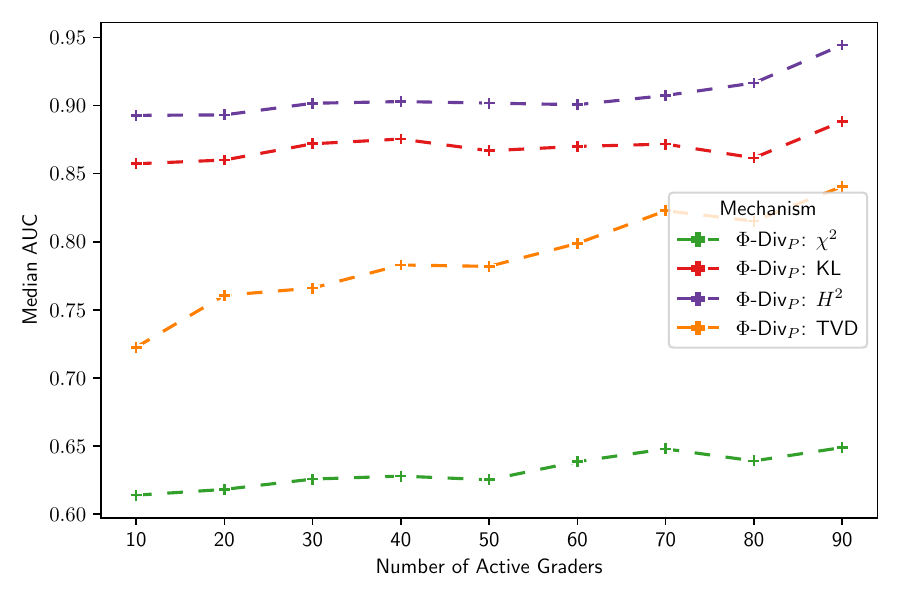}
  \captionsetup{width=0.8\textwidth}
  \caption{Median AUC scores for \pPhiDiv\, mechanisms as the number of active graders varies.}
  \label{subfig:unbias-be-pphidiv}
\end{subfigure}
\begin{subfigure}{.5\textwidth}
  \centering
  \includegraphics[width=0.75\linewidth]{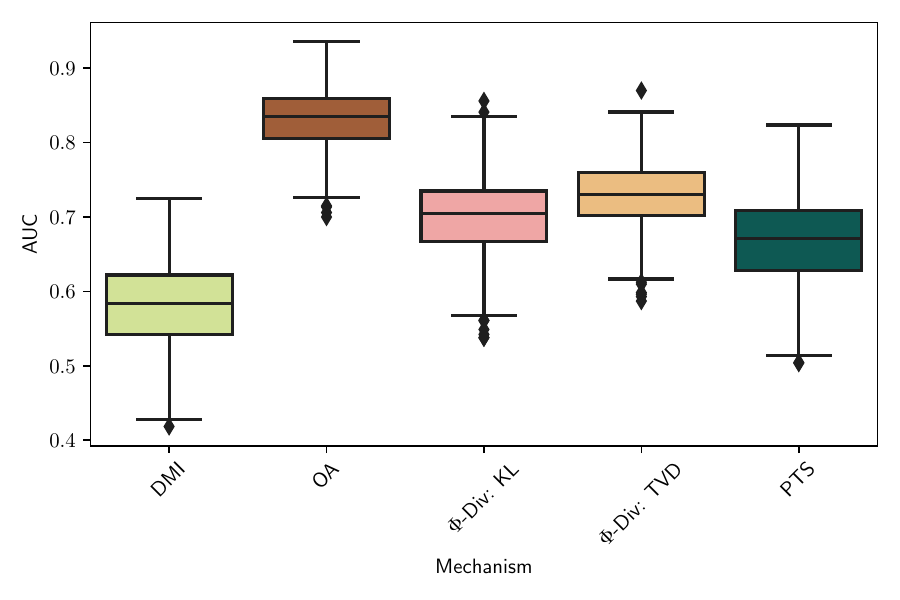}
  \captionsetup{width=0.8\textwidth}
  \caption{AUC scores for non-parametric mechanisms; 500 semesters, 50 active graders.}
  \label{subfig:unbias-be-nonparam}
\end{subfigure}%
\begin{subfigure}{.5\textwidth}
  \centering
  \includegraphics[width=0.75\linewidth]{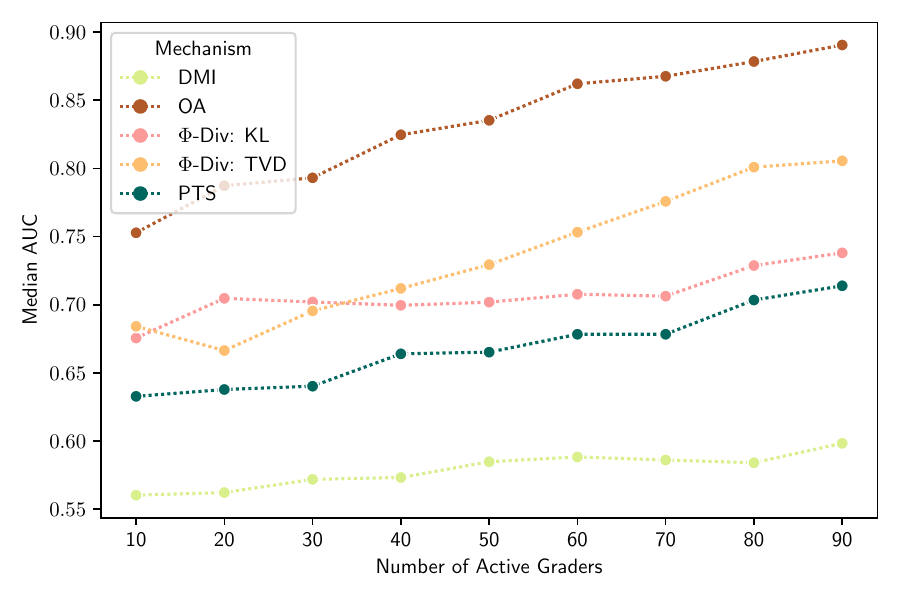}
  \captionsetup{width=0.8\textwidth}
  \caption{Median AUC scores for non-parametric mechanisms as the number of active graders varies.}
  \label{subfig:unbias-be-nonparam-box}
\end{subfigure}
\begin{subfigure}{.5\textwidth}
  \centering
  \includegraphics[width=0.75\linewidth]{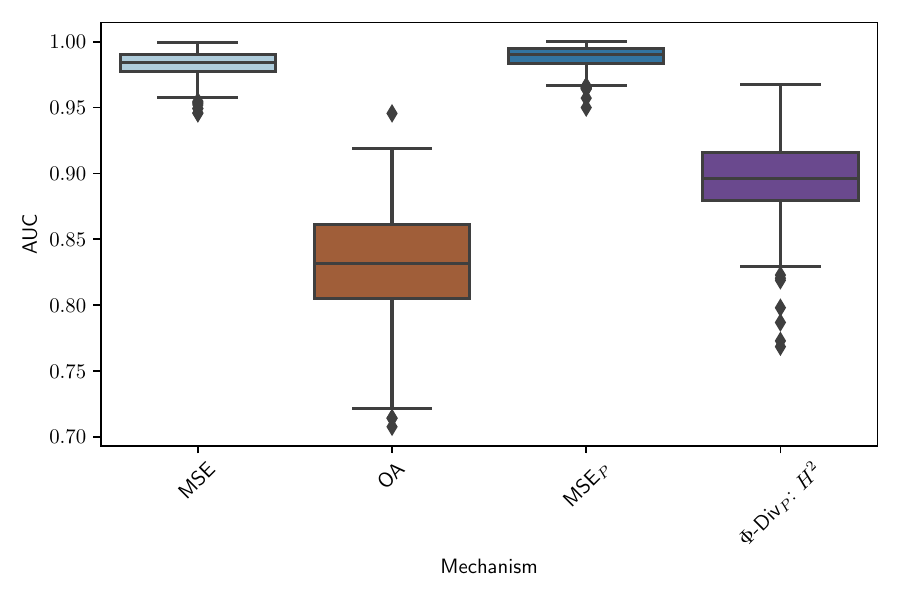}
  \captionsetup{width=0.8\textwidth}
  \caption{AUC scores for parametric mechanisms, baseline, and OA; 500 semesters, 50 active graders.}
  \label{subfig:unbias-be-best-box}
\end{subfigure}%
\begin{subfigure}{.5\textwidth}
  \centering
  \includegraphics[width=0.75\linewidth]{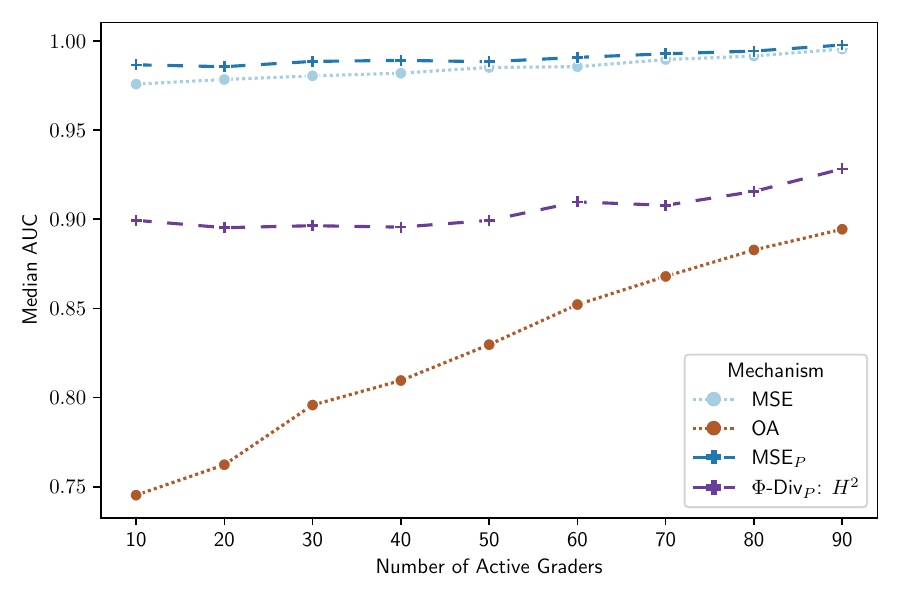}
  \captionsetup{width=0.8\textwidth}
  \caption{Median AUC scores for parametric mechanisms, baseline, and OA as the number of active graders varies.}
  \label{subfig:unbias-be-best}
\end{subfigure}
\caption{\textit{Binary Effort, Unbiased Agents.} }
\label{appendix-fig:unbias-be}
\end{figure}

\vspace{1 ex}
\noindent
\textbf{Binary Effort, Biased Agents.}
The results for this setting are shown in \Cref{appendix-fig:bias-be}. Unsurprisingly, we find that in the presence of biased agents, the performance of the mechanisms that do not attempt to correct for agent bias, degrades significantly compared to settings with unbiased agents. This includes the MSE baseline and the best-performing non-parametric peer prediction mechanisms. The amount of degradation varies depending on the mechanism.

For parametric mechanisms that do account for (and thus can correct for) bias, their relative performance compared to the baseline is also not equally affected by the presence of biased agents. The performance of the \pMSE\,mechanism degrades less significantly than the performance of the \pPhiDiv\, mechanisms. Further, while the best-performing \pPhiDiv\, mechanism, (\pPhiDiv: $H^2$) outperforms the baseline in this setting, the same does not hold for some other choices of $\Phi$-divergence. 

\begin{figure}
\centering
\begin{subfigure}{.5\textwidth}
  \centering
  \includegraphics[width=0.75\linewidth]{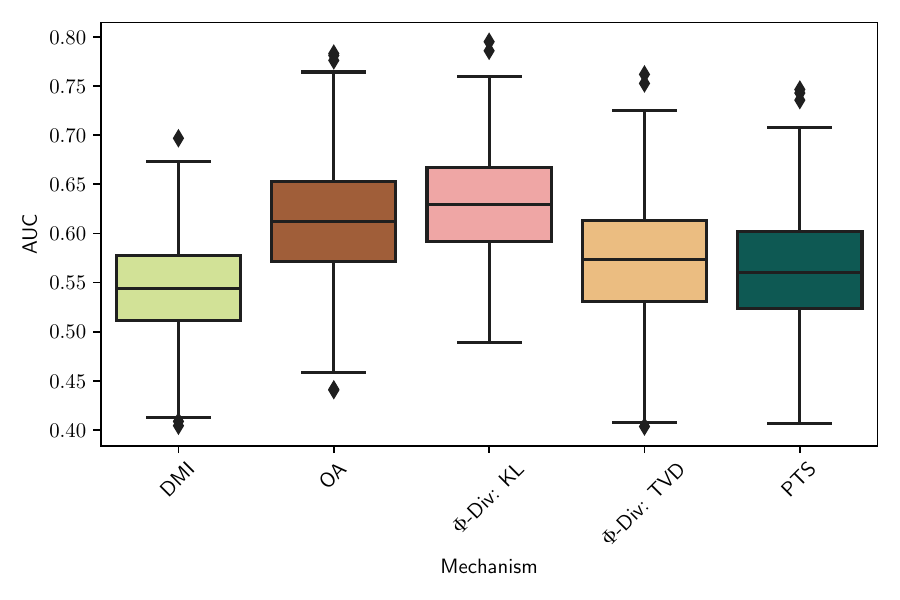}
  \captionsetup{width=0.8\textwidth}
  \caption{AUC scores for the non-parametric mechanisms and the baseline; 500 semesters, 50 active graders.}
  \label{subfig:bias-be-nonparam-box}
\end{subfigure}%
\begin{subfigure}{.5\textwidth}
  \centering
  \includegraphics[width=0.75\linewidth]{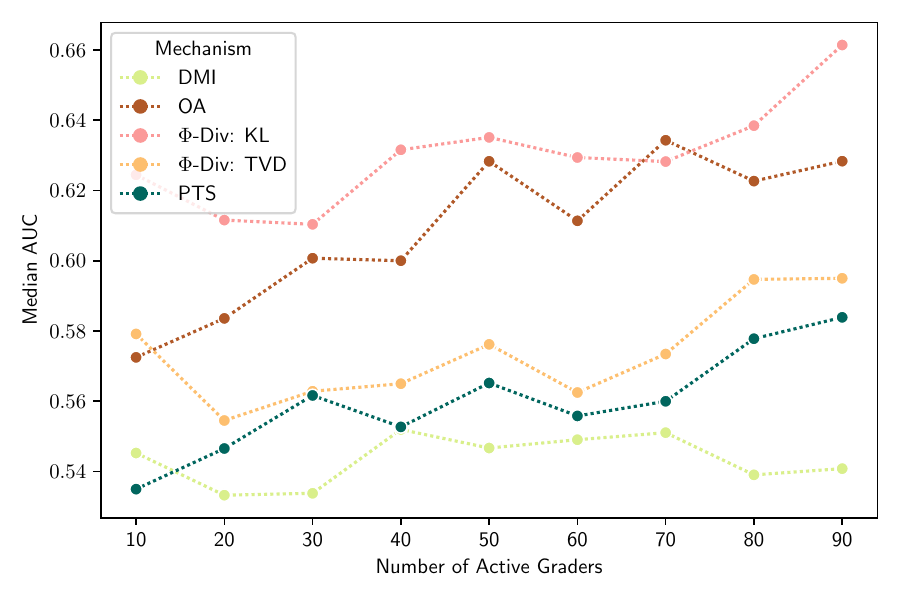}
  \captionsetup{width=0.8\textwidth}
  \caption{Median AUC scores for the non-parametric mechanisms and the baseline as the number of active graders varies.}
  \label{subfig:bias-be-nonparam}
\end{subfigure}
\begin{subfigure}{.5\textwidth}
  \centering
  \includegraphics[width=0.75\linewidth]{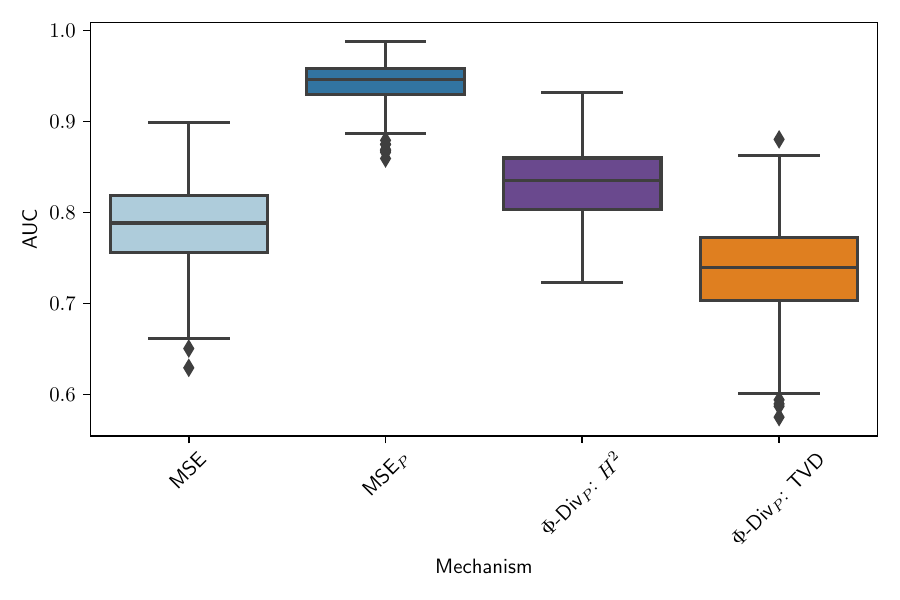}
  \captionsetup{width=0.8\textwidth}
  \caption{AUC scores for the parametric mechanisms and the baseline; 500 semesters, 50 active graders.}
  \label{subfig:bias-be-param-box}
\end{subfigure}%
\begin{subfigure}{.5\textwidth}
  \centering
  \includegraphics[width=0.75\linewidth]{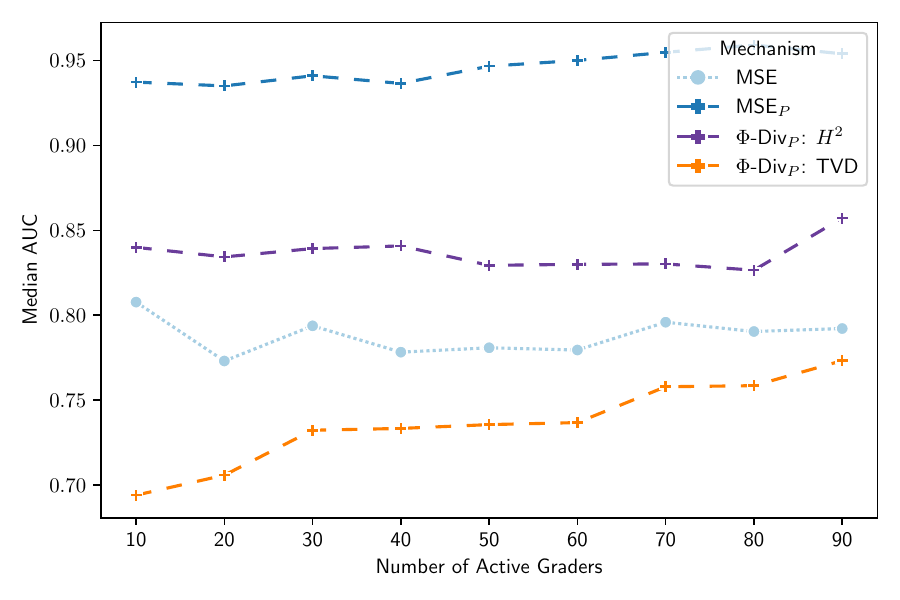}
  \captionsetup{width=0.8\textwidth}
  \caption{Median AUC scores for the parametric mechanisms and the baseline as the number of active graders varies.}
  \label{subfig:bias-be-param}
\end{subfigure}
\caption{\textit{Binary Effort, Biased Agents.} }
\label{appendix-fig:bias-be}
\end{figure}

\vspace{1 ex}
\noindent
\textbf{Continuous Effort, Biased Agents.}
The results for this setting---our most complex and realistic model, with continuous effort and biased agents---are shown in \Cref{appendix-fig:bias-ce}. We primarily focus on the three best-performing mechanisms for this section thus far: \pMSE, MSE, and \pPhiDiv: $H^2$ (\Cref{subfig:bias-ce-tau,subfig:bias-ce-tau-variances}). 

In \Cref{subfig:bias-ce-tau}, we see that as before \pMSE\,clearly has the best performance. \pPhiDiv: $H^2$ still outperforms the baseline, but not by much. The $\tau_B$ statistic in general does not have as clear of an interpretation as AUC, but because these two parametric mechanisms tend not to produce ties in the rewards, we are able to draw some conclusions about them. When there are no ties, the number of correctly ranked pairs is equal to $\frac{\tau_B + 1}{2}$. Thus, we can conclude, for example, that about half the time, \pMSE\,ranks at least 65\% of the pairs of agents appropriately with respect to their effort, because the median $\tau_B$ for \pMSE\, is about 0.3. Further, our experiments in the previous setting (\Cref{appendix-fig:bias-be}) show that for agents whose effort level is sufficiently distinct, like the levels of active and passive graders, \pMSE\,separates the groups with high accuracy. Consequently, it is likely that among the 35\% of pairs of agents who are not ranked correctly, few have differ drastically in their effort levels. 

In \Cref{subfig:bias-ce-tau-variances}, we plot the variances of the rankings assigned by the best-performing mechanisms for a fixed agent population with a fixed pool of submissions for each assignment. Although MSE performs best along this metric, the variance for all 3 mechanisms is negligible. For reference, nearly every variance recorded in the experiment is less than 0.006. A change in $\tau_B$ of 0.006 (when there are no ties) would correspond to a change in the relative ranking of about 15 out of the 4950 possible pairs.\footnote{The derivative of the number of concordant pairs with respect to $\tau_B$ is $\frac{1}{2}$, so a change of 0.006 in $\tau_B$ corresponds to a change of 0.003 in the number of concordant pairs.}

\begin{figure}
\begin{subfigure}[t]{.49\textwidth}
  \centering
  \includegraphics[width=0.75\linewidth]{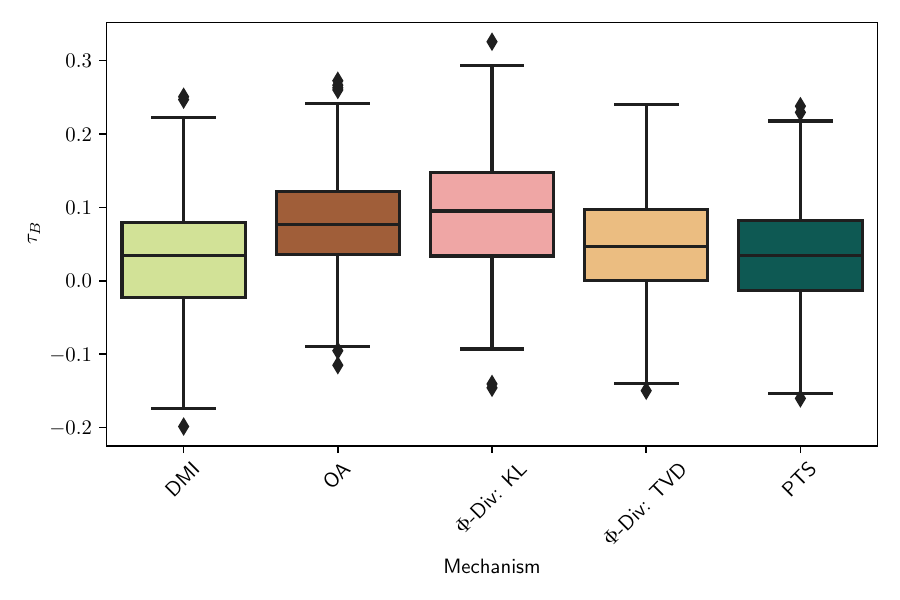}
  \captionsetup{width=0.8\textwidth}
  \caption{Kendall's $\tau_B$ scores for each of the non-parametric mechanisms; 500 semesters.}
  \label{subfig:bias-ce-tau-nonparam}
\end{subfigure} \hfill%
\begin{subfigure}[t]{.49\textwidth}
  \centering
  \includegraphics[width=0.75\linewidth]{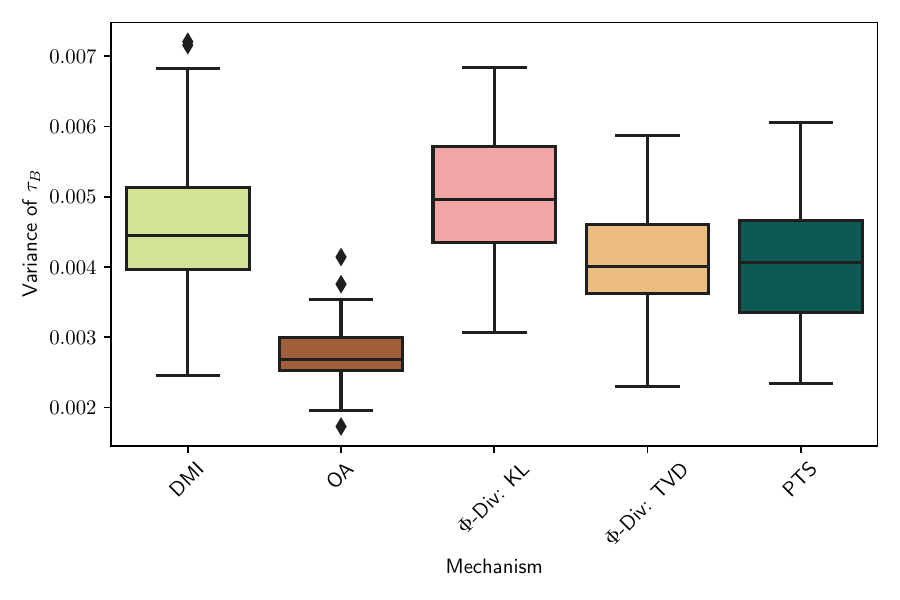}
  \captionsetup{width=0.8\textwidth}
  \caption{Variance of Kendall's $\tau_B$ scores; 50 iterations simulating a semester 50 times with the agent population and submissions fixed.}
  \label{subfig:bias-ce-tau-variances-nonparam}
\end{subfigure}
\begin{subfigure}[t]{.49\textwidth}
  \centering
  \includegraphics[width=0.75\linewidth]{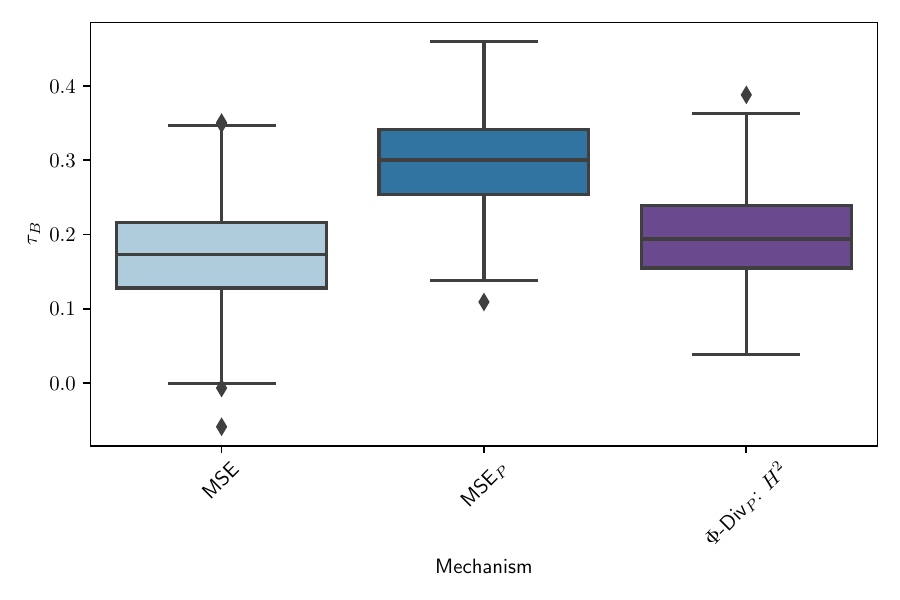}
  \captionsetup{width=0.8\textwidth}
  \caption{Kendall's $\tau_B$ scores for each of the best-performing mechanisms; 500 semesters.}
  \label{subfig:bias-ce-tau}
\end{subfigure} \hfill%
\begin{subfigure}[t]{.49\textwidth}
  \centering
  \includegraphics[width=0.75\linewidth]{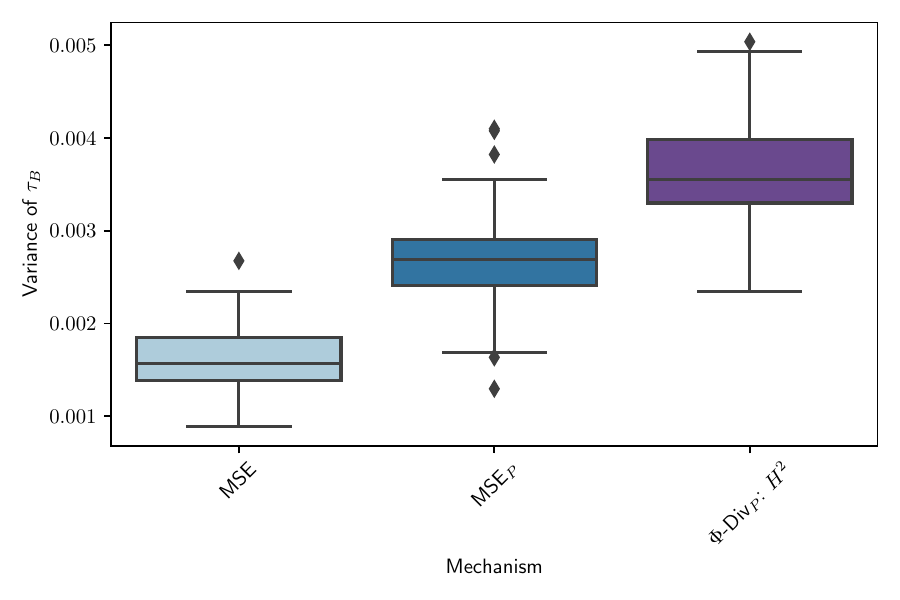}
  \captionsetup{width=0.8\textwidth}
  \caption{Variance of Kendall's $\tau_B$ scores; 50 iterations simulating a semester 50 times with the agent population and submissions fixed.}
  \label{subfig:bias-ce-tau-variances}
\end{subfigure}
\caption{\textit{Continuous Effort, Biased Agents.} }
\label{appendix-fig:bias-ce}
\end{figure}

\subsubsection{Conclusions} 
\label{subsubappendix:alternative-quality-conclusions}
By setting aside incentive concerns and investigating the ability of the mechanisms to measure agents according to the latent underlying quality of their reports (i.e., effort), we are able to demonstrate that out-of-the-box peer prediction mechanisms from the literature generally fail to exhibit high levels of measurement integrity. This is significant, because mechanisms that fail to exhibit measurement integrity in a stylized agent-based model setting are unlikely to exhibit it in a real-world deployment. These findings essentially corroborate our findings in \Cref{section:quantifying-mi}, where we consider a simpler notion of report quality. There are some differences, however---e.g., the best-performing choice of $\Phi$-divergence---which highlights the fact that the context-dependent choices that are made in defining an appropriate notion measurement integrity for a specific application matter for evaluating mechanisms. 

\subsection{Measurement Integrity in the Presence of Strategic Agents}
\label{subappendix:ranking-quality}
As mentioned above, when adopting the random variable view of report quality, strategic reporting influences the corresponding notion of measurement integrity by confounding the relationship between the quality of effort expended in observing a signal and the informative-ness of the resulting report. Our goal in this experiment is to quantify the effect of this confounding. How is measurement integrity (in terms of the quality of rankings of the agents according to their continuous effort parameters) affected when agents are allowed to report strategically? To explore this question, we focus on the 3 best-performing mechanisms from our experiments in \Cref{subappendix:alternative-quality-experiments} and on the \textit{informed strategies} from \Cref{subsection:strategies}.  

\subsubsection{Methods}
\label{subsubappendix:ranking-quality-methods}
For each strategy, we replicate the measurement integrity experiment in the \textit{Continuous Effort, Biased Agents} setting from \Cref{subappendix:alternative-quality-experiments} while varying the number of strategic agents from 0 to 100 in steps of 10. We simulate 100 semesters at each step.

\subsubsection{Results}
\label{subsubappendix:ranking-quality-results}
The results for this experiment are shown in \Cref{appendix-fig:strategic-bias-ce}. Note that we plot results in steps of size 20 instead of 10 for clarity.

Although generally in peer prediction, the focus is on deterring strategic manipulation of reports, not all strategic manipulations are necessarily equally damaging to the mechanism. Uninformed strategies are clearly detrimental, since the mechanism receives no information about the signals of agents following those strategies. However, for some informed strategies, it is possible that the mechanism might still be able to demonstrate relatively high measurement integrity even if those strategies become common among the agents. For example, it is reasonable to expect the \textit{Fix Bias} strategy to only minimally affect the performance of parametric mechanisms, which take bias into account. Indeed, we find that this is the case in \Cref{subfig:strategic-bias-ce-fix-bias}, which shows that the performance of all 3 mechanisms is relatively constant as the number of agents adopting the \textit{Fix Bias} strategy grows.

On the other hand, though, other strategies do significantly affect the performance of the mechanisms. In \Cref{section:robustness}, we singled out \textit{Hedge} as a particularly attractive strategy for agents participating in the MSE and \pMSE\,mechanisms. In \Cref{subfig:strategic-bias-ce-hedge}, we see that additionally, the number of agents adopting the \textit{Hedge} strategy has a clear effect on the measurement integrity of those mechanisms (although the effect is not monotonic). When 40 agents adopt the \textit{Hedge} strategy, for example, \pPhiDiv: $H^2$ arguably supplants \pMSE\,as the best mechanism. 

\begin{figure}
\begin{subfigure}{.49\textwidth}
  \centering
  \includegraphics[width=0.75\linewidth]{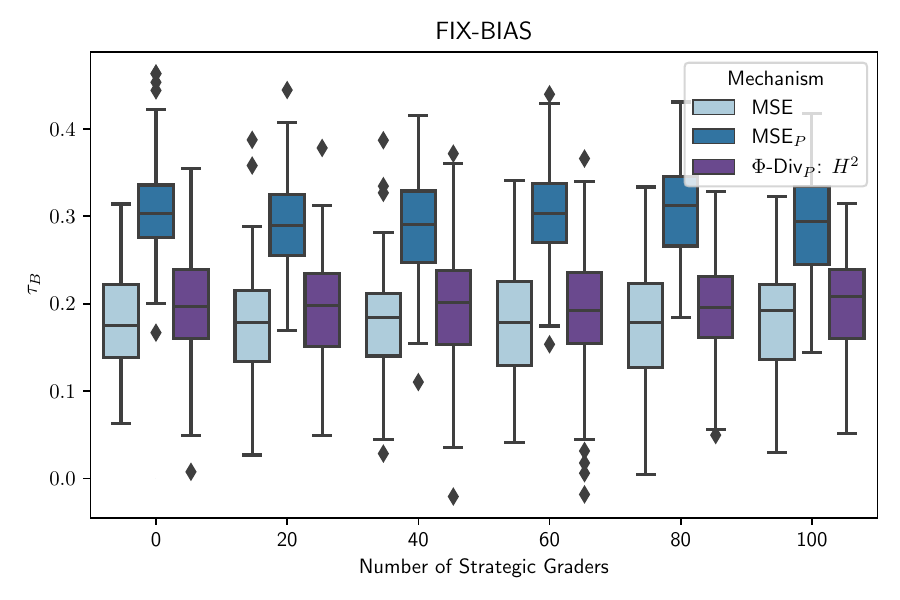}
  \captionsetup{width=0.8\textwidth}
  \caption{}
  \label{subfig:strategic-bias-ce-fix-bias}
\end{subfigure} \hfill %
\begin{subfigure}{.49\textwidth}
  \centering
  \includegraphics[width=0.75\linewidth]{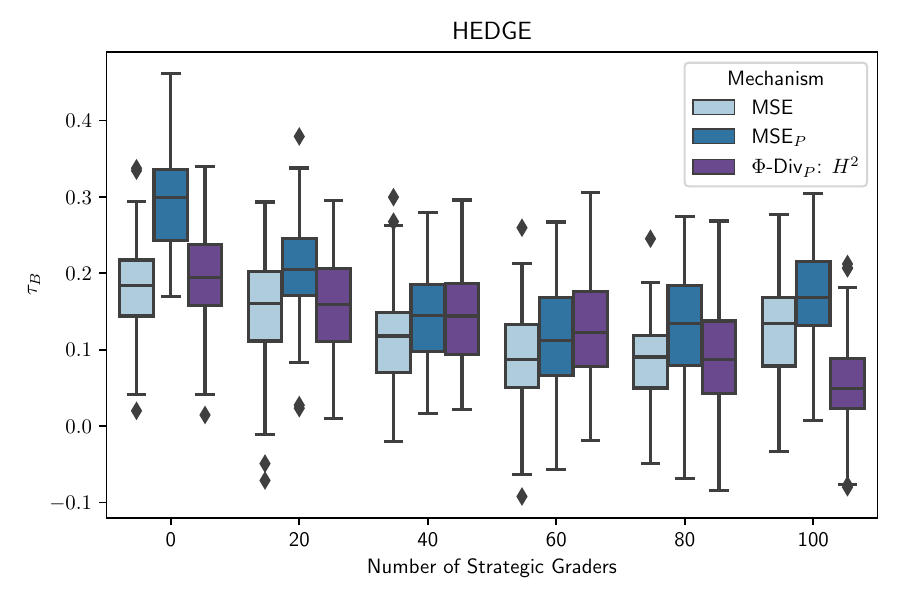}
  \captionsetup{width=0.8\textwidth}
  \caption{}
  \label{subfig:strategic-bias-ce-hedge}
\end{subfigure}
\begin{subfigure}{.49\textwidth}
  \centering
  \includegraphics[width=0.75\linewidth]{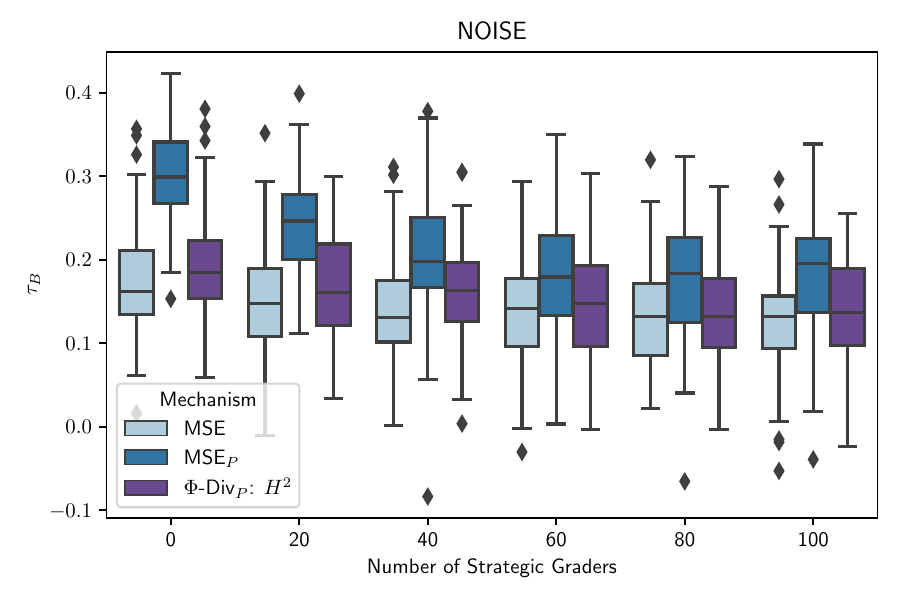}
  \captionsetup{width=0.8\textwidth}
  \caption{}
  \label{subfig:strategic-bias-ce-noise}
\end{subfigure} \hfill %
\begin{subfigure}{.49\textwidth}
  \centering
  \includegraphics[width=0.75\linewidth]{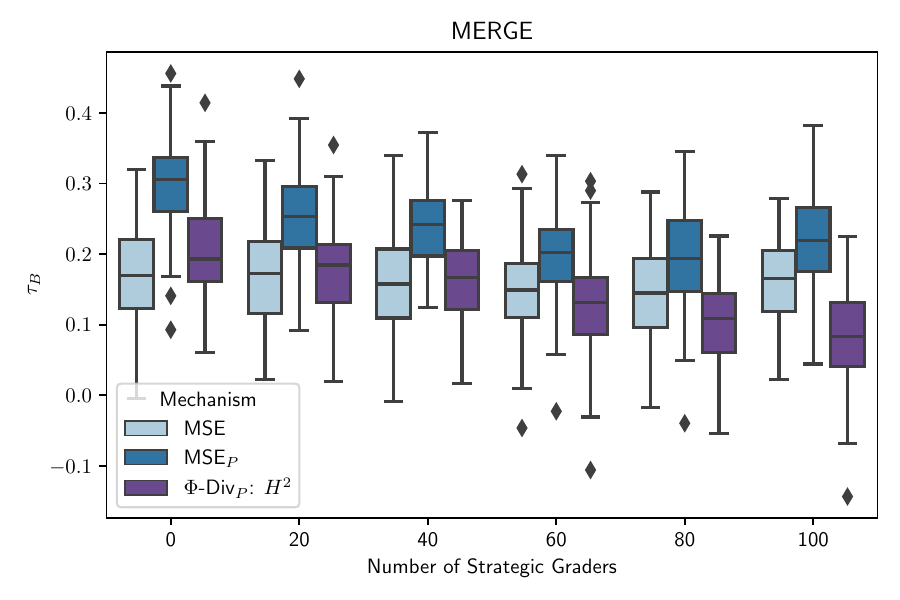}
  \captionsetup{width=0.8\textwidth}
  \caption{}
  \label{subfig:strategic-bias-ce-merge}
\end{subfigure}
\caption{\textit{Measurement Integrity in the Presence of Strategic Agents.} Kendall's $\tau_B$ for each of the best-performing mechanisms as the number of strategic agents varies; 100 semesters each.
}
\label{appendix-fig:strategic-bias-ce}
\end{figure}
 
\section{Additional Experimental Results}
\label{appendix:additional-results}
\setcounter{figure}{0}
In this section, we plot the results of the experiments from \Cref{section:quantifying-mi,section:robustness} for the remaining mechanisms that were not included in the body of the paper. 

\subsection{Measurement Integrity}
\label{subappendix:mi-results}
\setcounter{figure}{0}

\subsubsection{Computational Experiments with ABM}
\label{subsubappendix:mi-results-abm}
See \Cref{appendix-fig:quantifying-mi-abm-other} to see the results from the computational experiments for quantifying measurement integrity with ABM for all of the mechanisms that fail to outperform the OA mechanism, the baseline with the lowest performance. This includes the DMI mechanism, which is not present in the body of the paper.

\begin{figure}
\begin{subfigure}{\textwidth}
  \centering
  \includegraphics[width=0.4\linewidth]{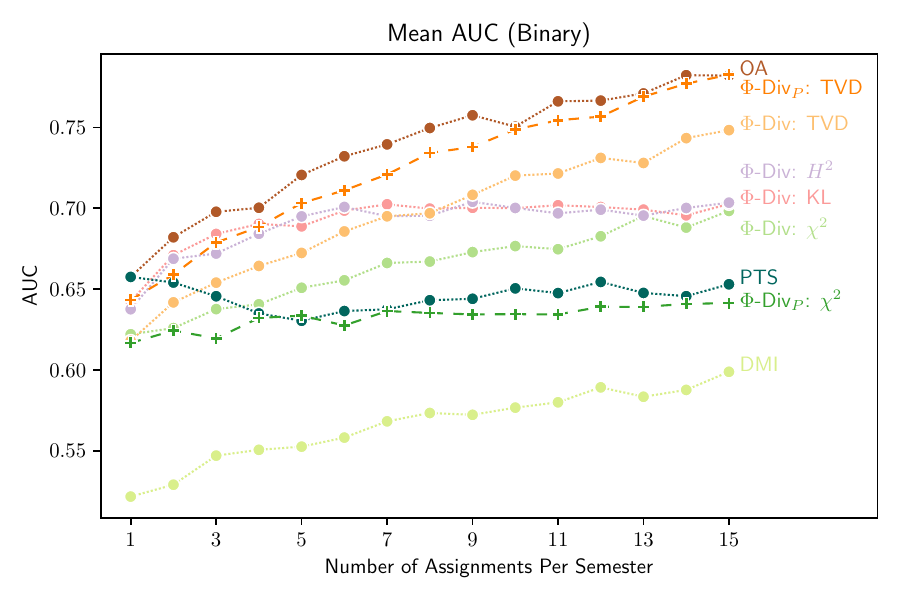}
  \includegraphics[width=0.4\linewidth]{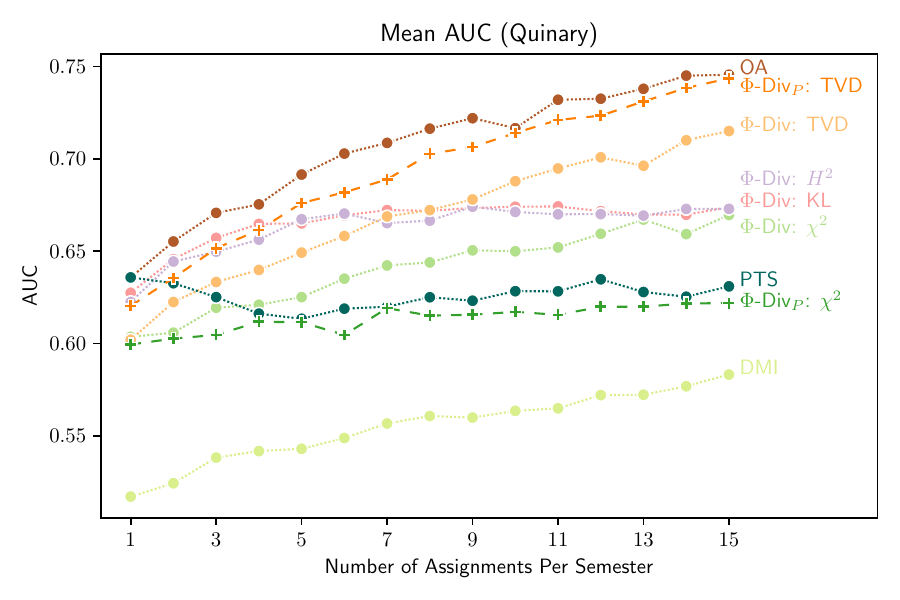}
  \hfill
  \includegraphics[width=0.4\linewidth]{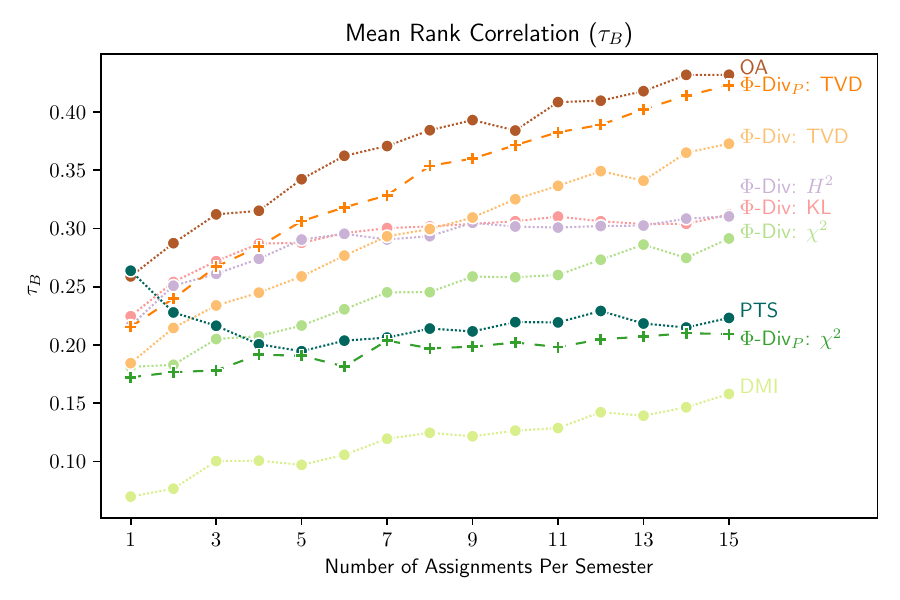}
  \includegraphics[width=0.4\linewidth]{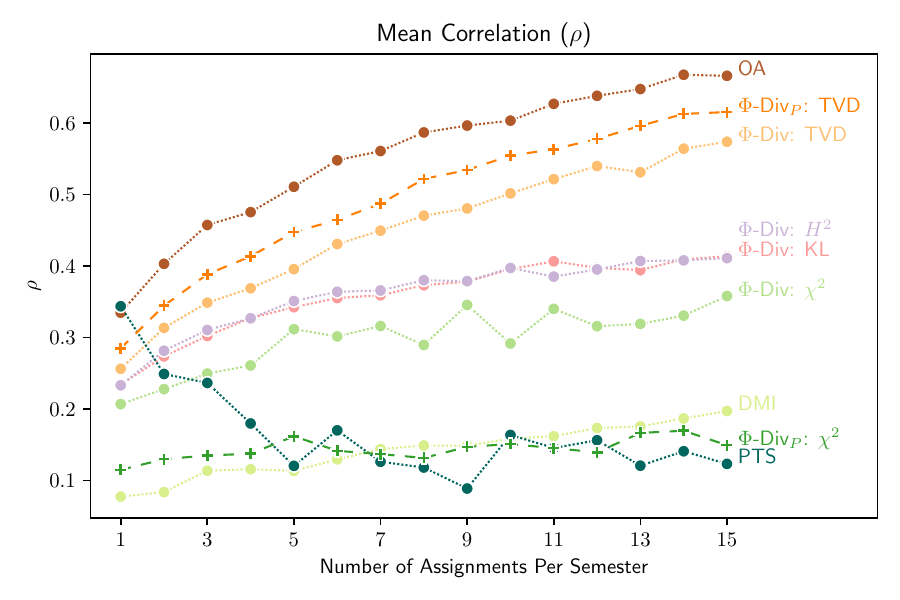}
\end{subfigure}
\caption{\textit{Quantifying Measurement Integrity with ABM.} Average values of evaluation metrics for measurement integrity as the number of assignments in a simulated semester grows. The average for each number of assignments is taken over 50 simulated semesters. }
\label{appendix-fig:quantifying-mi-abm-other}
\end{figure}

\subsubsection{Computational Experiments with Real Data}
\label{subsubappendix:mi-results-real}
See \Cref{appendix-fig:quantifying-mi-real-other-metrics} for the evaluation metrics---other than $\tau_B$, which is shown in \Cref{fig:quantifying-mi-real}---from the computational experiments for quantifying measurement integrity with the real peer grading data.

Repeating the analysis we did with $\tau_B$ in \Cref{subsubsection:quantifying-mi-real-results} by averaging the value of each evaluation metric for each mechanism across all quartiles and all semesters gives the following top 5 mechanisms:
\begin{itemize}
    \item AUC (Binary): \pMSE, MSE, \pPhiDiv: $H^2$, \pPhiDiv: KL, $\Phi$-Div: $H^2$ (followed closely by OA).
    
    \item AUC (Quinary): \pMSE, MSE, \pPhiDiv: KL, \pPhiDiv: $H^2$, OA (followed closely by $\Phi$-Div: $H^2$).
    
    \item AUC (Binary): \pMSE, MSE, \pPhiDiv: KL $H^2$, OA, $\Phi$-Div: $H^2$.
\end{itemize}

\begin{figure}
  \centering
  \includegraphics[width=0.32\linewidth]{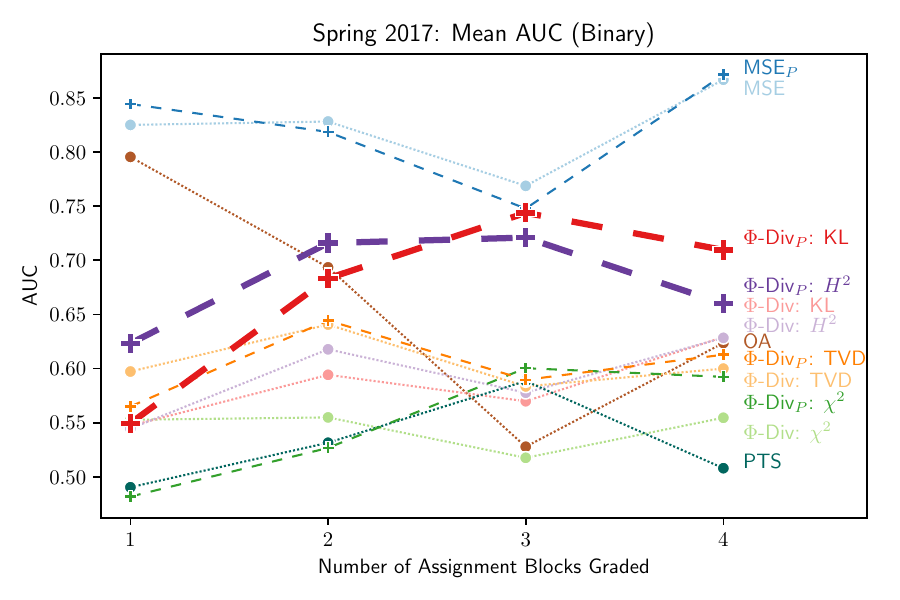}
  \hfill
  \includegraphics[width=0.32\linewidth]{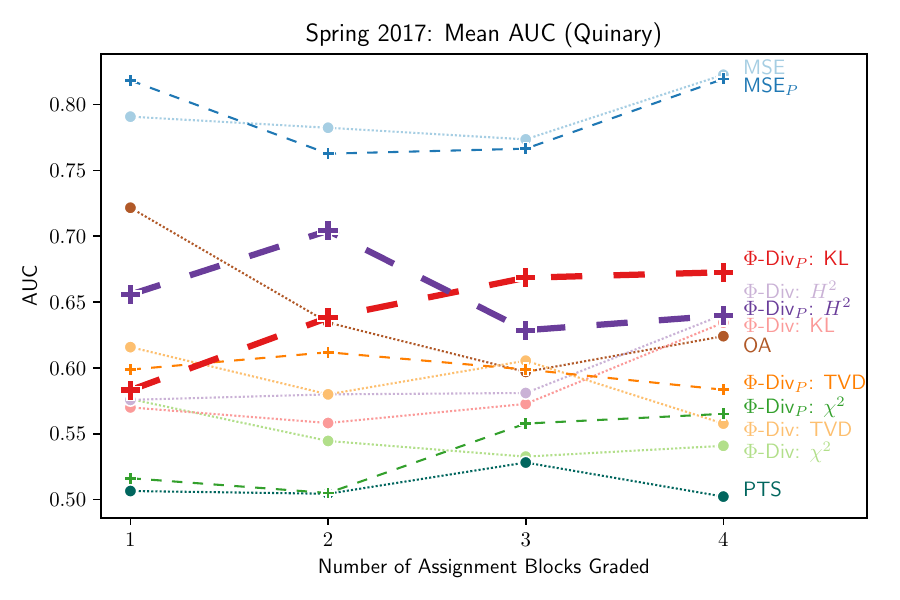}
  \hfill
  \includegraphics[width=0.32\linewidth]{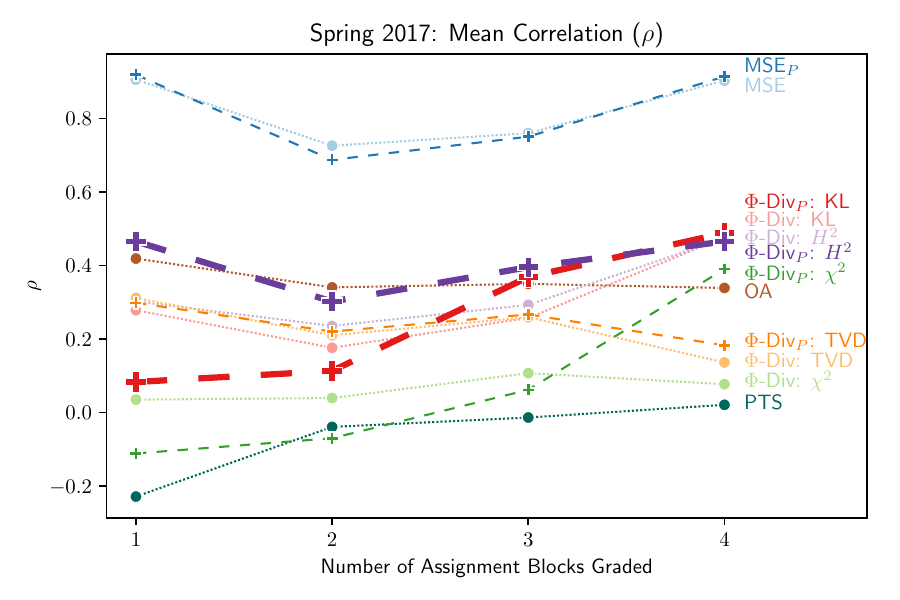}
  \hfill
  \includegraphics[width=0.32\linewidth]{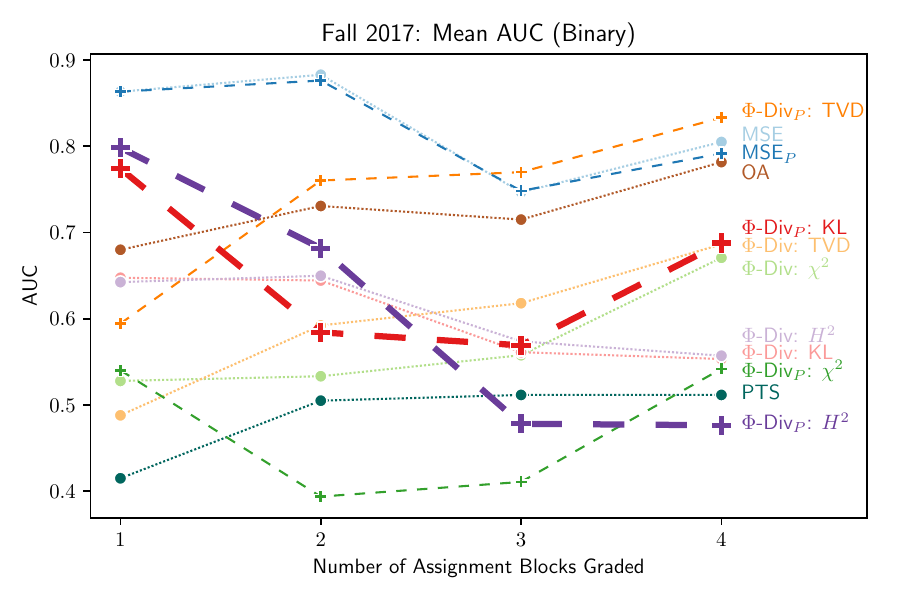}
  \hfill
  \includegraphics[width=0.32\linewidth]{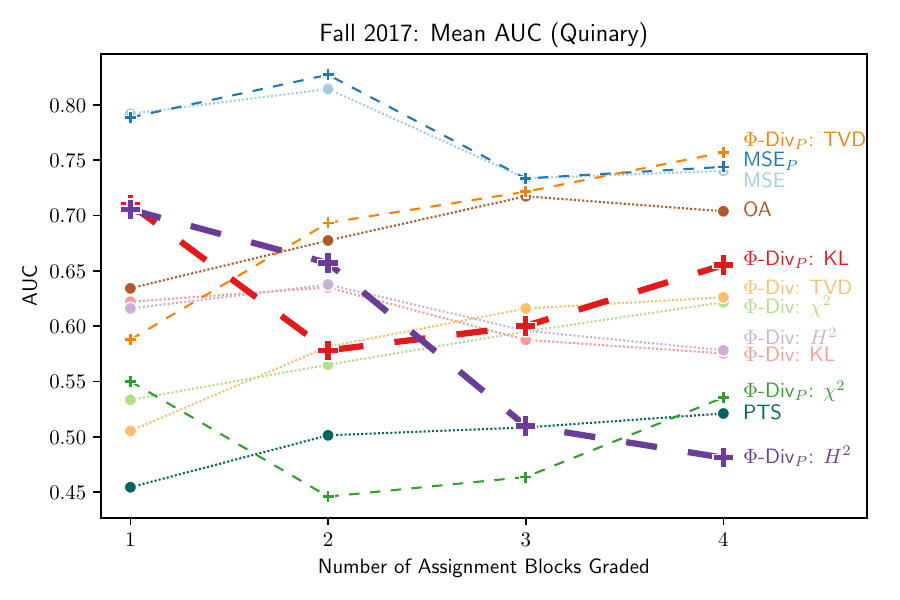}
  \hfill
  \includegraphics[width=0.32\linewidth]{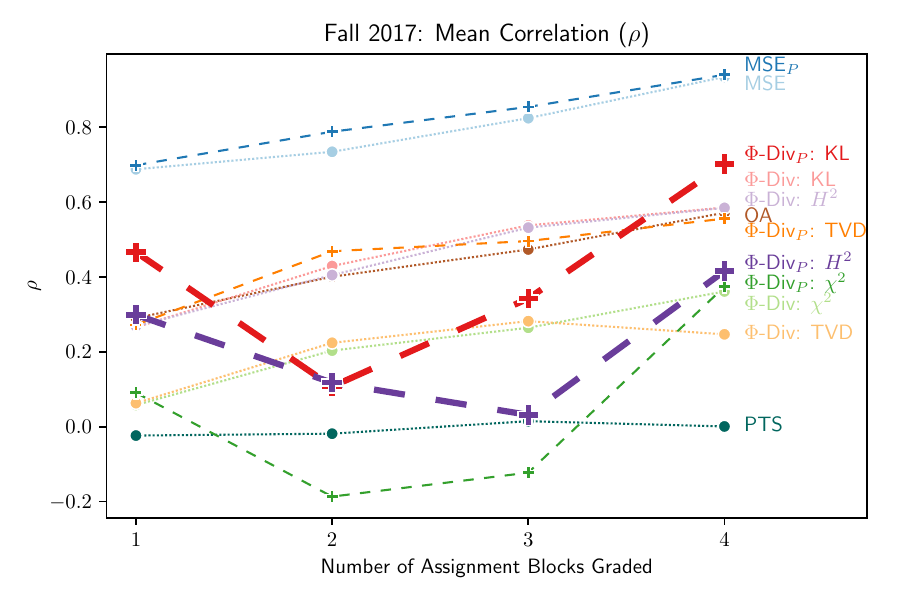}
  \hfill
  \includegraphics[width=0.32\linewidth]{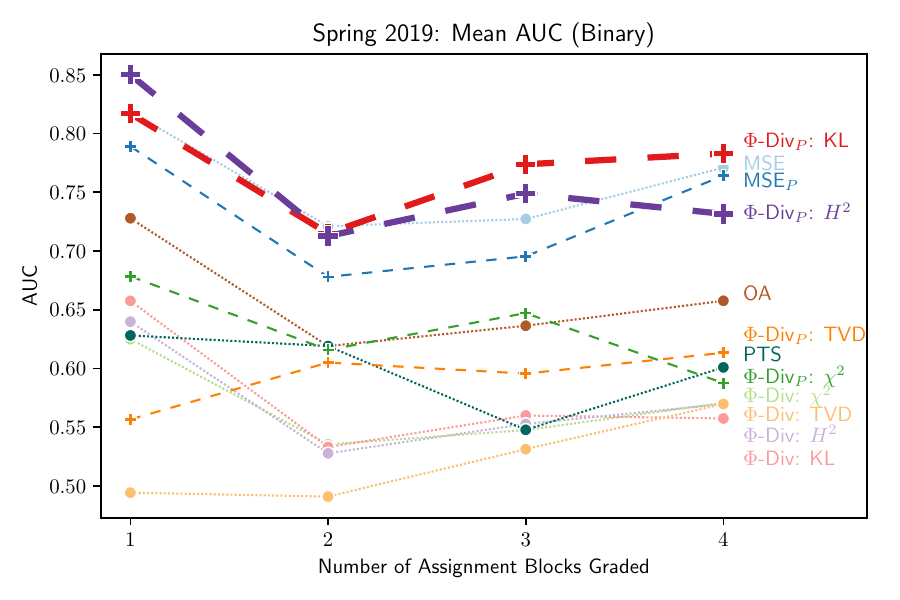}
  \hfill
  \includegraphics[width=0.32\linewidth]{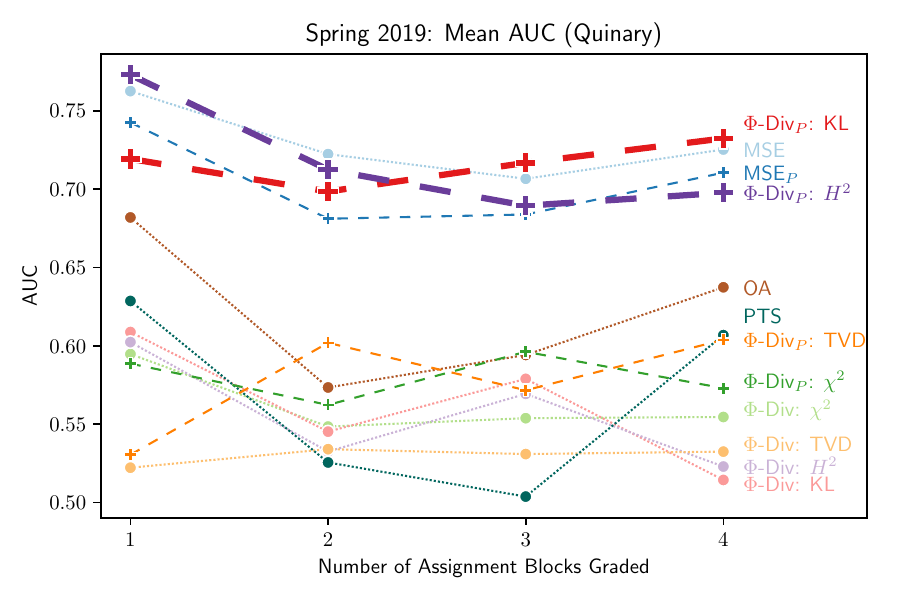}
  \hfill
  \includegraphics[width=0.32\linewidth]{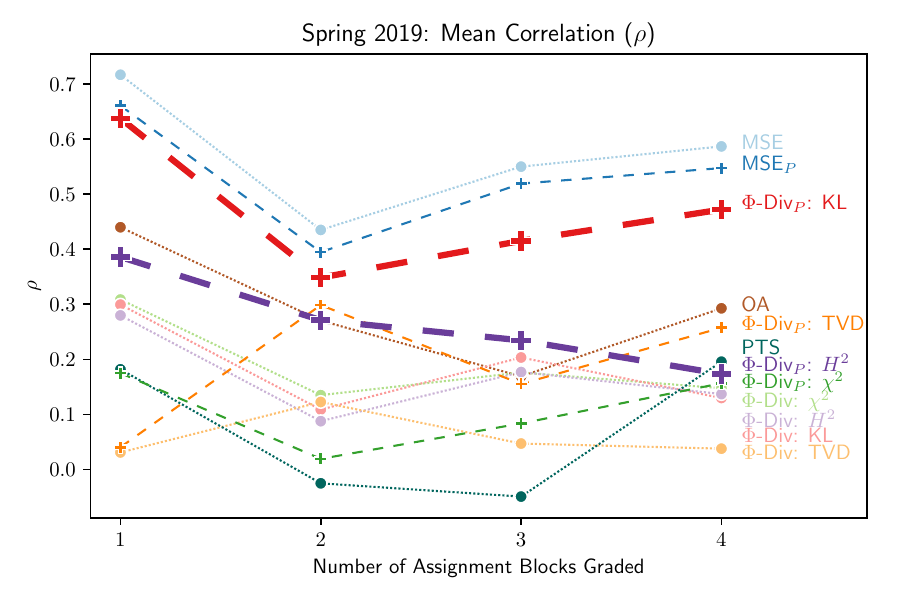}
  \hfill
  \includegraphics[width=0.32\linewidth]{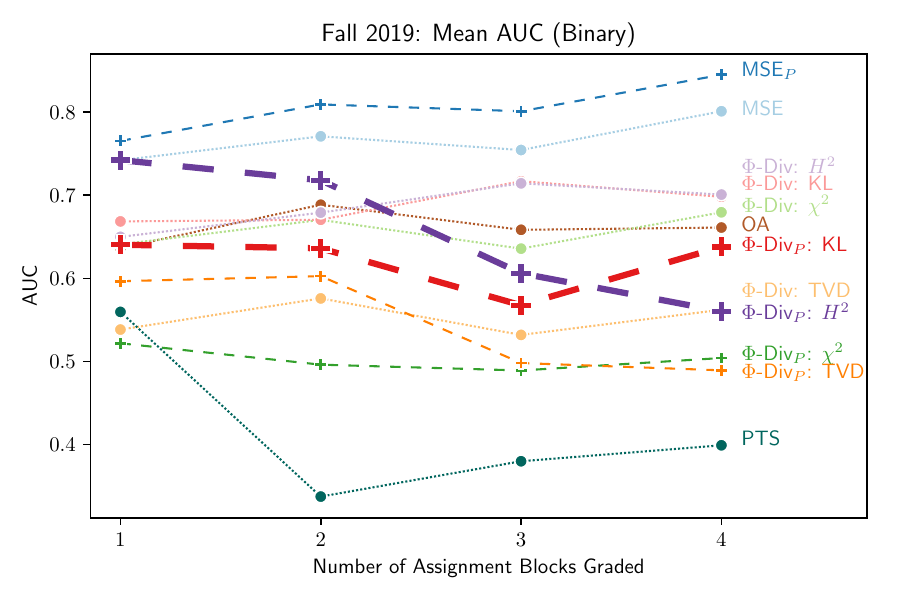}
  \hfill
  \includegraphics[width=0.32\linewidth]{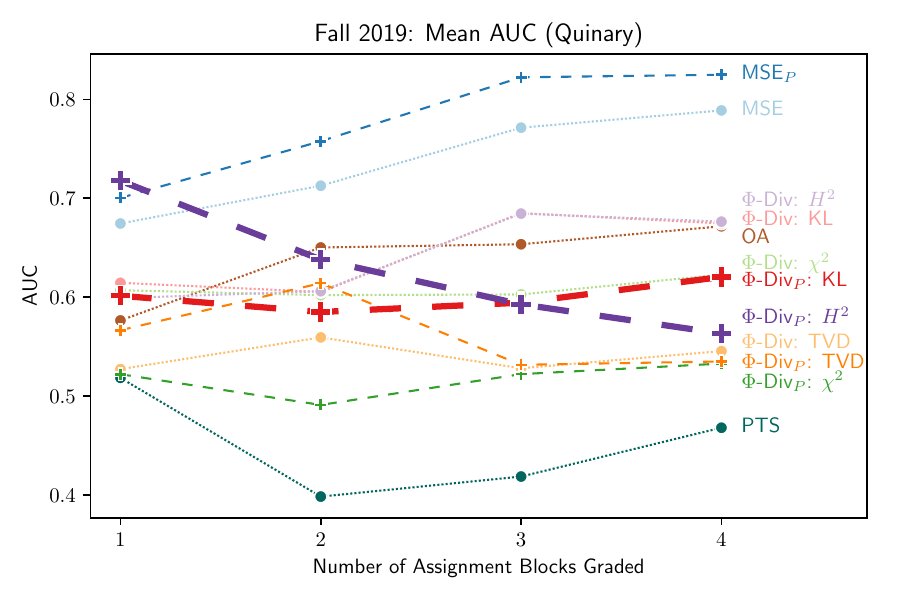}
  \hfill
  \includegraphics[width=0.32\linewidth]{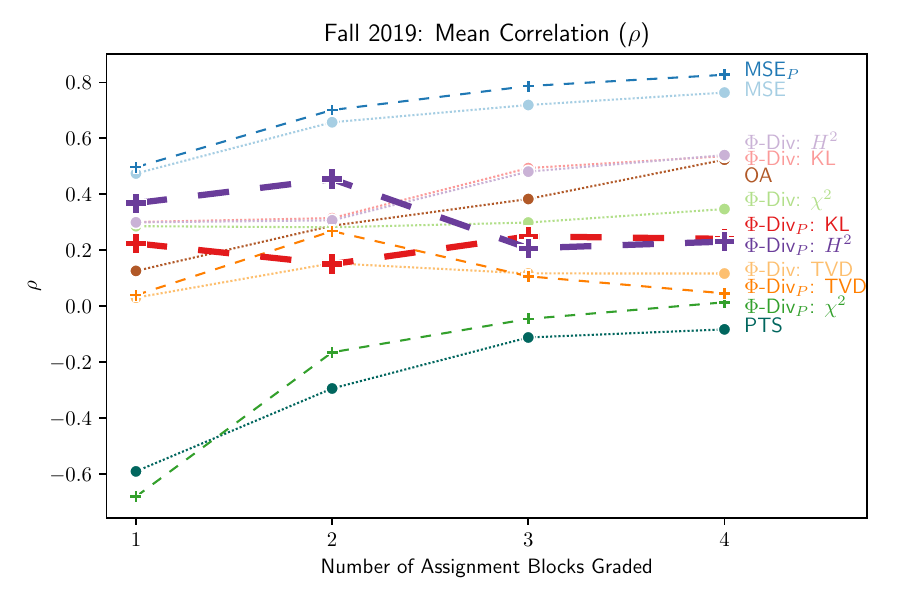}
  \hfill
  \caption{\textit{Quantifying Measurement Integrity with Real Data.}  For each semester, the average values of evaluation metrics over 50 iterations of each mechanism. In each iteration, the metrics are calculated within each quartile---according to the number of peer grades with which they are associated---of students.
  }
  \label{appendix-fig:quantifying-mi-real-other-metrics}
\end{figure}

\subsection{Robustness Against Strategic Reporting}
\label{subappendix:strategic-results}
\setcounter{figure}{0}

\subsubsection{Computational Experiments with ABM}
\label{subsubappendix:strategic-results-abm}
\begin{figure}
  \centering
  \includegraphics[width=0.32\linewidth]{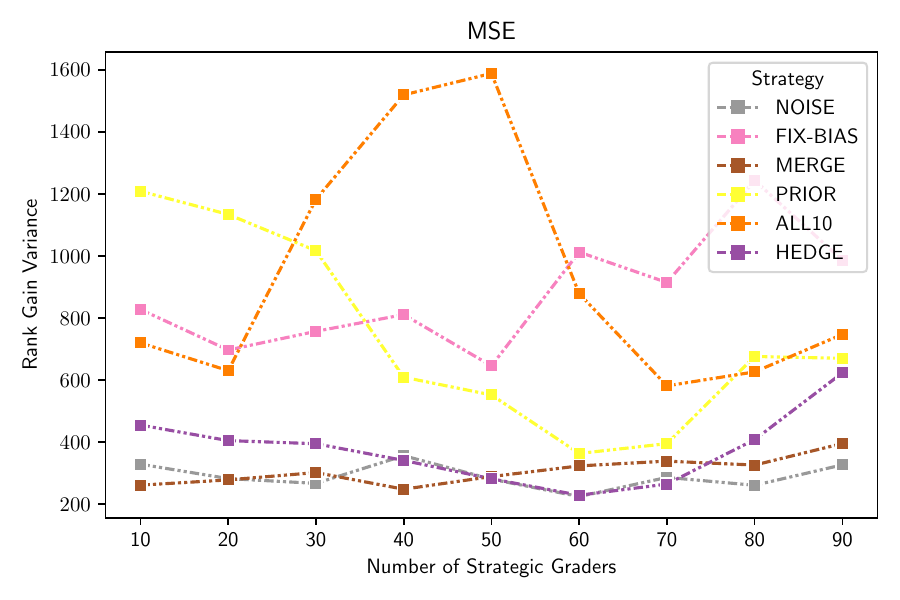}
  \hfill
  \includegraphics[width=0.32\linewidth]{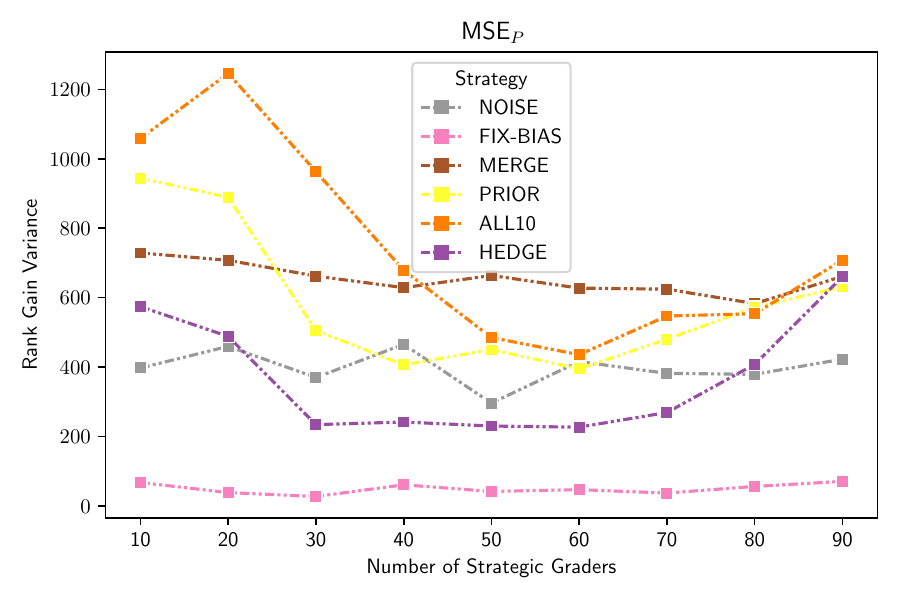}
  \hfill
  \includegraphics[width=0.32\linewidth]{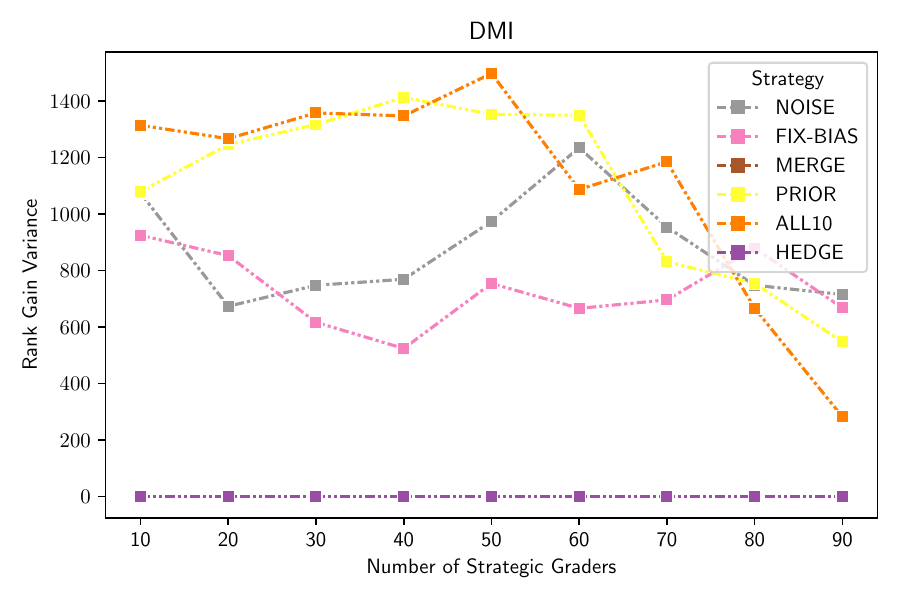}
  \hfill
  \includegraphics[width=0.32\linewidth]{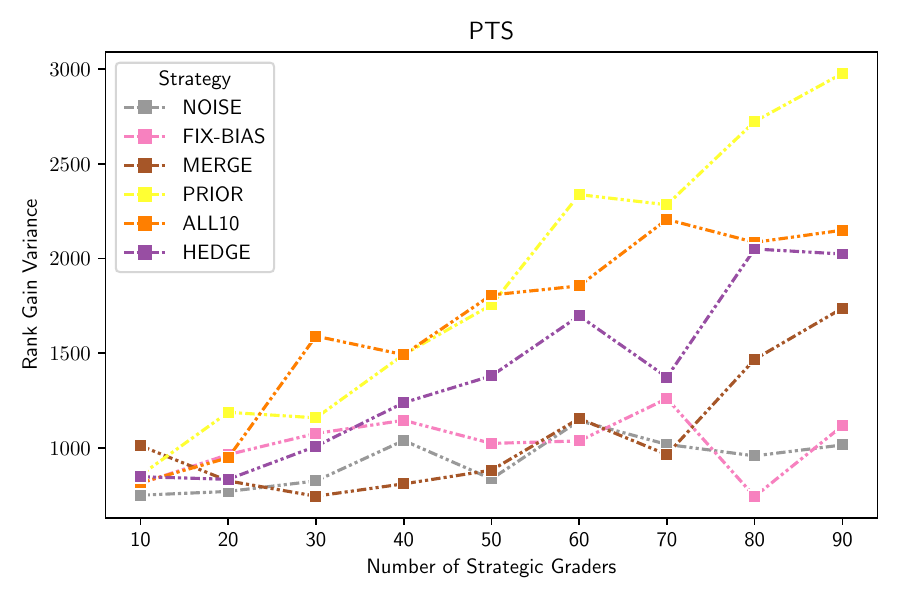}
  \hfill
  \includegraphics[width=0.32\linewidth]{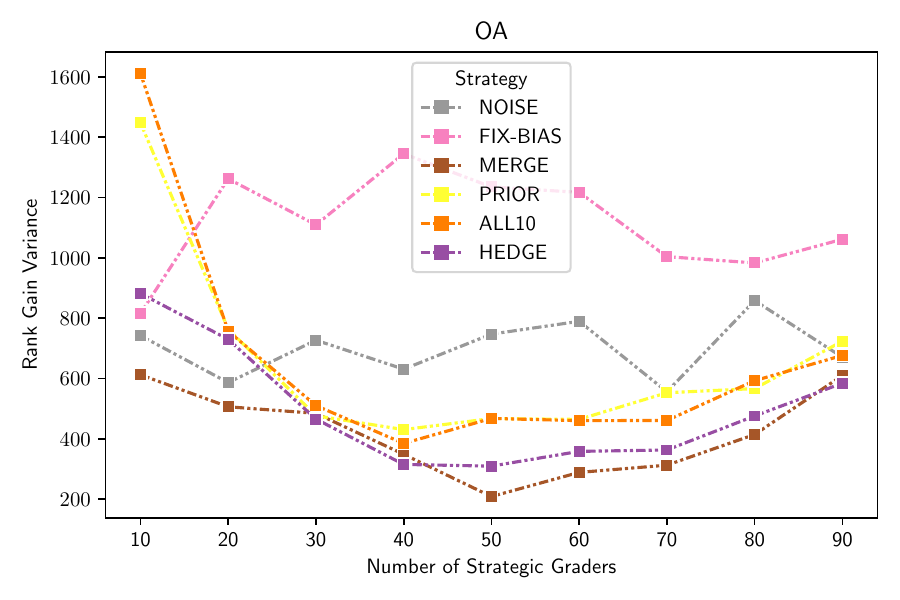}
  \hfill
  \includegraphics[width=0.32\linewidth]{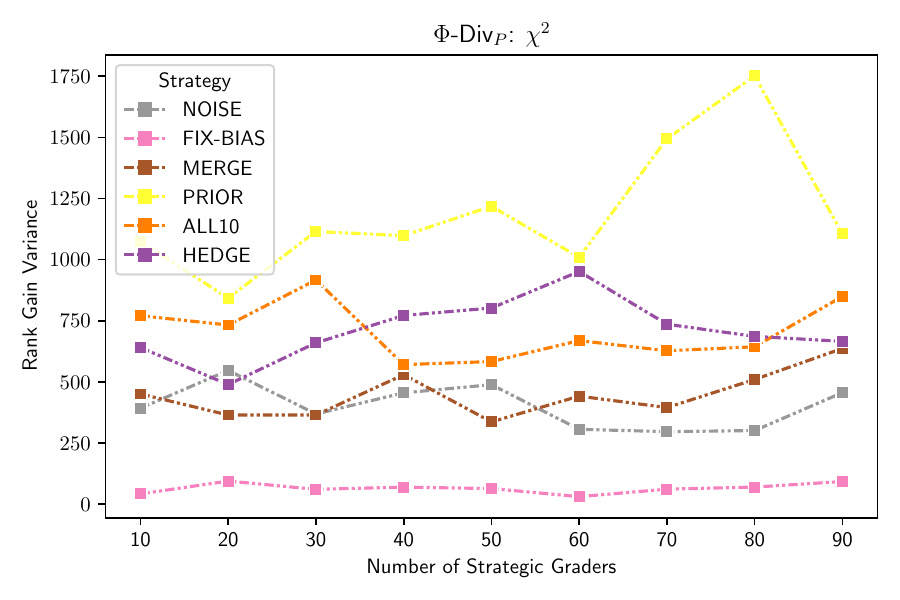}
  \hfill
  \includegraphics[width=0.32\linewidth]{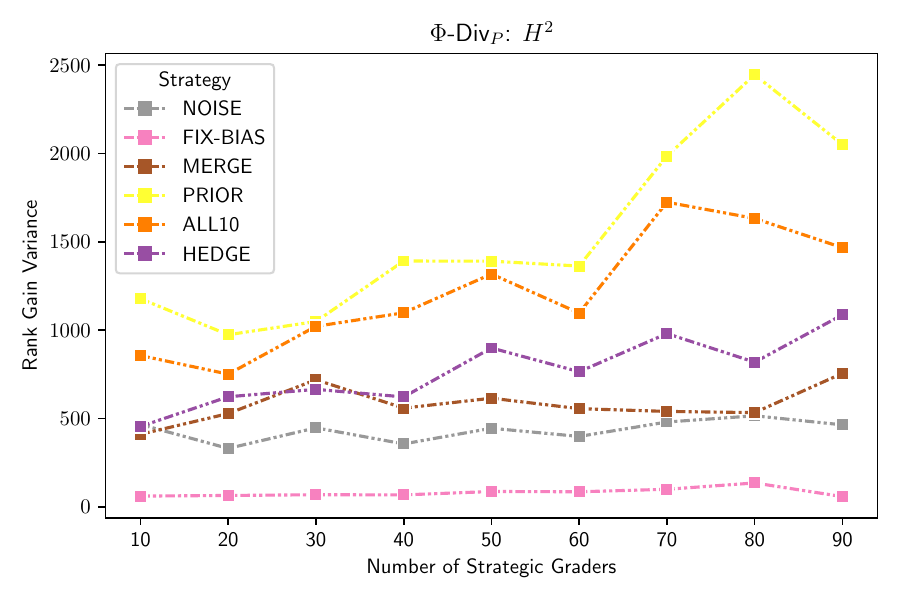}
  \hfill
  \includegraphics[width=0.32\linewidth]{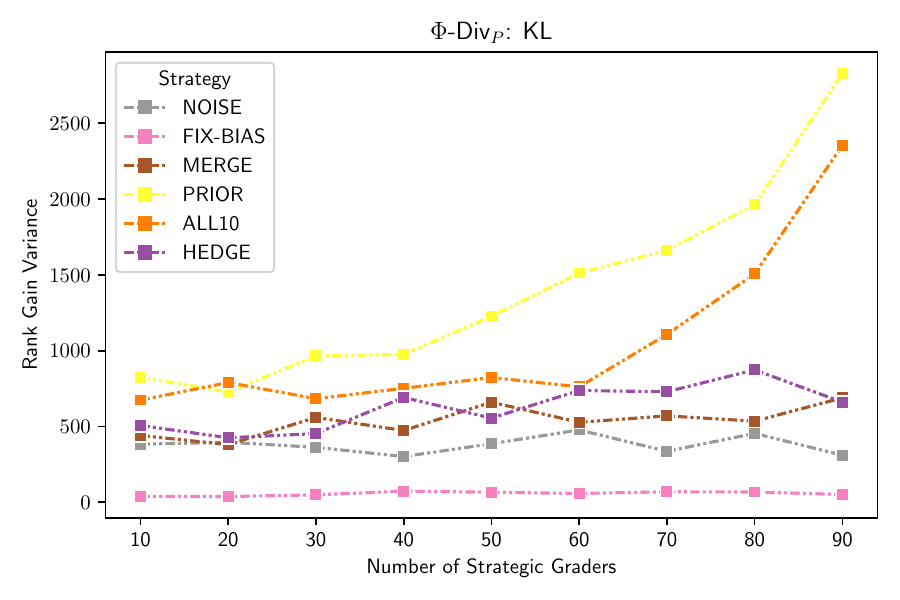} 
  \hfill
  \includegraphics[width=0.32\linewidth]{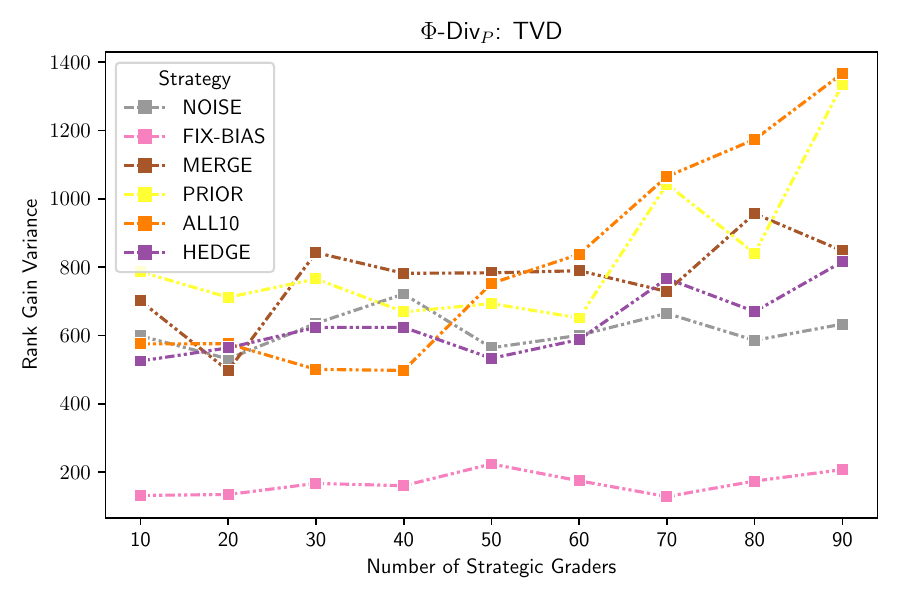}
  \hfill
  \includegraphics[width=0.32\linewidth]{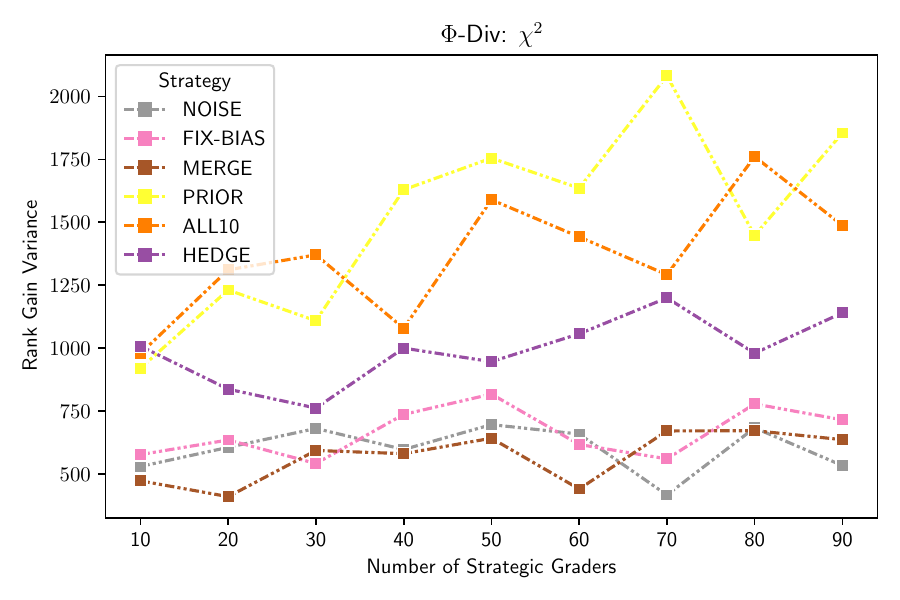}
  \hfill
  \includegraphics[width=0.32\linewidth]{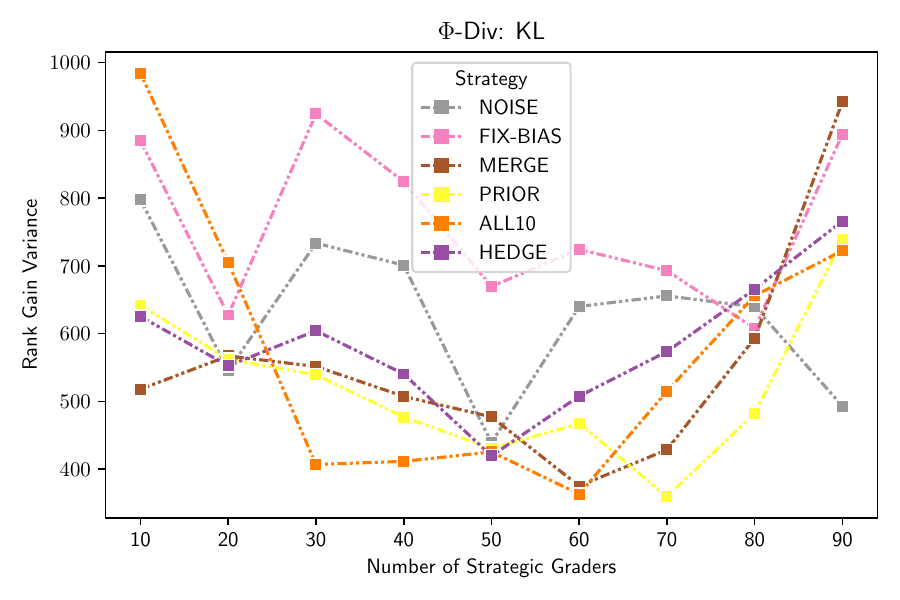}
  \hfill
  \includegraphics[width=0.32\linewidth]{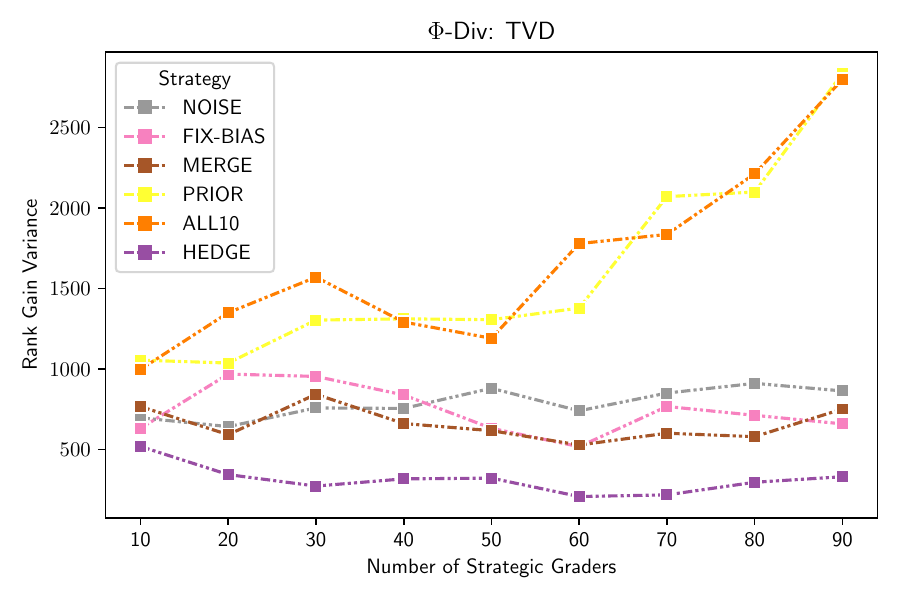}
  \hfill
  \includegraphics[width=0.32\linewidth]{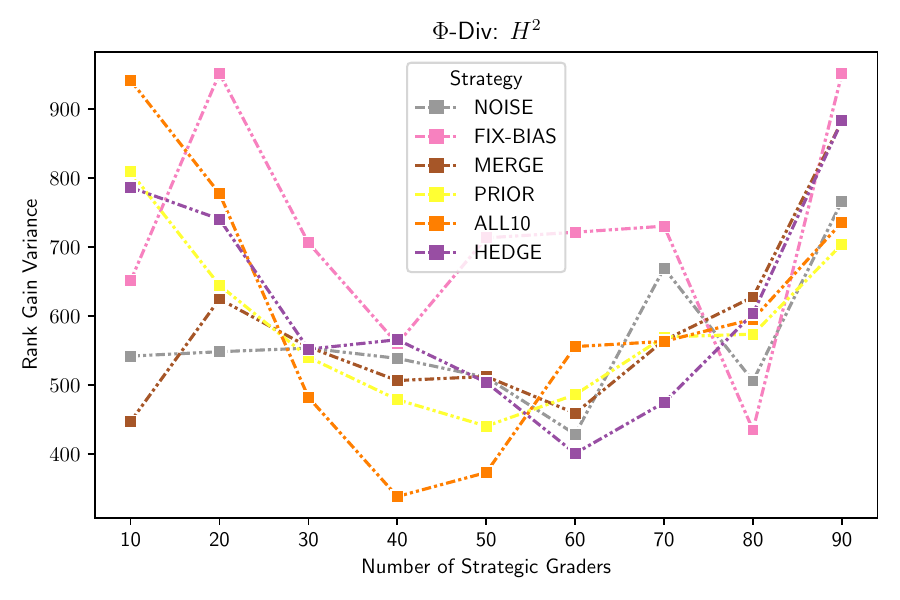}
  \caption{\textit{Quantifying Robustness with ABM.} Variance of the rank gain achieved by a single student deviating from truthful to strategic reporting.}
  \label{appendix-fig:deviation-rank-variance}
\end{figure}

See \Cref{appendix-fig:deviation-mean-rank-gain} for the individual perspective on deviation measured by the mean of the rank gain achieved by a single student deviating from truthful to strategic reporting. We also record the \textit{variance} of the rank gain for each mechanism (\Cref{appendix-fig:deviation-rank-variance}) . Note that for the DMI mechanism, the effect of the strategies is different, because of the mapping down to only 2 report options. This explains why several of the strategies are completely neutral under DMI; they don't affect the value of the report after the mapping.

One particularity of our approach in these experiments is that we consider only homogeneous strategy profiles (all non-truthful agents adopt the same strategy), which could potentially limit the generalizability of our results. This concern is somewhat mitigated by the fact that robustness in our experiments is more or less binary. The degree to which mechanisms reward or punish strategic behavior varies, as shown in \Cref{fig:2D-tradeoff}, but mechanisms tend to either be susceptible to strategic behavior---rewarding it to some degree in nearly all cases that we consider in our experiments, thereby having a negative $x$-coordinate in \Cref{fig:2D-tradeoff}---or are robust against it---punishing it to some degree in all cases or in nearly all cases except for a few extreme ones, thereby having a positive $x$-coordinate in \Cref{fig:2D-tradeoff}. This consideration of strategy profiles is also not a concern in our experiments with the real data (\Cref{subsubappendix:strategic-results-real}), since using real grades allows us to experiment with actual strategy profiles (whatever they may have been) followed by real students using peer grading in a course. 

\begin{figure}
  \centering
  \includegraphics[width=0.32\linewidth]{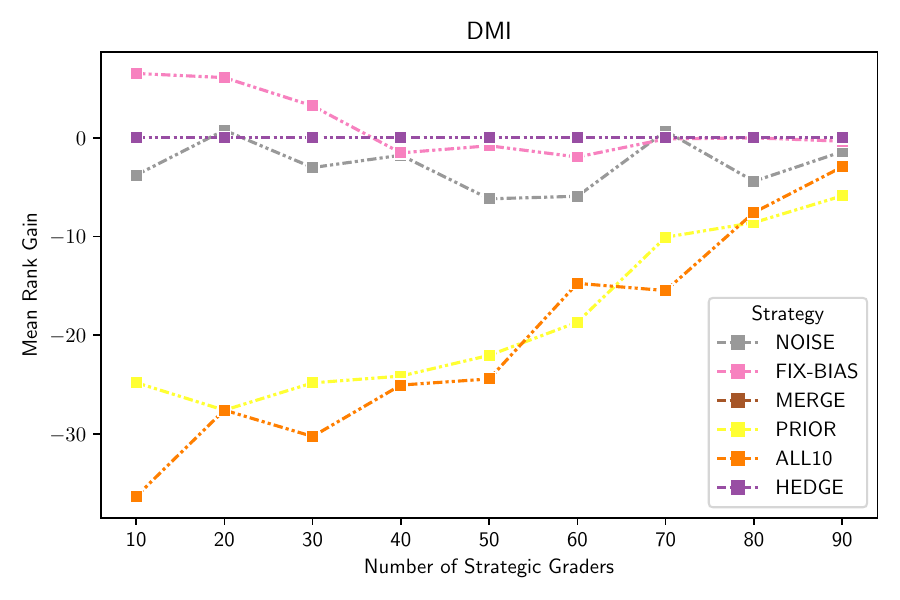}
  \hfill
  \includegraphics[width=0.32\linewidth]{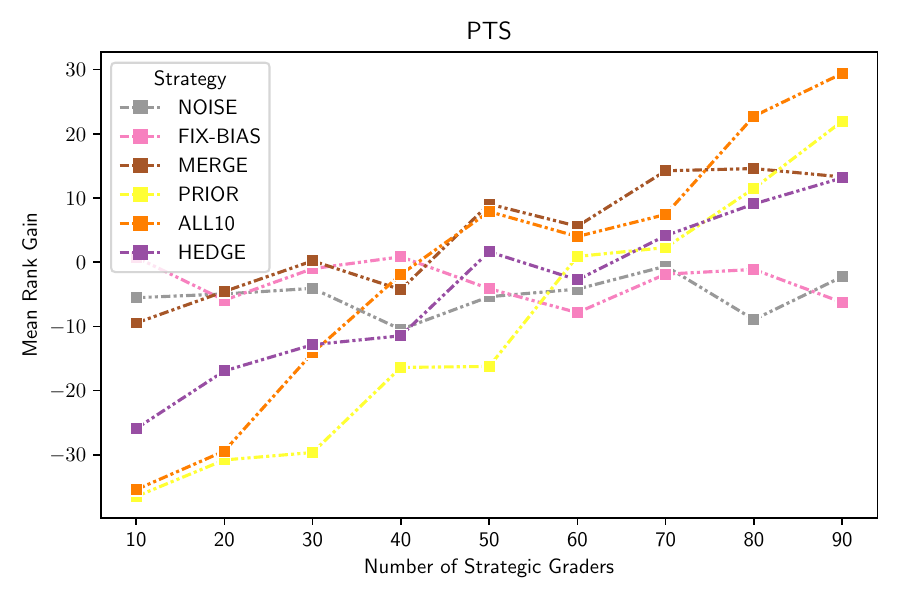}
  \hfill
  \includegraphics[width=0.32\linewidth]{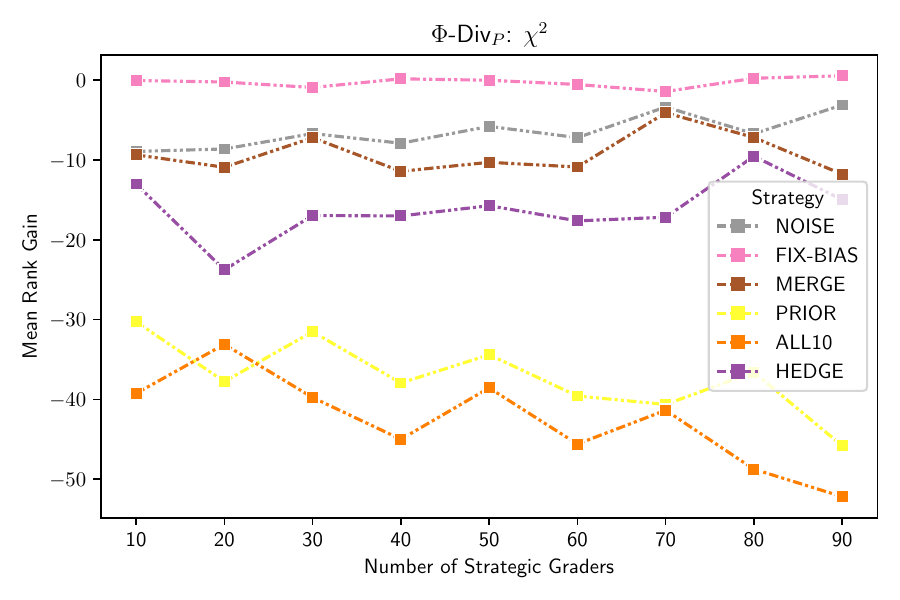} 
  \hfill
  \includegraphics[width=0.32\linewidth]{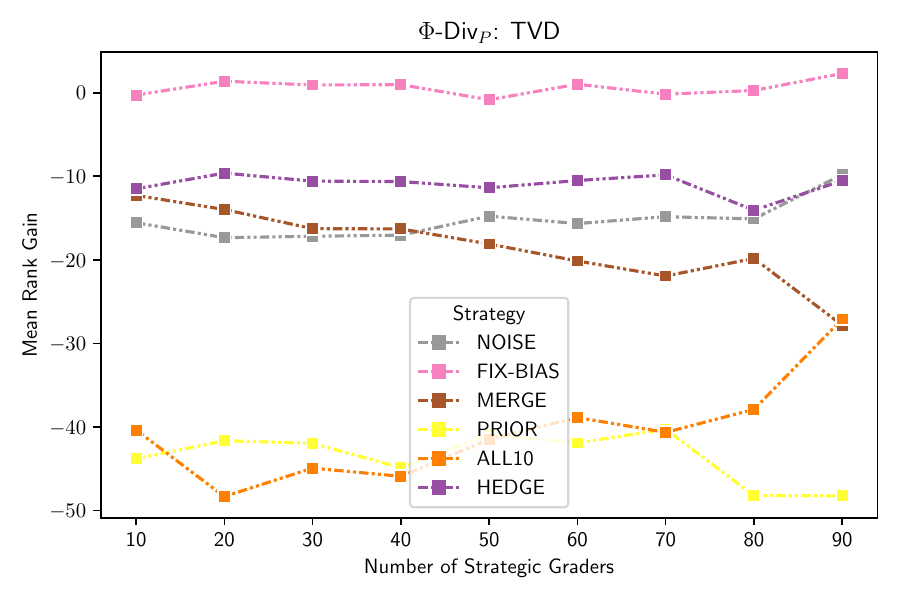} 
  \hfill
  \includegraphics[width=0.32\linewidth]{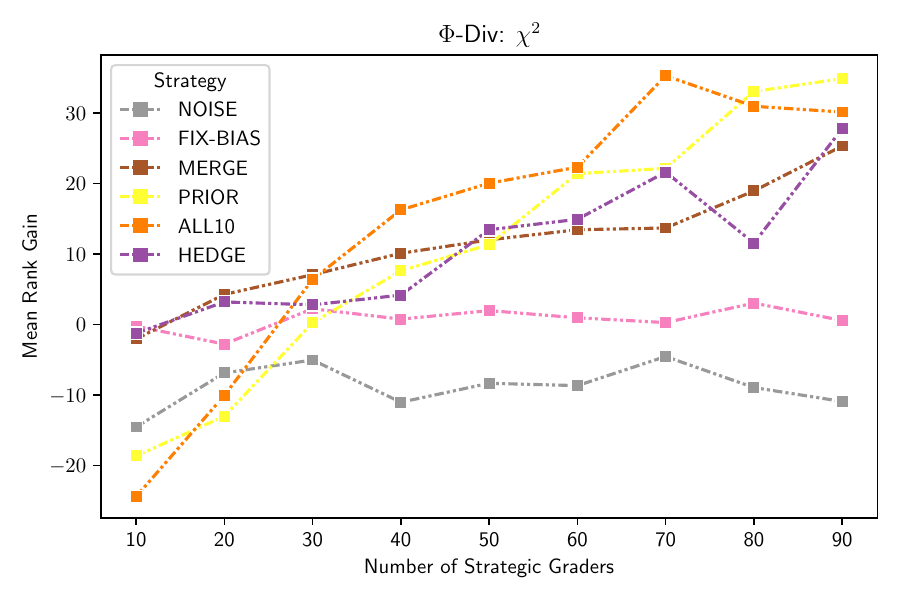} 
  \hfill
  \includegraphics[width=0.32\linewidth]{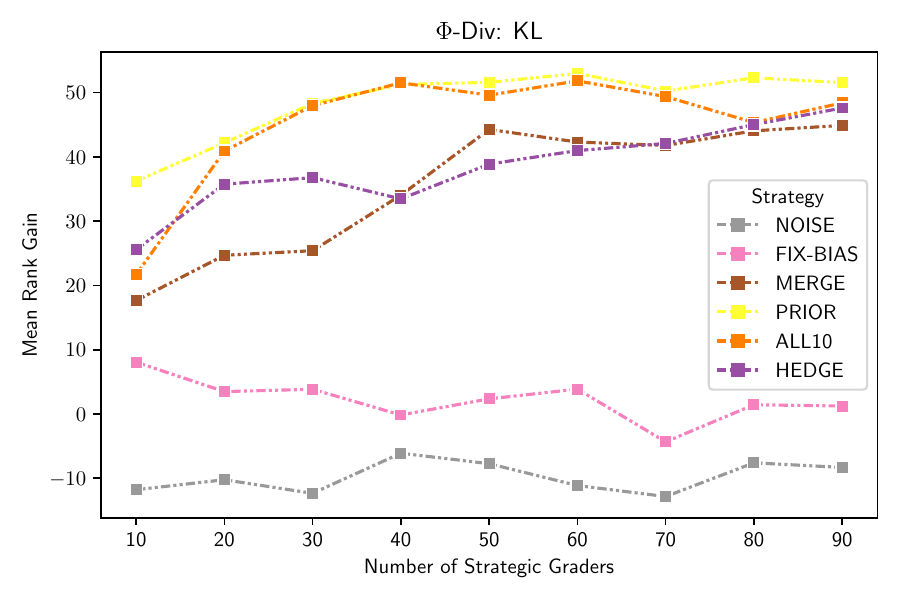}
  \hfill
  \includegraphics[width=0.32\linewidth]{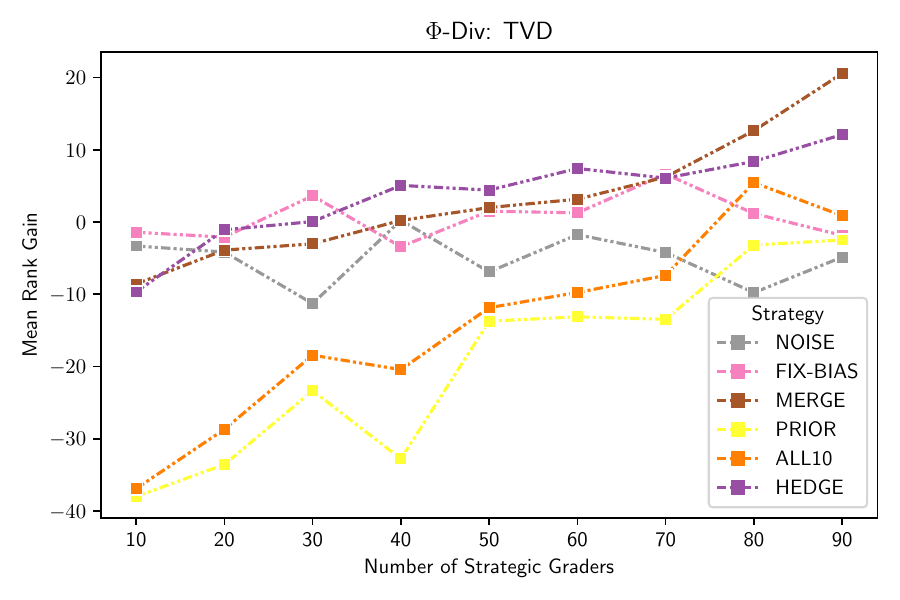}
  \includegraphics[width=0.32\linewidth]{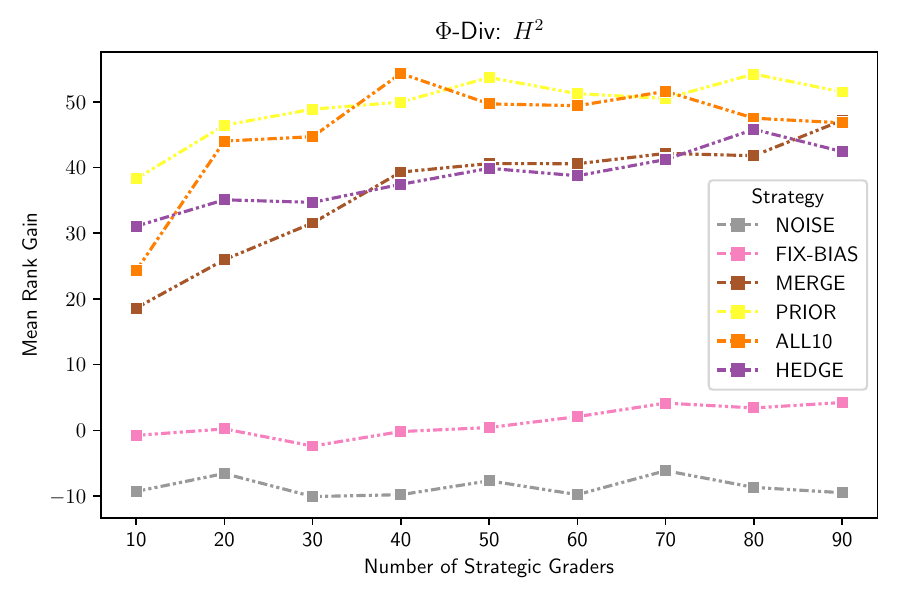} 
  \caption{\textit{Quantifying Robustness with ABM.} Average rank gain achieved by a single student deviating from truthful to strategic reporting. 
  }
  \label{appendix-fig:deviation-mean-rank-gain}
\end{figure}

In our simulated experiments (described in \Cref{subsubsection:strategic-abm-methods}), during the first reward assignment of each iteration, we also record the AUC resulting from a consideration of each mechanism's rewards as scores with which to classify the agents as either truthful or strategic. This gives a more population-level view of the incentives for reporting truthfully or strategically. In this case, AUC can be interpreted as the probability that a uniformly random truthful agent is ranked higher than a uniformly random strategic agent. The results are shown in \Cref{appendix-fig:strategic-payment-auc}.\footnote{For clarity, the plots are shown in steps of size 20 instead of 10.} This population-level view tells the same story that is depicted by \Cref{fig:deviation-gain} and \Cref{appendix-fig:deviation-mean-rank-gain}, corroborating our individual-level analysis.

\begin{figure}[ht]
  \centering
  \includegraphics[width=0.32\linewidth]{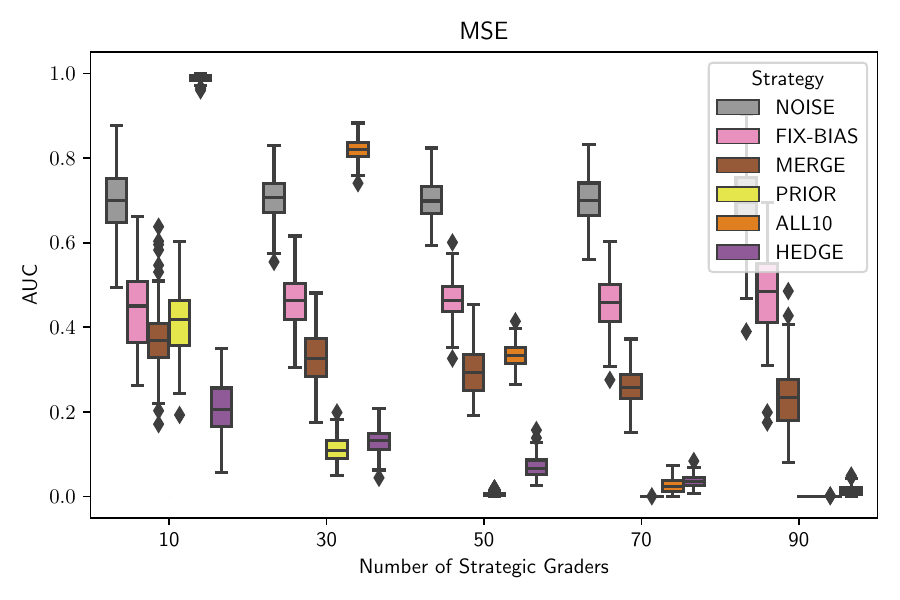}
  \hfill
  \includegraphics[width=0.32\linewidth]{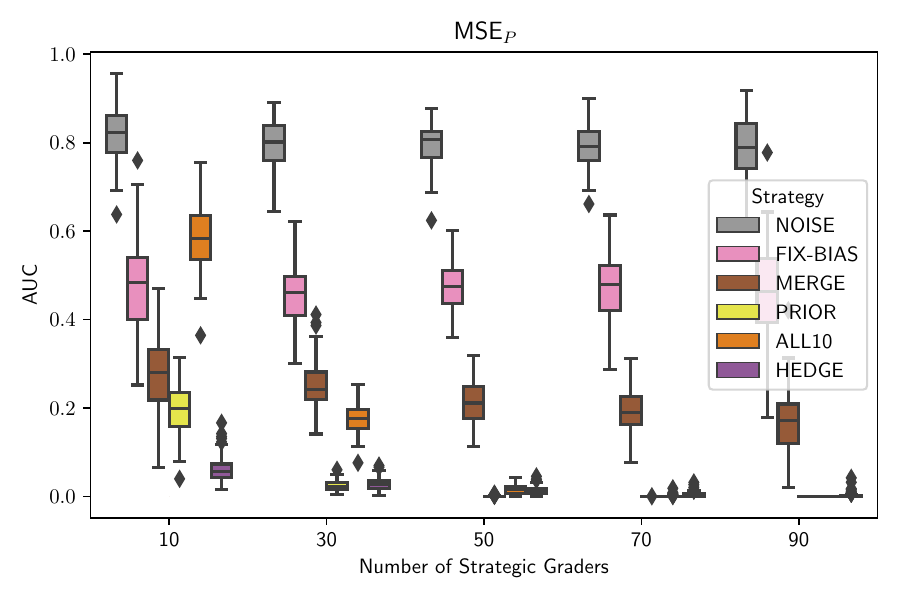}
  \hfill
  \includegraphics[width=0.32\linewidth]{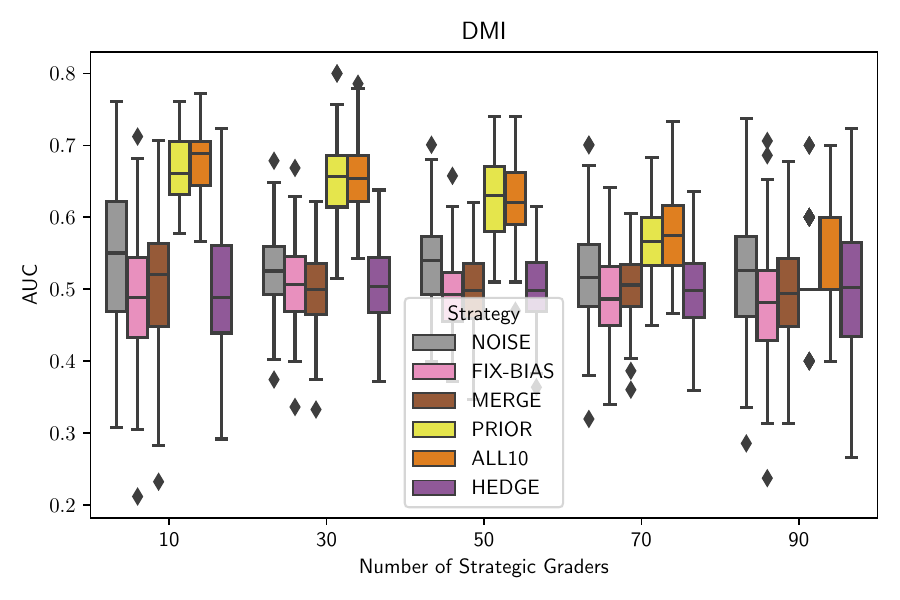}
  \hfill
  \includegraphics[width=0.32\linewidth]{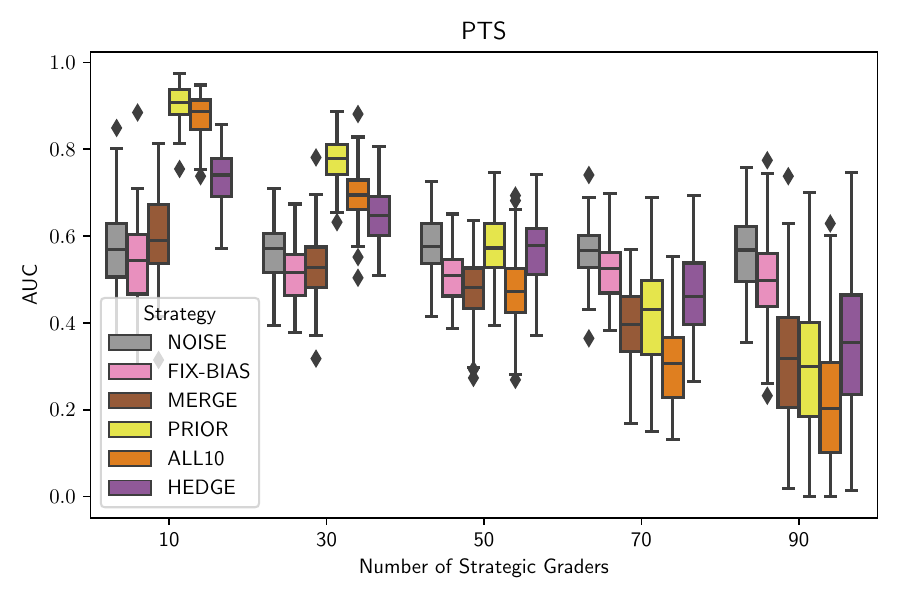}
  \hfill
  \includegraphics[width=0.32\linewidth]{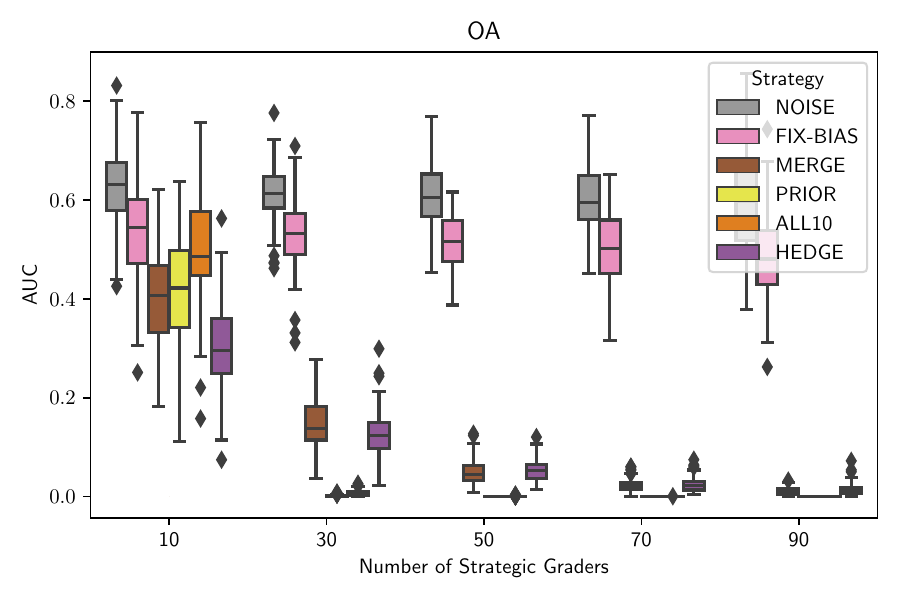}
  \hfill
  \includegraphics[width=0.32\linewidth]{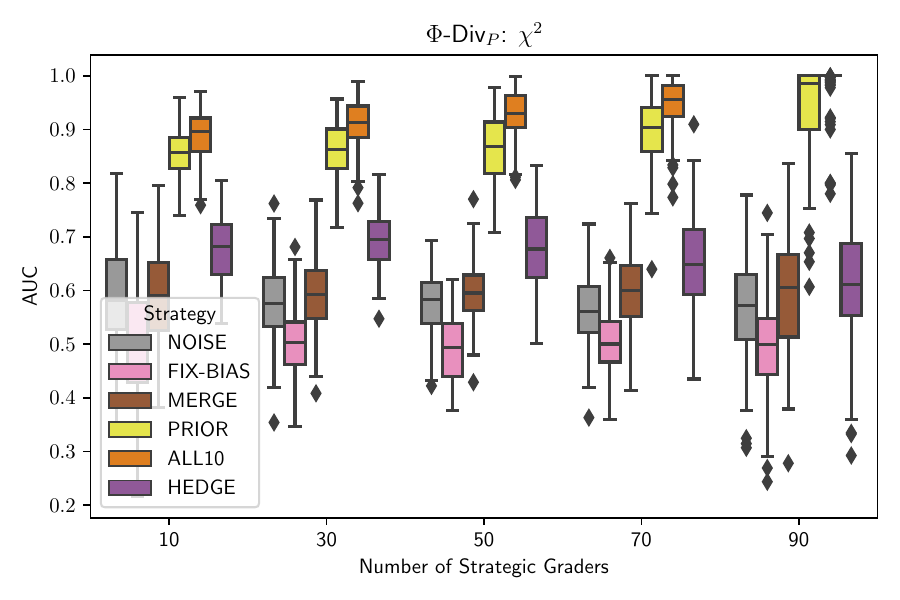}
  \hfill
  \includegraphics[width=0.32\linewidth]{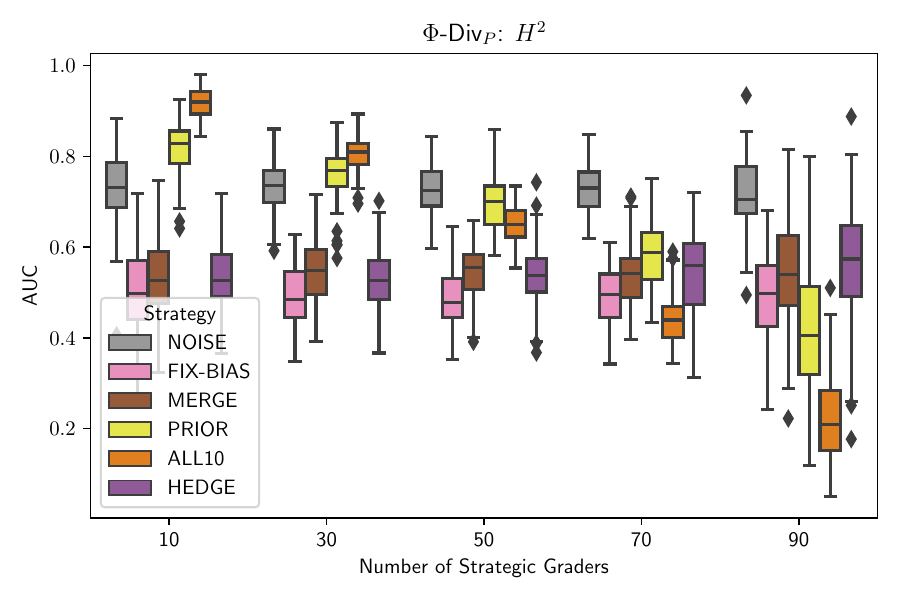}
  \hfill
  \includegraphics[width=0.32\linewidth]{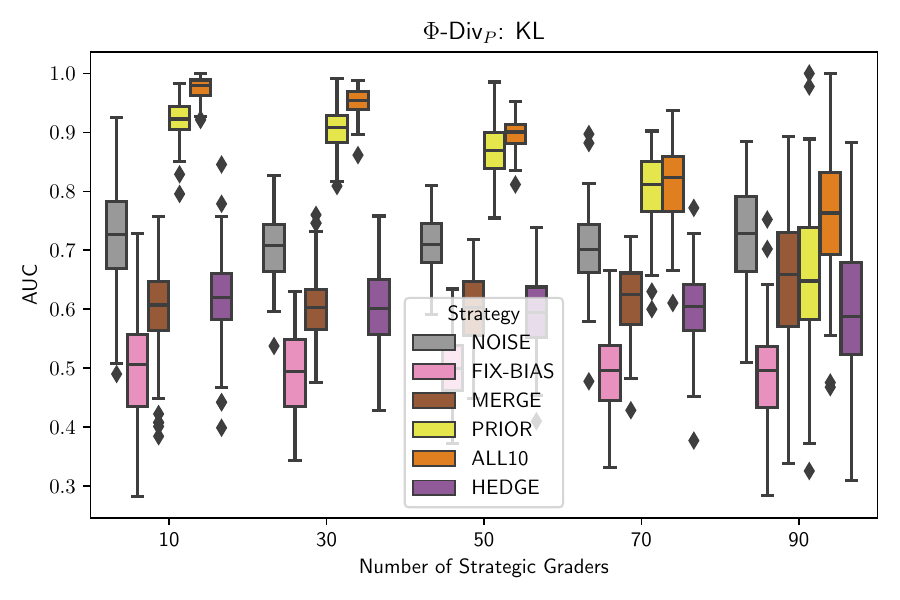}
  \hfill
  \includegraphics[width=0.32\linewidth]{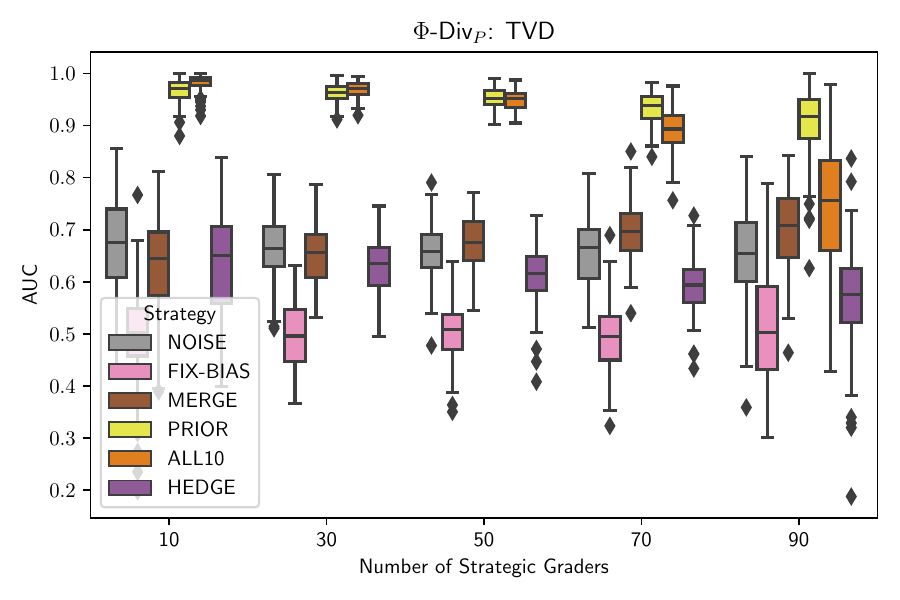}
  \hfill
  \includegraphics[width=0.32\linewidth]{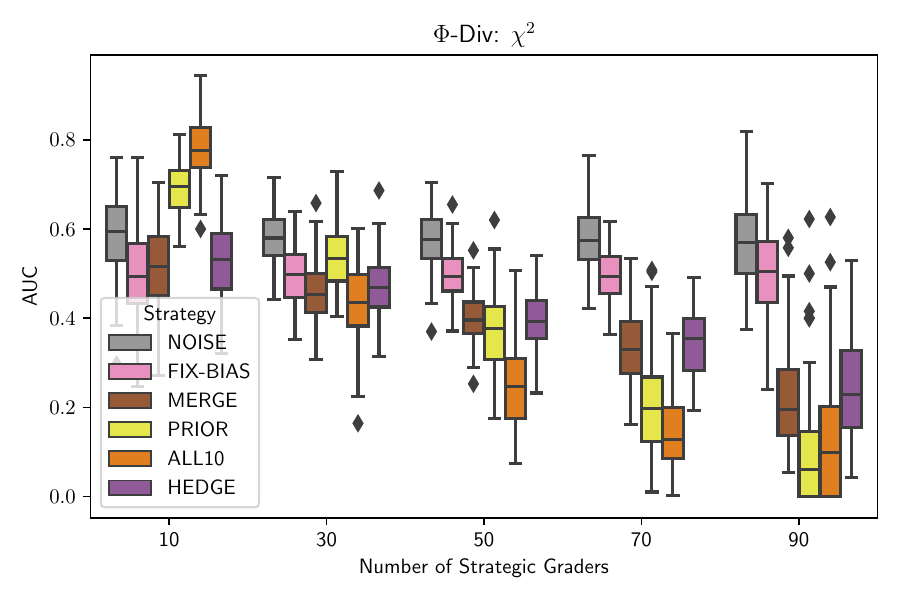}
  \hfill
  \includegraphics[width=0.32\linewidth]{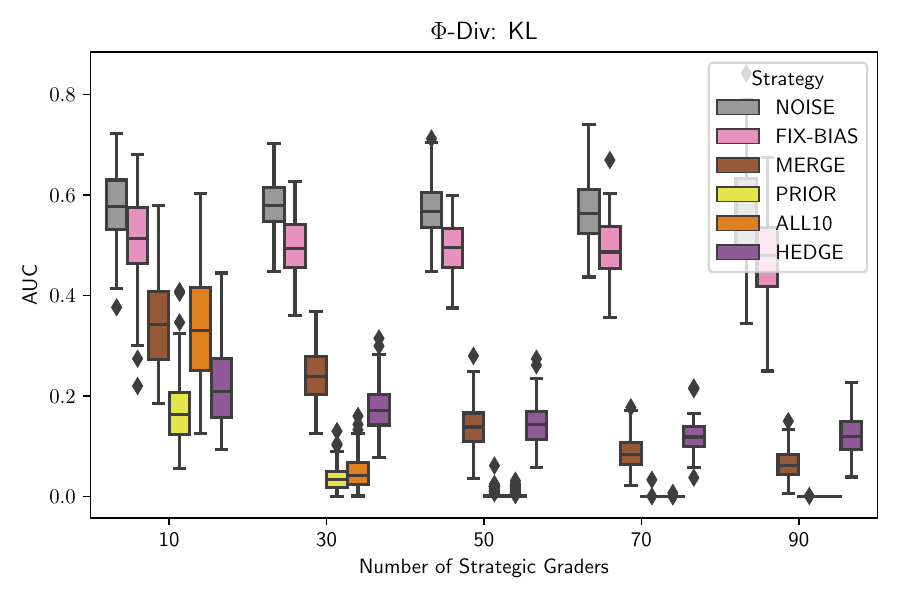}
  \hfill
  \includegraphics[width=0.32\linewidth]{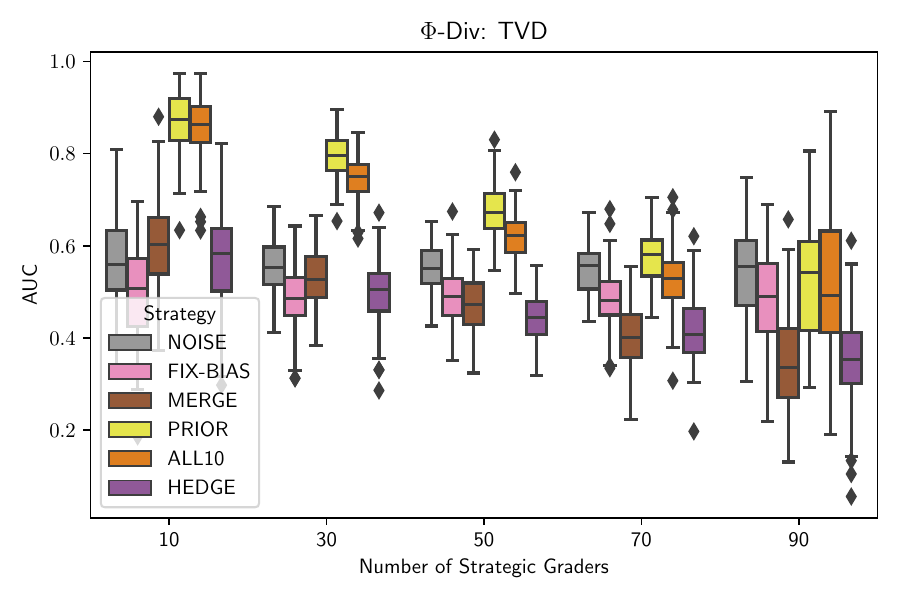}
  \hfill
  \includegraphics[width=0.32\linewidth]{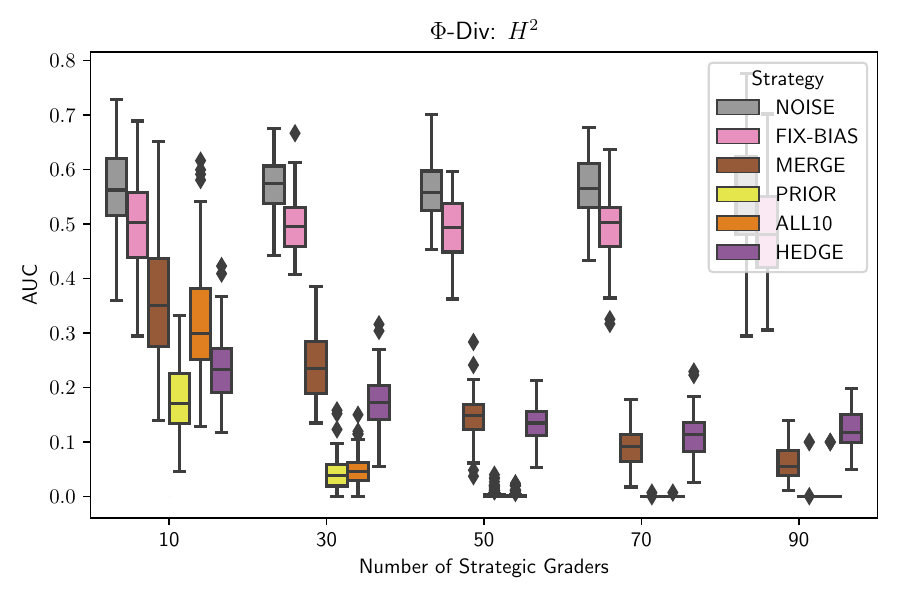}
\caption{\textit{Quantifying Robustness with ABM.} Comparing the rewards between strategic and truthful agents using AUC.}
\label{appendix-fig:strategic-payment-auc}
\end{figure}

\subsubsection{Computational Experiments with Real Data} \hfill
\label{subsubappendix:strategic-results-real}

\noindent
\textbf{Methods.}
In our experiments with simulated data, we explored how incentives for deviating from truthful reporting changed as the number of agents adopting a particular strategy increased. In the real data, however, we do not need to impose a strategy profile on the population of agents---the data were already generated according to some actual strategy profile adopted by the real students. As a result of this key difference, we modify the experiment described in \Cref{subsubsection:strategic-abm-methods} in the following manner.

For each semester in the real data, for each strategy, and for each mechanism:
    \begin{enumerate}
        \item The students, submissions, and reports for that semester are loaded from the data.
        
        \item For each student $s$:
        
        \begin{enumerate}[i.]
        
            \item Rewards are assigned twice according to the mechanism with a fixed random seed.
            In the first assignment, assignment of payments occurs without any modifications to the data.
            In the second assignment reports from student $s$ are modified according to the prescribed strategy. Due to the fixed random seed, every other factor is consistent with the first reward assignment.
        
            \item The gain in rank achieved by student $s$, i.e difference in the ranks according to the two reward assignments computed by the mechanism, is recorded.
        \end{enumerate}
        
    \item The mean and variance of the gain in rank over all students is computed for each mechanism, for each semester.
\end{enumerate}

Since we don't have access to a student's latent bias in the real data (and the true scores are noisy as a reference point) we do not consider the \textit{Fix Bias} strategy in these experiments.

\begin{figure}
  \centering
  \hfill
  \includegraphics[width=0.32\linewidth]{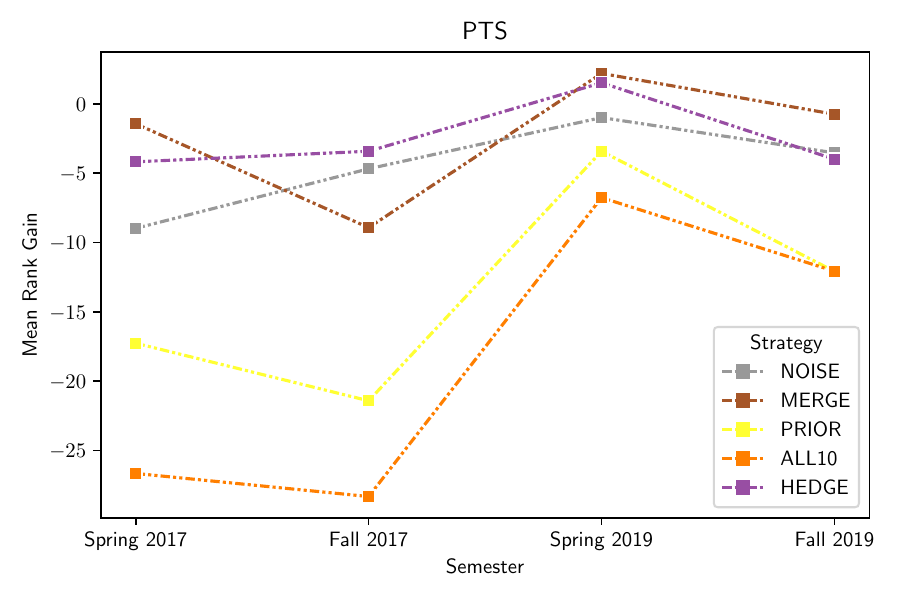}
  \hfill
  \hfill
  \includegraphics[width=0.32\linewidth]{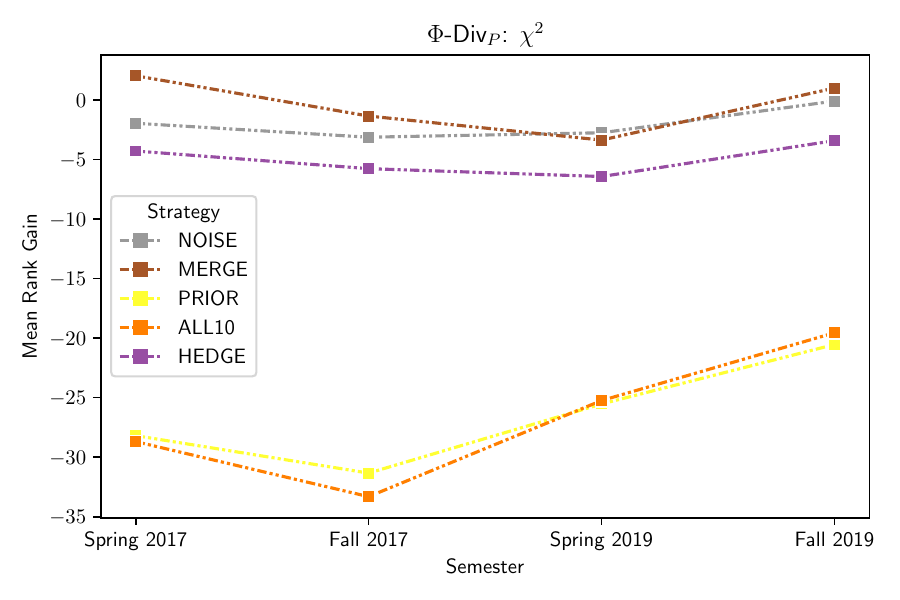}
  \hfill
  \includegraphics[width=0.32\linewidth]{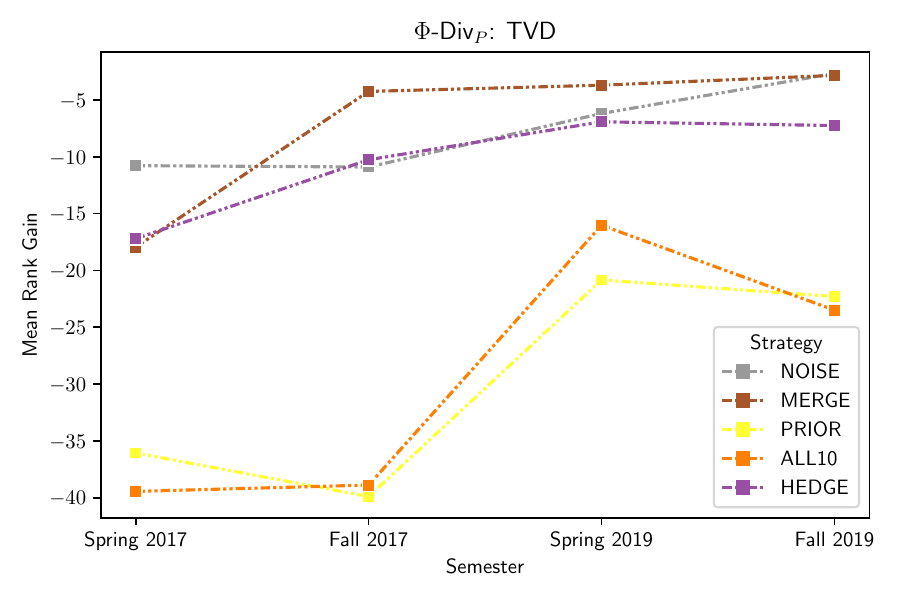}
  \hfill
  \includegraphics[width=0.32\linewidth]{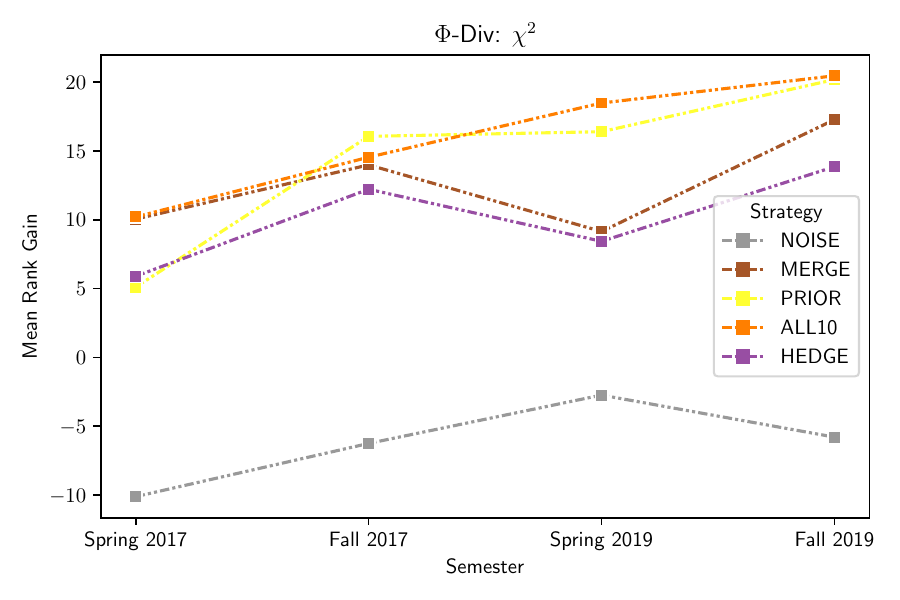}
  \hfill
  \includegraphics[width=0.32\linewidth]{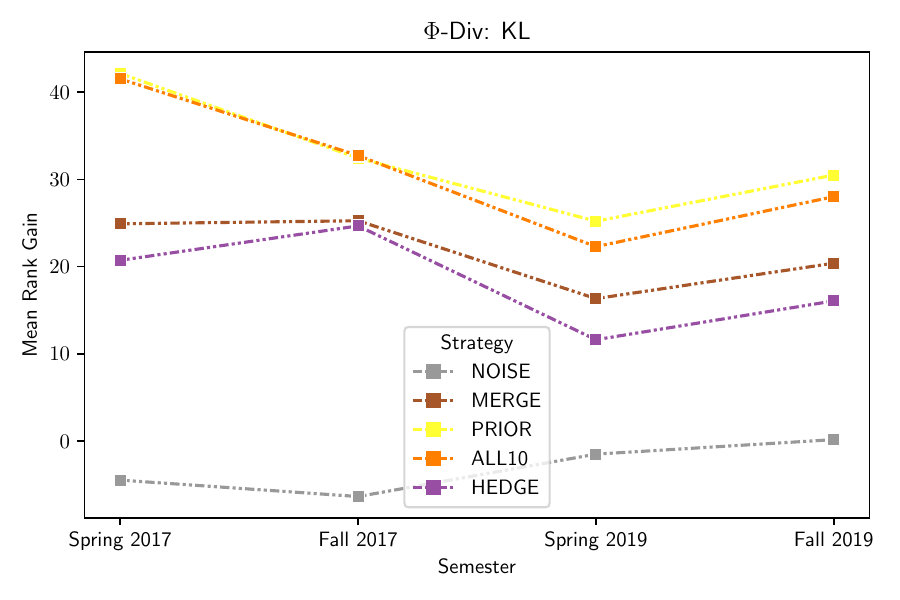}
  \hfill
  \includegraphics[width=0.32\linewidth]{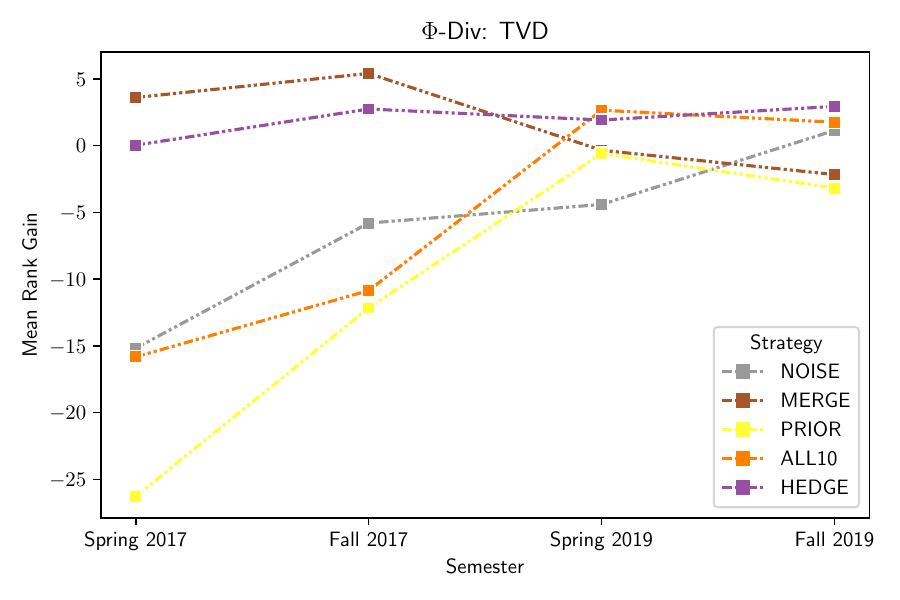}
  \hfill
  \includegraphics[width=0.32\linewidth]{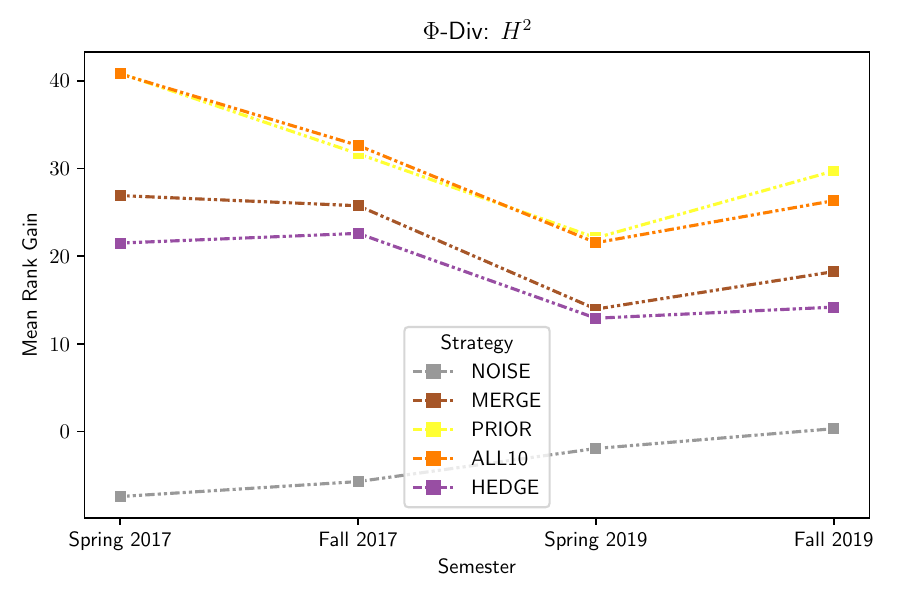}
  \caption{\textit{Quantifying Robustness with Real Data.} Average rank gain achieved by a single student deviating from truthful to strategic reporting.}
  \label{appendix-fig:mean-rank-gain-real}
\end{figure}

\begin{figure}
  \centering
  \includegraphics[width=0.32\linewidth]{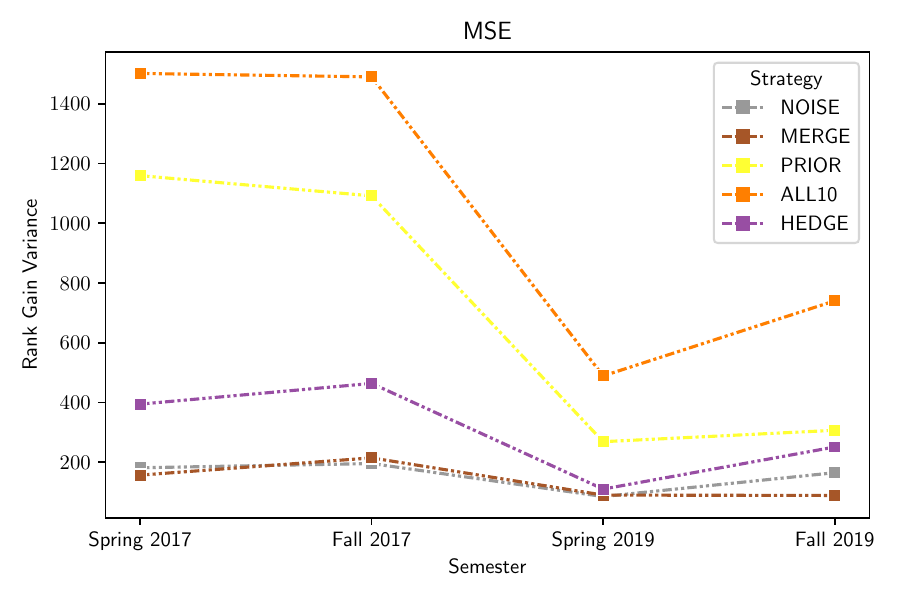}
  \hfill
  \includegraphics[width=0.32\linewidth]{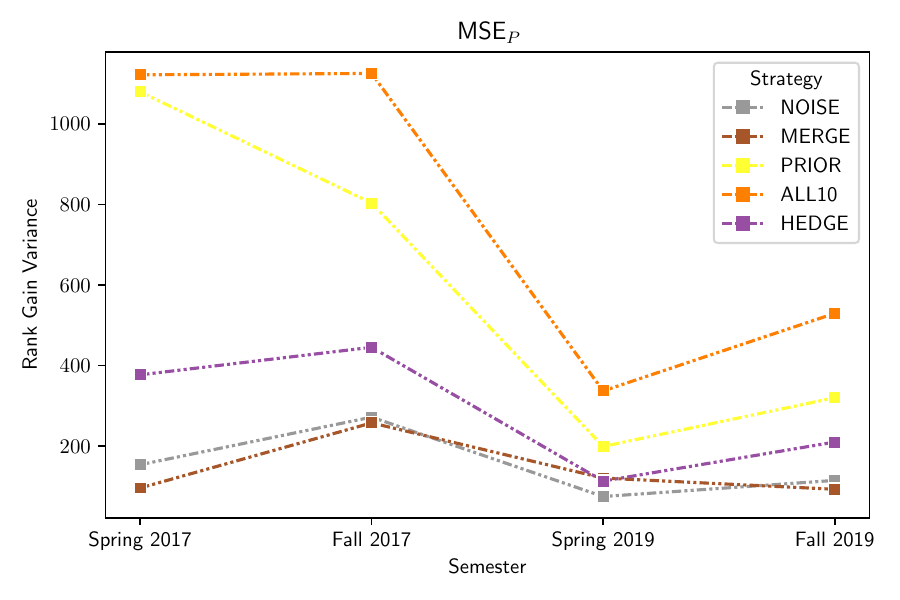}
  \hfill
  \includegraphics[width=0.32\linewidth]{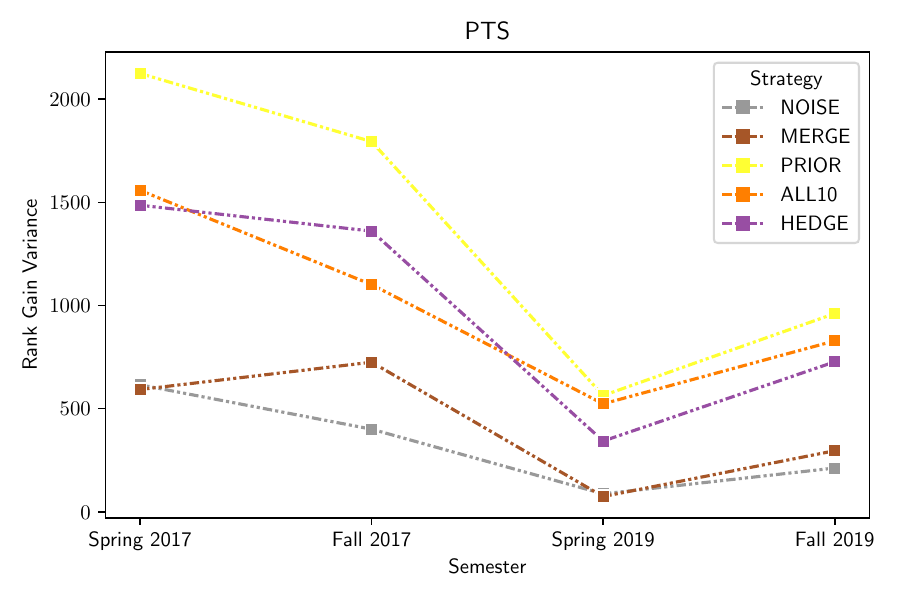}
  \hfill
  \includegraphics[width=0.32\linewidth]{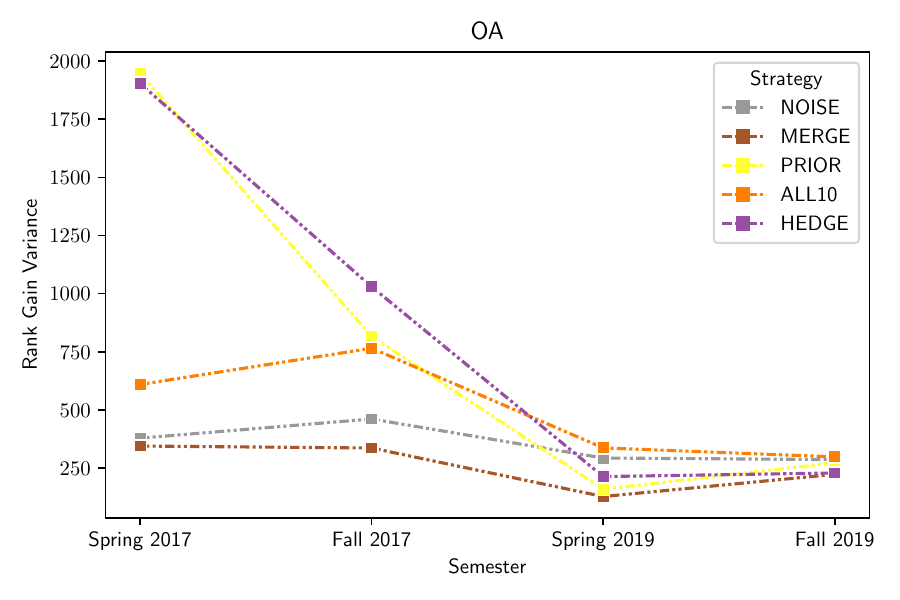}
  \hfill
  \includegraphics[width=0.32\linewidth]{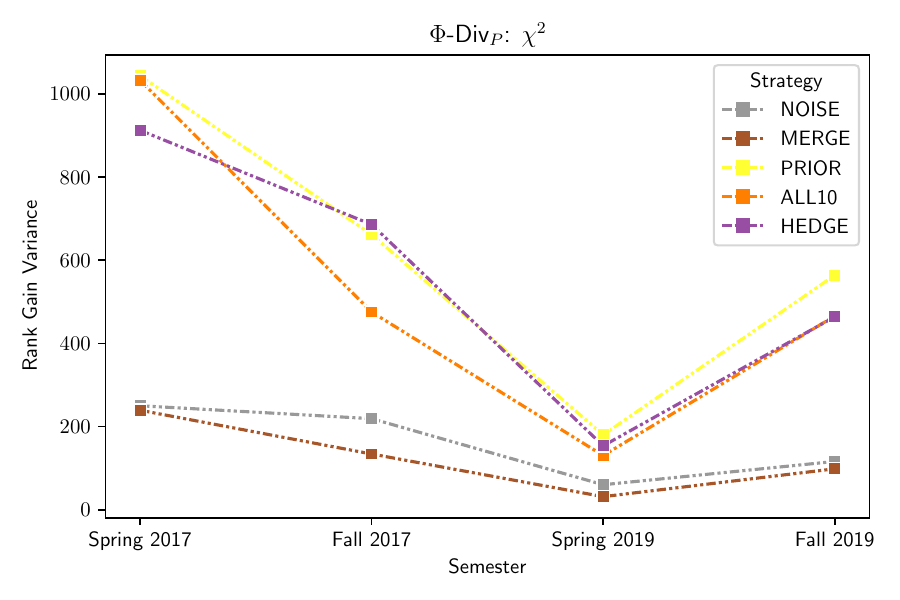}
  \hfill
  \includegraphics[width=0.32\linewidth]{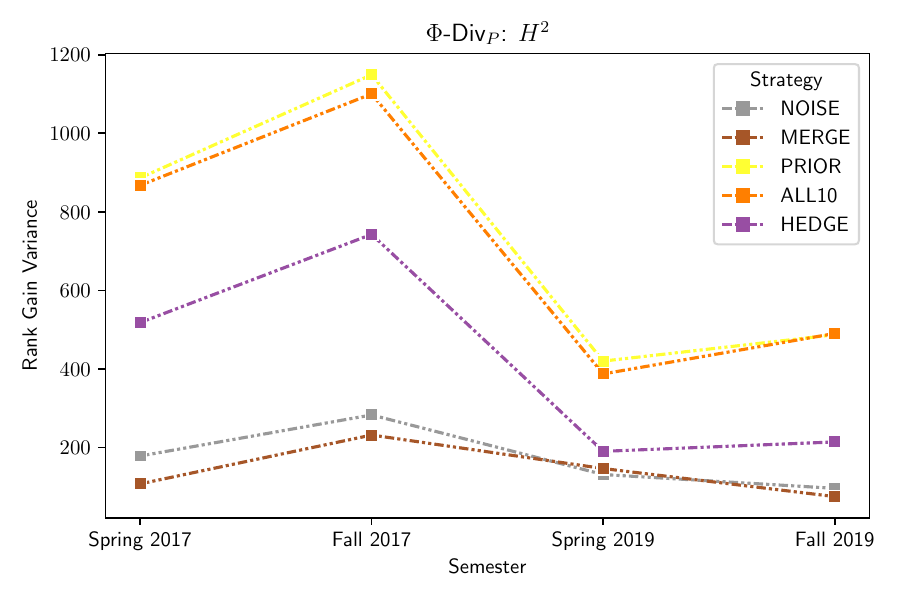}
  \hfill
  \includegraphics[width=0.32\linewidth]{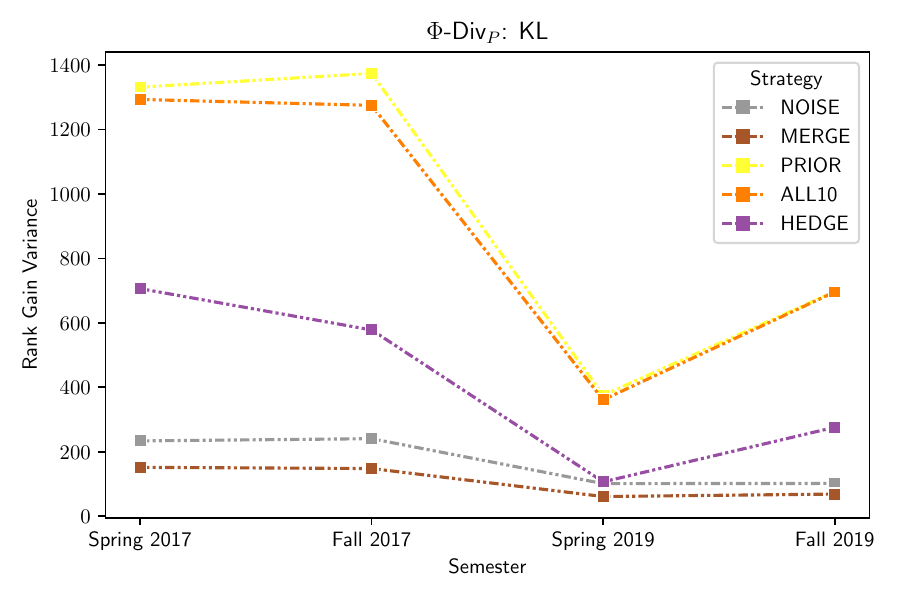} 
  \hfill
  \includegraphics[width=0.32\linewidth]{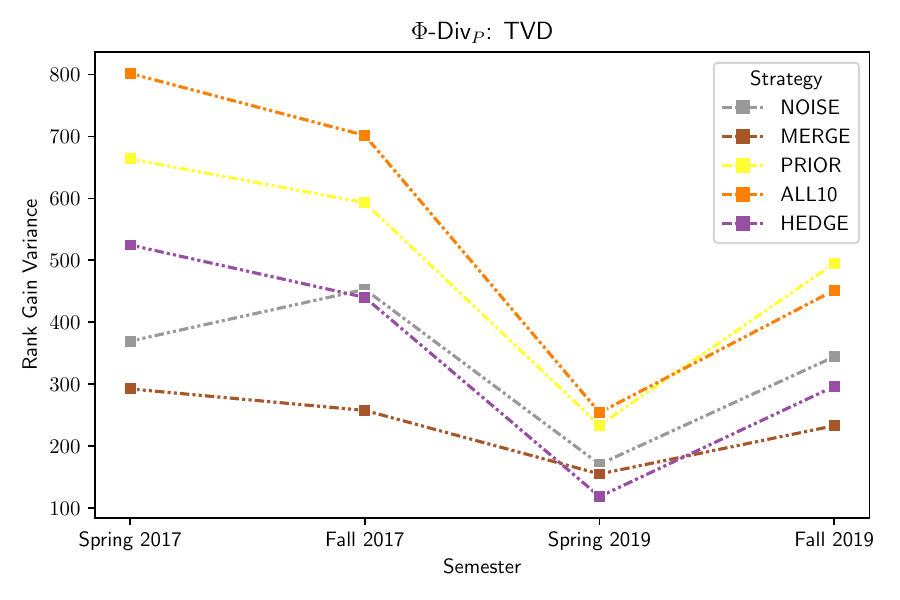}
  \hfill
  \includegraphics[width=0.32\linewidth]{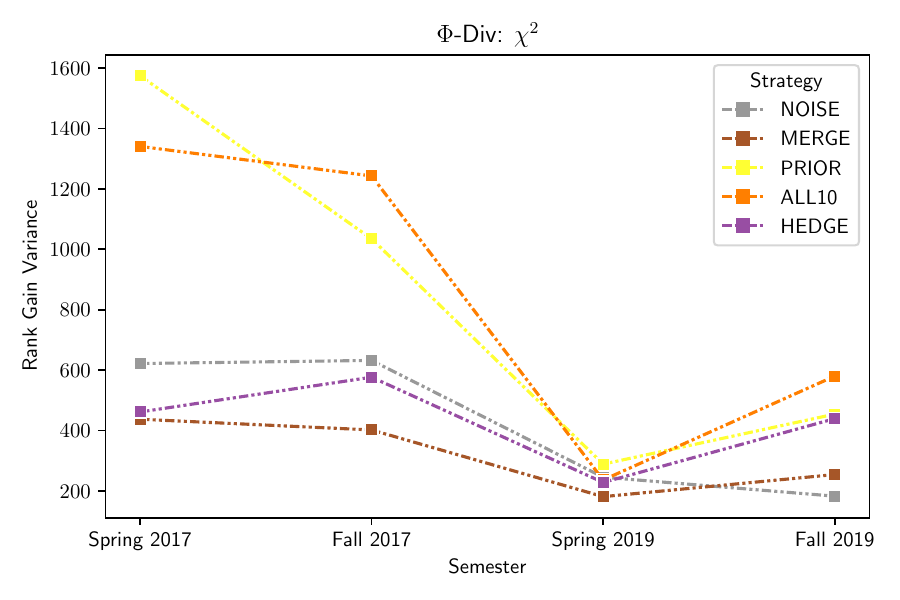}
  \hfill
  \includegraphics[width=0.32\linewidth]{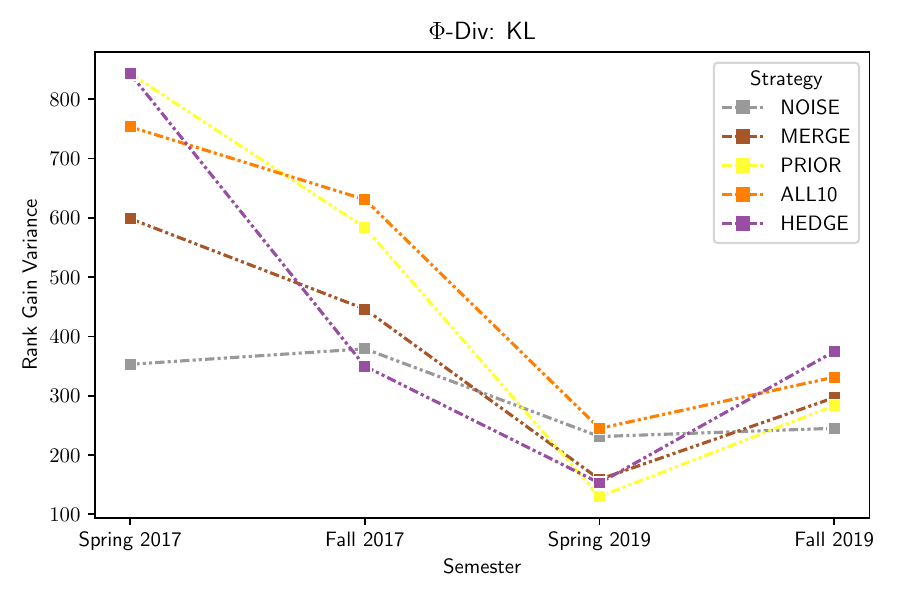}
  \hfill
  \includegraphics[width=0.32\linewidth]{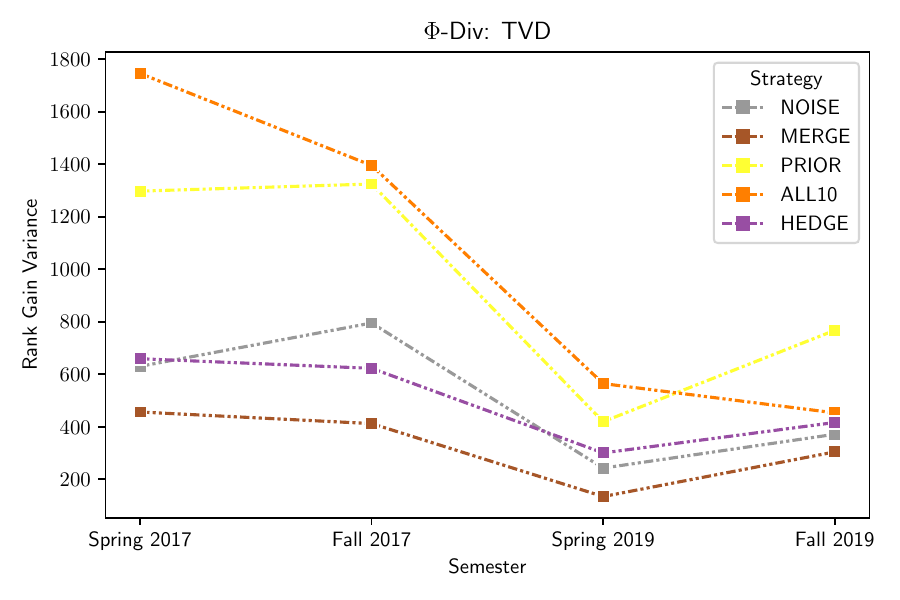}
  \hfill
  \includegraphics[width=0.32\linewidth]{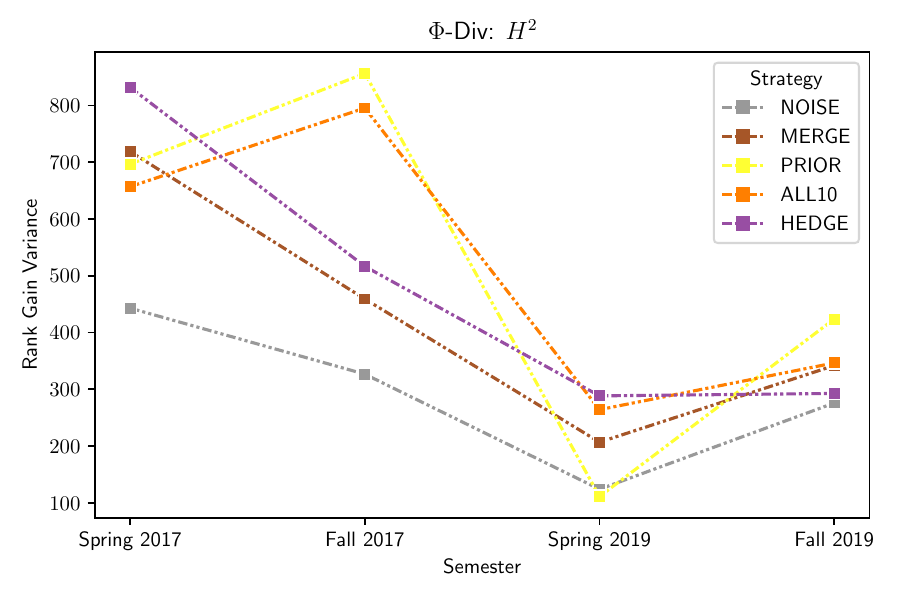}
  \caption{\textit{Quantifying Robustness with Real Data.} Variance of the rank gain achieved by a single student deviating from truthful to strategic reporting.}
  \label{appendix-fig:rank-gain-variance-real}
\end{figure}

\vspace{1 ex}
\noindent
\textbf{Results.} The results of this experiment involve the mean gain (\Cref{appendix-fig:mean-rank-gain-real}) and the variance of the gain (\Cref{appendix-fig:rank-gain-variance-real}) over the population of students achieved by each student deviating (one at a time) to each strategy in each semester.

As in our experiments with measurement integrity, this experiment with the real data largely corroborates our analogous experiments with ABM. Each \pPhiDiv\,mechanism, the PTS mechanism, and to some extent the non-parametric $\Phi$-Div: TVD mechanism, are robust against strategic reporting for many or all semesters and strategies, whereas the remaining mechanisms are consistently susceptible to various kinds of strategic behavior.

Unlike in our computational experiments with ABM, though, certain seemingly unlikely strategy profiles (e.g. where nearly every student reports 10 regardless of their signal), do not come into play in these experiments with the real data, since there is only one ``strategic'' agent considered at a time. All of the other students reports (though they may also have resulted from some unknown strategy) are taken as given. As a result, certain strategies like \textit{Report All 10s} are somewhat less potent than in the experiments with ABM.


\subsection{Estimating Ground Truth Scores}
\label{subappendix:estimation-results}
\setcounter{figure}{0}

Here, we provide evidence for the utility of our parameter estimation procedure that is described in \Cref{subsubappendix:implementation-parametric}. 

In the \textit{Continuous Effort, Unbiased Agents} setting with truthfully reporting agents, we simulate the grading of 1000 assignments with 100 submissions each and record the mean squared error of the estimation of the ground truth scores for each of the following two methods:
\begin{enumerate}
    \item \textit{Consensus Grade.} Estimates the true score of a submission as the mean of the graders' reports.
    
    \item \textit{Parameter Estimation Procedure, No Bias (Procedure-NB).} Estimates the true score of a submission using the parameter estimation procedure from \Cref{subsubappendix:implementation-parametric}, but without estimating agent biases. All agent biases are assumed to be 0 and the \textbf{Update} step in which biases are estimated is skipped in each iteration of the procedure.
\end{enumerate}

We do the same in the \textit{Continuous Effort, Biased Agents} setting for each of the following 3 methods:
\begin{enumerate}
    \item \textit{Consensus Grade.} Estimates the true score of a submission as the mean of the graders' reports.
    
    \item \textit{Parameter Estimation Procedure, No Bias (Procedure-NB).} Estimates the true score of a submission using the parameter estimation procedure from \Cref{subsubappendix:implementation-parametric}, but without estimating agent biases. All agent biases are assumed to be 0 and the \textbf{Update} step in which biases are estimated is skipped in each iteration of the procedure.
    
    \item \textit{Parameter Estimation Procedure (Procedure).} Estimates the true score of a submission using the parameter estimation procedure from \Cref{subsubappendix:implementation-parametric} (including estimating agent biases).
\end{enumerate}

The results of both experiments are plotted in \Cref{appendix-fig:estimation}. We find that our estimation procedure improves over the consensus grade in both cases. We also, once again, see the value of modeling the bias of agents in \Cref{appendix-subfig:estimation-bias}.

\begin{figure}
    \centering
    \begin{subfigure}{0.4\textwidth}
    \includegraphics[width=\linewidth]{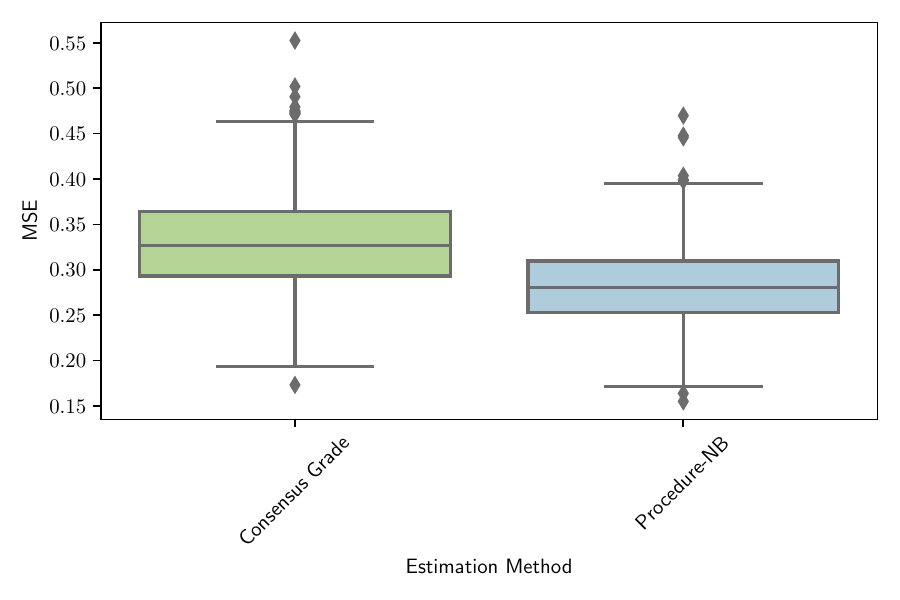}
    \captionsetup{width=0.9\textwidth}
    \caption{\textit{Continuous Effort, Unbiased Agents.}}
    \label{appendix-subfig:estimation-no-bias}
    \end{subfigure}\hfill
    \begin{subfigure}{0.4\textwidth}
     \includegraphics[width=\linewidth]{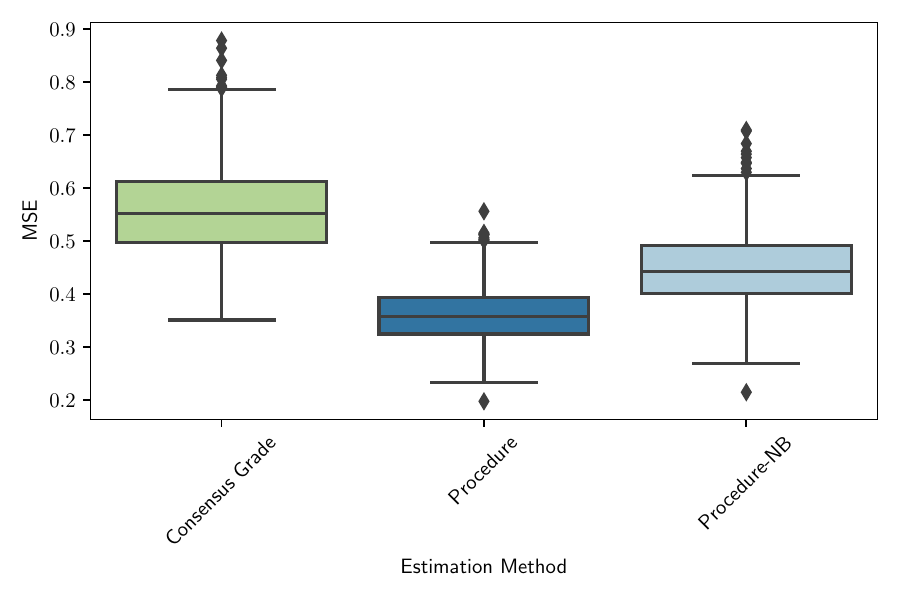}
    \captionsetup{width=0.9\textwidth}
    \caption{\textit{Continuous Effort, Biased Agents.}}
    \label{appendix-subfig:estimation-bias}
    \end{subfigure}
\caption{\textit{Estimating Ground Truth Scores.} Mean squared errors for the estimation of ground truth scores on 1000 assignments with 100 submissions each.}
\label{appendix-fig:estimation}
\end{figure}

\end{document}